\documentclass[12pt,a4paper]{article}
\usepackage{anyfontsize} 
\usepackage{amssymb}
\usepackage{amsmath}
\usepackage{amsfonts}
\usepackage{graphicx}
\usepackage{epstopdf}
\usepackage{pdflscape}
\usepackage{float}
\usepackage{latexsym}
\usepackage{color}
\usepackage[T1]{fontenc}
\usepackage[utf8]{inputenc}
\usepackage{authblk} 
\usepackage{makecell}
\usepackage{float}
\usepackage{titlesec}
\usepackage{comment}
\usepackage{tabularx}
\usepackage{xcolor}
\usepackage{tcolorbox}
\usepackage{algorithm}
\tcbuselibrary{raster}
\usepackage{algpseudocode}
\algrenewcommand{\alglinenumber}[1]{\textbf{Step #1:}}
\usepackage{array}
\newcolumntype{Y}{>{\raggedright\arraybackslash}X}
\newcolumntype{C}{>{\centering\arraybackslash}X}
\newcolumntype{Z}{>{\hsize=1.1\hsize\linewidth=\hsize}X}
\usepackage{float}
\titlelabel{\thetitle.\quad}
\usepackage{placeins}
\usepackage{xcolor,colortbl,booktabs,siunitx}
\usepackage[round]{natbib}

\usepackage{xr-hyper}
\usepackage[
  colorlinks=true,
  linkcolor=blue,
  citecolor=blue,
  urlcolor=blue
]{hyperref}

\usepackage{soul}
\usepackage[round]{natbib}

\definecolor{cadmiumyellow}{rgb}{1.0, 0.96, 0.0}

\definecolor{Highlight}{HTML}{FFF59C}
\newcolumntype{h}{>{\columncolor{Highlight}}c}

\usepackage[margin=0.8in]{geometry}
\def\orcid#1{\kern.08em\href{https://orcid.org/#1}{\includegraphics[keepaspectratio,width=0.7em]{orcid.png}}}

\usepackage{amsthm}
\newtheorem{theorem}{Theorem}[section]
\newtheorem{corollary}{Corollary}[theorem]
\newtheorem{lemma}[]{Lemma}
\newtheorem{result}[]{Result}

\newtheorem{example}{Example}
\theoremstyle{definition}
\newtheorem{definition}{Definition}[]

\theoremstyle{remark}
\newtheorem{remark}{Remark}
\usepackage{setspace}
\title{Power-optimal allocation in multi-arm cluster randomized trials}

\author[]{Muneeb Ahmad Wani}
\author[]{Satya Prakash Singh\thanks{Corresponding author.
Email: \texttt{singhsp@iitk.ac.in}.}}
\affil{Department of Statistics and Data Science, Indian Institute of Technology Kanpur, 208016, India.}

\date{}

\begin{document}

\maketitle
\begin{abstract} 
This paper discusses a power-based framework for optimal allocation in multi-arm cluster randomized trials with continuous or binary outcomes, allowing cluster sizes and within-cluster correlations to differ across treatment groups. We consider optimal cluster allocation with prespecified cluster sizes as well as joint optimization of cluster allocations and sizes. Using generalized estimating equations to account for correlated outcomes and a quasi-score test to assess treatment homogeneity, we derive locally optimal designs, which depend on specified parameter values. Since locally optimal designs can be highly sensitive to changes in the parameter values, we also propose maximin designs, which maximize the minimum power over a prescribed parameter space and provide robust allocations under parameter uncertainty. We analytically characterize the maximin designs and the least favorable configurations of treatment means and provide procedures for determining the sample size required to attain a target power. Applications to the WASH Benefits Bangladesh and SCORE trials illustrate gains in calculated power and reductions in the required sample size relative to the original trial designs. \\

\end{abstract}

\section{Introduction}
Cluster randomized trials (CRTs) are widely used in public health, medicine, education, and community-based research to evaluate interventions when randomization at the individual level is infeasible, undesirable, or susceptible to contamination \citep{Donner_2000}. The natural unit of intervention may be a village, school, health facility, or community. Unlike individually randomized trials, CRTs may generate correlated responses within clusters, and this dependence can substantially affect both the precision of treatment comparisons and the power of hypothesis tests \citep{Breukelen_2015, Singh_2015, Singh_2016, hayes2017cluster, Lewis_2021}. 

At the planning stage, decisions concerning the number of clusters, their sizes, and their allocation across treatment groups are integral components of the design of an efficient CRT. Designing a CRT becomes substantially more challenging when a trial contains more than two treatment groups. Multi-arm CRTs are increasingly used to compare several intervention strategies simultaneously, often with a common control or with multiple active interventions. For example, the Water, Sanitation, and Hygiene (WASH) Benefits Bangladesh trial compared seven intervention conditions and enrolled 720 clusters, with outcomes including both binary measures and continuous measures of child growth \citep{Luby2018}. A similar multi-arm WASH benefits trial was also conducted in Kenya \citep{null2018effects}. Similarly, the Schistosomiasis Consortium for Operational Research and Evaluation (SCORE) studies \citep{Ouattara2022,phillips2020evaluating,phillips2017assessing} have used multi-arm cluster randomized designs to compare alternative strategies for controlling schistosomiasis in several African countries; the Côte d'Ivoire trial of \cite{Ouattara2022}, for example, compared four intervention strategies. The Sanitation Hygiene Infant Nutrition Efficacy (SHINE) trial \citep{team2015sanitation}, \cite{humphrey2019independent} is a multi-arm CRT to evaluate intervention conditions such as the independent and combined effects of improved water, sanitation, hygiene, and complementary feeding. These studies illustrate that multi-arm CRTs are not merely theoretical extensions of two-arm trials but arise naturally in contemporary intervention research. These examples also show that such trials can involve appreciable differences in both cluster sizes and arm-specific intracluster correlations (ICCs). We shall re-analyze the WASH benefits Bangladesh trial and SCORE Côte d'Ivoire trial in the numerical section. 

A conventional approach to designing a multi-arm CRT is to allocate approximately equal numbers of clusters to the treatment groups. Although attractive for its simplicity and transparency, such a balanced allocation may not always be statistically efficient. The information contributed by a cluster depends not only on the number of individuals it contains but also on the extent of within-cluster correlation. Thus, treatment arms with different cluster sizes or different ICCs can contribute substantially different amounts of information per cluster. Heterogeneity in ICCs across treatment arms is well documented in the methodological literature; \cite{crespi2009using}, for example, developed a second-order GEE framework specifically to accommodate heterogeneous ICCs in CRTs. Treatment-dependent ICCs have subsequently been incorporated into sample-size and power calculations for binary CRTs \citep{wu2017maximin, kennedy2022power}. The possibility that ICCs may also depend on cluster size further weakens the practical justification for treating all clusters as statistically equivalent.

A substantial literature has considered optimal allocation in two-arm CRTs. For continuous outcomes, \cite{candel2015sample} considered optimal number of clusters when cluster sizes, variances, and ICCs differ between treatment arms. \citet{van2018efficient, van2021maximin} developed robust designs under treatment-dependent costs and variances, while \citet{copas2021optimal} considered power-optimal allocation of clusters and individuals when allocation and cluster size may differ between the two arms. Related developments for binary outcomes include \citet{wu2017maximin, singh2023some, kennedy2022power}. Much of this literature, however, formulates optimality in terms of efficient estimation of a treatment effect, typically through minimization of its variance or maximization of an efficiency measure.

For a trial whose primary objective is hypothesis testing, a power based optimality criterion should be used. \citet{Baldi2021} studied optimal designs for testing equality of treatment means in heterogeneous, independent treatment groups using a Wald-type test. For correlated observations, \citet{LiuLiang1997}  developed the quasi-score test and corresponding sample-size methodology for comparing two groups. These results suggest a natural framework for constructing power-oriented designs for multi-arm CRTs, but they do not provide a general solution when treatment arms differ simultaneously in cluster size and within-cluster correlation. This article develops such a framework. We consider a parallel-group CRT with multiple treatment arms under a marginal GEE model \citep{LiangZeger1986} and formulate the design problem directly in terms of the power of the test of treatment homogeneity. We allow the ICC and cluster size to vary across treatment arms, thereby accommodating heterogeneous information contributions that arise naturally in multi-arm CRTs. The resulting noncentrality parameter provides a unified criterion for studying allocation efficiency for both continuous and binary outcomes. In contrast to approaches based solely on treatment-effect variance, our formulation targets the inferential objective of the trial directly: maximizing the ability to detect departures from the global null hypothesis.

We consider two complementary approaches to this problem. First, we derive locally optimal designs that maximize power for specified values of the treatment effects, cluster sizes, and ICCs. Such designs provide useful insight into how the optimal allocation responds to heterogeneity in the information content of different treatment arms. Their practical limitation, however, is that the treatment effects required to construct them are generally unknown at the design stage. This sensitivity is particularly consequential in multi-arm settings, where a locally optimal design can concentrate most or all clusters on a subset of treatment groups and can consequently perform poorly when the anticipated treatment effects are misspecified.

Second, we develop maximin designs that protect against uncertainty in the treatment effects by maximizing the minimum power over a prespecified parameter space. This provides a robust alternative to locally optimal allocation and is especially relevant when reliable prior information about the magnitude and pattern of treatment effects is unavailable. We further extend the framework to allow optimization of cluster sizes in addition to the allocation of clusters among treatment arms. Thus, the proposed methodology treats cluster allocation and cluster size as complementary design variables rather than addressing them separately.

The practical importance of the proposed approach is illustrated using real multi-arm CRTs. In particular, the WASH Benefits and SCORE studies provide settings in which the treatment arms differ in observed outcomes, cluster sizes, and estimated ICCs. Reconsidering these studies from a design perspective demonstrates how departures from balanced allocation can translate into gains in power and reductions in the number of clusters required to achieve a prespecified power. For example, the analysis of the WASH Benefits trial in the manuscript shows that allowing both allocation proportions and cluster sizes to vary can improve the power relative to the original allocation, while the SCORE application illustrates the potential gains from an unequal allocation when the information contributions of the treatment arms differ.

The paper is organized as follows. The model formulation along with the hypothesis setup is presented in Section 2. Various design criteria are introduced in Section 2. In Section 3 we present the locally and maximin optimal designs for the continuous response. In Section 4 we present the counterpart optimal designs in case of a binary response. In Section 5 several real life applications of the proposed designs are presented both for continuous and binary response. Finally, we conclude with Section 6, where we present some conclusions from the present work and provide directions for the future work. 

\section{Model formulation and design criterion}
\subsection{Model} Suppose we have $K+1$ treatments in a cluster randomized trial and let $\pi_{\ell}$ denote the proportion of clusters assigned to the $\ell$--th treatment, where $\ell\in\mathcal{S} = \{0,1,\hdots,K\}$. Conventionally, $\ell=0$ represents a control group. Let $y_{ij}$ be the response of the $j$-th unit in the $i$-th cluster, $j=1,2,\ldots,n_{i}$, $i=1,2,\ldots, m$. Here, $n_i$ is the size of the $i$-th cluster and $ m$ is the total number of clusters. Also, let $\boldsymbol{x}_{i}$ be the treatment-specific indicator vector, i.e., $\boldsymbol{x}_{i}=\boldsymbol{e}_{0}^{\rm T} = (0,\ldots,0)$ for clusters receiving the control and $\boldsymbol{x}_{i}=\boldsymbol{e}_{\ell}^{\rm T}$ for clusters receiving the $\ell$--th treatment, where $\boldsymbol{e}_1,\ldots,\boldsymbol{e}_K$ represent the standard basis of $\mathbb{R}^K$. Defining the marginal model with $\boldsymbol{x}_{i}$ as the only covariate 
\begin{align}\label{model.1}
g(\mathbb{E}(y_{ij}|\boldsymbol{x}_{i})) =\lambda + \boldsymbol{x}_{i}\boldsymbol{\psi},\qquad j=1,2,...,n_{i}, i=1,2,...,m 
\end{align}
where $g(\cdot)$ is the link function that associates the expected value of $y_{ij}$ with the covariate $\boldsymbol{x}_{i}$, $\lambda\in \mathbb{R}$ represents the control effect (nuisance parameter by the terminology of \cite{LiuLiang1997}) and $\boldsymbol{\psi}=\left(\psi_{1},\psi_{2},...,\psi_{K}\right)^{\rm T} \in \mathbb{R}^K$, with $\psi_{\ell}$ denotes the effect of the $\ell$--th treatment ($\ell\neq0$). Let $\boldsymbol{\theta}=(\theta_{0},\theta_{1},\ldots,\theta_{K})^{\rm T}\in \mathbb{R}^{K+1}$ be the vector of treatment means with $\theta_{\ell}$ being the expected value of the observation in the $
\ell$-th treatment arm, i.e, $\theta_{\ell}=E(y_{ij}|x_{i}=e_{\ell}).$ Suppose that the purpose of the study is to test the hypothesis of homogeneity of the treatment means. This hypothesis can be expressed as 
\begin{equation} \label{hyp.1}
 H_{0}: \boldsymbol{\theta}\in \mathbf{\Theta}_0\quad \text{versus}\quad H_{1}: \boldsymbol{\theta}\in \mathbf{\Theta}_1,
\end{equation}
where $\mathbf{\Theta}_{0}=\{\boldsymbol{\theta}\in \mathbb{R}^{K+1}: \theta_0=\theta_1=\theta_2=\cdots=\theta_K\},$ $\mathbf{\Theta}_1=\mathbf{\Theta}\setminus \mathbf{\Theta}_0,$ and $\mathbf{\Theta}=\{\boldsymbol{\theta}\in \mathbb{R}^{K+1}\}.$
Note, we can directly write the model in terms of treatment means as $g(\theta_{0})=\lambda$ and $g(\theta_{\ell})=\lambda+\psi_{\ell}, \ell=1,2,\hdots,K$. Equivalently, the hypothesis in \eqref{hyp.1} can be written as 
\begin{equation}\label{hyp.2}
\boldsymbol{H}_{0}: \boldsymbol{\psi}=\boldsymbol{\psi}^{(0)} \quad \text{versus}\quad \boldsymbol{H}_{1}: \boldsymbol{\psi}=\boldsymbol{\psi}^{(1)}
\end{equation}
where $\boldsymbol{\psi}^{(0)}=\boldsymbol{0}$ and $\boldsymbol{\psi}^{(1)}\neq\boldsymbol{0}$.

Recently, \cite{Baldi2021} developed optimal designs for studies whose primary objective is to test the equality of treatment means in an ANOVA framework using the Wald test. In the context of generalized linear models with correlated observations, \cite{LiuLiang1997} proposed the Quasi-Score Test (QST) statistic for hypothesis testing and derived sample size formulae for two-arm studies comparing a control and an intervention group. In this work, we employ the QST to test the hypothesis of homogeneity of treatment means and derive optimal allocations of clusters across multiple treatment arms. The QST is based on the Generalized Estimating Equations (GEE) framework, which avoids the need for full likelihood specification, and the correlation among observations within the same cluster is modelled through a working correlation structure\citep{LiangZeger1986}. 

Let $\boldsymbol{y}_i = (y_{i1}, \ldots, y_{in_i})^{\rm T}$ be a vector of correlated responses from the $i$--th cluster and $V_{i}$ be the covariance matrix of $\mathbf{y}_{i}$. The true covariance matrix of $\mathbf{y}_i$ is generally unknown, in the GEE approach, we typically write the covariance matrix $V_{i}$ as $V_i=A_{i}^{1/2}R(\rho_{i})A_{i}^{1/2}$, where $A_{i}$ is the diagonal matrix of variances and $R(\boldsymbol{\rho}_{i})$ is the called working correlation matrix for the $i$-th cluster, which depends on the association parameter(s) $\rho_{i}$. The association parameter(s) $\rho_{i}$ represents the measure of similarity between responses in the $i$-th cluster. Following \cite{kennedy2022power}, we refer to the association parameters(s) $\rho_{i}$ as the ICC(s) throughout the paper. 
Furthermore, we assume that the ICCs and cluster sizes are the same for clusters within any treatment arm, but may vary between clusters in different treatment arms, i.e. $n_{i}=n_{\ell}$ and $\rho_{i}=\rho_{\ell}$ if the $i$-th cluster is allocated to the $\ell$-th treatment arm. \cite{crespi2009using} and \cite{omar2000analysis} discussed the implications of this assumption in detail. \cite{singh2023some,wu2017maximin,kennedy2022power} also assumed treatment dependent ICCs. Furthermore, it was noted by \cite{crespi2009using} that the ICCs may also depend on the cluster sizes; however, most of the design work on CRTs assumes an equal cluster size for all clusters, which may not hold in general. We relax this assumption, by assuming the treatment dependent cluster sizes. 

The QST statistic follows a (central) $\chi^2$ distribution with $K$ degrees of freedom (df) under the null in \eqref{hyp.2} as the number of clusters '$\mathrm{m}$' goes to infinity, whereas under the alternative it follows a non-central $\chi^{2}_{K}(\phi)$ distribution asymptotically with $K$ degrees of freedom (df) and $\phi$ being the non-centrality parameter (NCP) \citep{LiuLiang1997,CoxHinkley1974,SelfMauritsen1988}. Note, that the NCP $\phi$ is context-specific; its form depends on the underlying data distribution and will be defined explicitly in each setting where it is required. 

\subsection{Design criteria}

Before formally defining the design criteria, we introduce the main parameter vectors that play a crucial role in the design selection procedure.
The first is the vector of the cluster allocation proportions $\boldsymbol{\pi}=(\pi_0,\ldots,\pi_K)^{\rm T}$, where $\pi_\ell$
represents the probability that $\boldsymbol{x}_i$ is equal to
$\boldsymbol{e}_\ell^{\rm T}$ \citep{LiuLiang1997}. The second is
the vector of arm-specific response means
$\boldsymbol{\theta}=(\theta_0,\theta_1,\ldots,\theta_K)^{\rm T}
\in\mathbb{R}^{K+1}$.
For binary responses, $\theta_\ell$ denotes the response probability in $\ell$--th arm. We also write
$\boldsymbol{n}=(n_0,\ldots,n_K)^{\rm T}$ for the vector of cluster sizes, where $n_\ell$ is the common cluster size within the
$\ell$--th treatment arm. By an optimal design, we mean a choice of cluster allocations,
and possibly cluster sizes, that optimizes an appropriate design
criterion. We consider two cases. In the first case, the vector of cluster
sizes $\boldsymbol{n}$ is specified in advance, and only the
allocation proportions $\pi_{\ell}$'s are obtained optimally. 
The corresponding design space is $\Pi=\left\{\boldsymbol{\pi}:\sum_{\ell=0}^{K}\pi_\ell=1,\quad\pi_\ell\geq 0,\quad \ell=0,\ldots,K\right\}$.
In the second case, both the cluster allocation proportions 
and the cluster sizes are optimized. The cluster sizes are treated as continuous variables during optimization, as in \citet{liu2017optimal,liu2023overview}. We also impose an additional constraint
$\sum_{\ell=0}^{K}\pi_\ell n_\ell=\eta$, where $\eta=N/m$ is the average cluster size and $N$ is total number of
observations. The joint design space in this case is $\Xi=\left\{(\boldsymbol{\pi},\boldsymbol{n}):\boldsymbol{\pi}\in(0,1)^{K+1},\quad
\boldsymbol{n}\in(0,\infty)^{K+1},\quad\sum_{\ell=0}^{K}\pi_\ell=1,\quad
\sum_{\ell=0}^{K}\pi_\ell n_\ell=\eta
\right\}.$ Unless stated otherwise, no specific structure is assumed for the arm-specific working correlation matrices when cluster sizes are fixed. However, for joint optimization of cluster allocation proportions and cluster sizes, we assume an exchangeable working correlation structure throughout. Since our primary interest lies in experimental designs tailored to hypothesis testing, we adopt a power-based design criterion. Power-based optimality criteria have been considered by
\citet{Singh_2019} and \citet{ASO_2025_SJS} for multiple
treatment comparisons with normal responses, and by
\citet{singh_JSPI_2026} for binary response data.
For fixed degrees of freedom and type I error, the power function of the QST test statistic depends on the design and model parameters only through the
NCP. Furthermore, this power is a monotone function of the NCP. Therefore, we shall use the NCP as the
design criterion. Note that, the NCP depends on $\boldsymbol{\pi}$,
$\boldsymbol{\theta}$, $\boldsymbol{n}$, the total number of
clusters $m$, and the vector of ICCs
$\boldsymbol{\rho}=(\rho_0,\ldots,\rho_K)^{\rm T}$, where
$\rho_\ell$ denotes the common ICC within the $\ell$-th treatment
arm. Consequently, we shall use the notation
$\phi(\boldsymbol{\pi},\boldsymbol{\theta},m,
 \boldsymbol{n},\boldsymbol{\rho})$
for the NCP to make these dependencies explicit.
We shall use two types of design criteria: (i) local
optimality and (ii) maximin optimality. Each is considered with
either prespecified cluster sizes or joint optimization of cluster
allocations and cluster sizes, giving the four definitions below.

\begin{definition}
A locally optimal design (LOD) $\boldsymbol{\pi}_{\rm LOD}$ with prespecified cluster sizes is obtained by maximizing
the NCP with respect to $\boldsymbol{\pi}$ for given
$(\boldsymbol{\theta},m,\boldsymbol{n},\boldsymbol{\rho})$; that is,
\begin{equation}
\label{Def.LOD}
\boldsymbol{\pi}_{\rm LOD}=
\arg\max_{\boldsymbol{\pi}\in\Pi}
\phi(\boldsymbol{\pi},\boldsymbol{\theta},m,
 \boldsymbol{n},\boldsymbol{\rho}).
\end{equation}
\end{definition}

\begin{definition}
A locally optimal design (LOD) $(\boldsymbol{\pi}_{\rm LOD},\boldsymbol{n}_{\rm LOD})$ with jointly optimized cluster allocations and cluster sizes
is obtained by maximizing the NCP with respect to
$\boldsymbol{\pi}$ and $\boldsymbol{n}$ for given
$(\boldsymbol{\theta},m,\boldsymbol{\rho})$
and average cluster size $\eta$; that is,
\begin{equation}
\label{Def.LOD.joint} 
(\boldsymbol{\pi}_{\rm LOD},\boldsymbol{n}_{\rm LOD})=
\arg\max_{(\boldsymbol{\pi},\boldsymbol{n})\in\Xi}
\phi(\boldsymbol{\pi},\boldsymbol{\theta},m,
 \boldsymbol{n},\boldsymbol{\rho}).
\end{equation}
Here, the joint design space $\Xi$ incorporates the constraint
$\sum_{\ell=0}^{K}\pi_\ell n_\ell=\eta$.
\end{definition}
The LOD is usually not preferred due to its sensitivity with the change of parameter values $\boldsymbol{\theta}$. Example~\ref{example.2} illustrates this sensitivity and more examples illustrating the drawbacks of LODs can be found in \citet{ASO_2025_SJS} and
\citet{singh_JSPI_2026}. Another reasonable approach is to find an optimal design that optimizes the design criterion under the worst case scenario of the unknown parameters ($\boldsymbol{\theta}$ in our settings). This approach leads to a maximin (MM) design, which maximizes
the minimum NCP over a prespecified set of alternative
treatment-mean vectors $\Delta_{\delta}$. As with local optimality, we consider both prespecified cluster
sizes and joint optimization of cluster allocations and
cluster sizes.

\begin{definition} For given $(m,\boldsymbol{n},\boldsymbol{\rho})$, a
maximin (MM) design $\boldsymbol{\pi}_{\rm MM}$ satisfies the following condition 
\begin{equation}\label{Eq.MM.design}
\phi(\boldsymbol{\pi}_{\rm MM},\boldsymbol{\theta}_{\rm LFC},m,\boldsymbol{n},\boldsymbol{\rho})
=
\max_{\boldsymbol{\pi}\in\Pi}\;
\min_{\boldsymbol{\theta}\in\mathbf{\Delta}_{\delta}}
\phi(\boldsymbol{\pi},\boldsymbol{\theta},m,\boldsymbol{n},\boldsymbol{\rho}),
\end{equation}
where $\boldsymbol{\theta}_{\rm LFC}$ denotes the \emph{least favorable configuration} (LFC) of the parameter vector $\boldsymbol{\theta}$.
\end{definition}

\begin{definition}
For given $(m,\boldsymbol{\rho})$ and average cluster size
$\eta$, a maximin (MM) design $(\boldsymbol{\pi}_{\rm MM},\boldsymbol{n}_{\rm MM})$ satisfies the following condition 
\begin{equation} 
\phi(\boldsymbol{\pi}_{\rm MM},\boldsymbol{\theta}_{\rm LFC},m,\boldsymbol{n}_{\rm MM},\boldsymbol{\rho})=\max_{(\boldsymbol{\pi},\boldsymbol{n})\in\Xi}\;
\min_{\boldsymbol{\theta}\in\mathbf{\Delta}_{\delta}}
\phi(\boldsymbol{\pi},\boldsymbol{\theta},m,\boldsymbol{n},\boldsymbol{\rho}),
\end{equation}
\end{definition}

The specification of $\Delta_{\delta}$ depends on the
particular problem under consideration and will be defined
separately for each case.
MM--designs have been explored by
\citet{Singh_2019} for ordered experiments,
and by \citet{ASO_2025_SJS} and
\citet{singh_JSPI_2026} for multiple treatment comparisons
with normal and binary responses, respectively.
\section{Designs for continuous responses}
 Optimal designs have been derived for the simple control versus intervention CRT in case of continuous response. For fixed but unequal cluster sizes in the control and intervention arms, \citet{candel2015sample} derived the optimal numbers of clusters under heterogeneous ICCs and variances by minimizing the variance of the treatment effect estimator within a linear mixed-model framework. \cite{van2018efficient} derived the robust maximin designs for cluster randomized trials in case of continuous outcome under treatment dependent costs and treatment dependent variances by maximizing the minimum efficiency of treatment effects estimator. The resultant designs are generalizations of the designs proposed by \cite{liu2003statistical} which assume equal cluster sizes in both arms. Subsequently \cite{van2021maximin} obtained the MM--designs by using the relative efficiency as design criterion rather than the efficiency criterion used by \cite{van2018efficient}. While the aforementioned designs primarily focus on the efficient estimation of the treatment effect, \cite{copas2021optimal} recently proposed power-optimal designs for determining the optimal allocation of clusters and individuals across the control and intervention arms of a CRT by maximizing the power function based on normal distribution.
To our knowledge, no one has explored the optimal designs in terms of optimal allocation of clusters in different arms when the objective of the study is hypothesis testing instead of the estimation of the parameters in a multi-arm CRT. This work bridges this gap, wherein we propose optimal allocation of clusters in a multi-arm CRT by considering the 
treatment dependent cluster sizes and ICCs.

Using the identity link function for the continuous response, the marginal model in \eqref{model.1} takes the form 
\begin{equation}
\mathbb{E}(y_{ij}|\boldsymbol{x}_{i}) =\lambda + \boldsymbol{x}_{i}^{\rm T}\boldsymbol{\psi},\qquad j=1,2,...,n_{i},\;i=1,2,...,m. 
\end{equation}
Following \cite{LiuLiang1997}, we also assume homoscedastic outcomes and let $\sigma^2$ denote the common variance in different treatment arms. Consequently, the expression of the NCP in this case is obtained as

\begin{align}\label{cont.ncp}
\phi(\boldsymbol{\pi},\boldsymbol{\theta},m,\boldsymbol{n},\boldsymbol{\rho})
&= \frac{m}{\sigma^2}\left\{\sum\limits_{\ell=0}^{K-1}\pi_{\ell}r_{\ell}(\theta_{\ell}-\theta_K)^2-\frac{\left[\sum\limits_{\ell=0}^{K-1}\pi_{\ell}r_{\ell}(\theta_{\ell}-\theta_{K})\right]^{2}}{\sum\limits_{\ell=0}^{K}\pi_{\ell}r_{\ell}}\right\} 
\end{align}
where, $r_{\ell}=\boldsymbol{1}_{n_{\ell}}^{\rm T}\mathrm{R}(\boldsymbol{\rho}_\ell)^{-1}\boldsymbol{1}_{n_{\ell}}$ and $\mathrm{R}(\boldsymbol{\rho}_\ell)$ is the working correlation matrix for the clusters in the $\ell$-th treatment arm. 

The NCP in \eqref{cont.ncp} generalizes the NCP obtained in \cite{LiuLiang1997} for two groups with equal cluster sizes and ICCs in both arms to $K+1$ groups for unequal cluster sizes and ICCs. Further, the NCP in \eqref{cont.ncp} is a generalization of the NCP obtained by \cite{Baldi2021} for the Wald's test in a one-way-ANOVA setting with heterogeneous uncorrelated groups. 

\subsection{Locally optimal designs}
We first consider the locally optimal design (LOD). For the one-way ANOVA model with homogeneous treatment groups and a common variance, the LOD is a two-point design that allocates all observations to the pair of treatment groups whose mean responses are maximally separated \citep{Singh_2019}, \cite{ASO_2025_SJS}, \cite{singh_JSPI_2026}. In the heterogeneous case, \cite{Baldi2021} derived a two-point design in which the selected treatment groups are determined by a condition involving both the mean and variance parameters. Using a similar approach, we have derived the LOD for the present model, as stated in the following theorem.
\begin{theorem}\label{cont.lod.1}
 Suppose that 
\begin{equation}\label{Eq.LOD.Cond}
 \frac{(\theta_{s}-\theta_{t})^2}{\left(\sqrt{1/r_s}+\sqrt{1/r_t}\right)^2}=\underset{i\neq j}{\max}\left\{\frac{(\theta_{i}-\theta_{j})^2}{\left(\sqrt{1/r_i}+\sqrt{1/r_j}\right)^2}\right\}. 
\end{equation}
Then the components of the LOD $(\boldsymbol{\pi}_{\rm LOD})$ are given by
\begin{flalign}
\pi_{s}=\frac{\sqrt{1/r_{s}}}{\sqrt{1/r_{s}}+\sqrt{1/r_t}},\; \pi_{t}=1-\pi_{s},
\;\text{and}\; \pi_{\ell}=0, \forall \quad \ell\in\mathcal{S}\setminus\{s,t\} \label{cont.lod.eq.1}
\end{flalign}
If more than one pair satisfies condition \eqref{Eq.LOD.Cond}, then any convex combination of the corresponding two-point LODs is also an LOD. 
 \end{theorem}
The proof follows that of Theorem~2.1 in \citet{Baldi2021},
with the arm-specific variances adjusted to account for the
corresponding factors $r_\ell$.
 \begin{corollary}\label{cont.lod.corollary.1}
 When $r_\ell=r$ for all $\ell\in \mathcal{S}$ and $\theta_{0}\leq\theta_{1}\leq... \leq\theta_{K}$, with at least one strict inequality, then $\boldsymbol{\pi}_{\rm LOD}$ is given by 
\begin{flalign*}
\pi_{0}=\pi_{K}=1/2, \;\text{and}\; \pi_{\ell}=0, \forall \; \ell\in\mathcal{S}\setminus\{0,K\}.
\end{flalign*}
\end{corollary}
The increasing order of the mean parameters considered in Corollary \ref{cont.lod.corollary.1} is a common feature of many dose--response studies in practice testing for trends \citep{Terpstra_2003}, \cite{ASO_2025_SJS}. In this scenario an LOD is a two-point design allocating equal proportion of clusters to the extreme groups.

In certain scenarios, it is also of interest to optimally allocate the cluster sizes, rather than treating them as fixed quantities \citep{van2021maximin}. In this case, the LOD is obtained by maximizing the NCP with respect to both $\boldsymbol{\pi}$ and $\boldsymbol{n}$ over the design space $\Xi$. The following theorem presents the LOD for a simple two group comparison 

\begin{theorem}\label{cont.lod.2}
For two experimental groups (control versus treatment, K = 1), the components of LOD, $(\boldsymbol \pi_{LOD},\boldsymbol n_{LOD})=(\pi_0,\pi_1,n_0,n_1)$ are given by
\begin{eqnarray}
\begin{aligned}
\pi_{0}&=\frac{\sqrt{\rho_{0}}}{\sqrt{\rho_{0}}+\sqrt{\rho_{1}}},\; \qquad \pi_{1}=1-\pi_{0}, \\
n_{0}&=\frac{\eta\sqrt{1-\rho_{0}}}{\pi_{0}(\sqrt{1-\rho_{0}}+\sqrt{1-\rho_1})},\;\;and \;\; n_{1}=\frac{\eta\sqrt{1-\rho_{1}}}{\pi_{1}(\sqrt{1-\rho_{0}}+\sqrt{1-\rho_1})}.\label{cont.lod.eq.2}
\end{aligned}
\end{eqnarray}
 \end{theorem}
 \begin{remark}
The LOD in \eqref{cont.lod.eq.2} coincides with the unconstrained optimal design of \cite{copas2021optimal} under the linear mixed-effects model. As noted by \cite{copas2021optimal}, a higher ICC in the intervention arm leads to allocation of more clusters but fewer individuals per cluster to that arm, reflecting the greater value of increasing the number of clusters when the ICC is high.
\end{remark}

 Clearly, $n_{\ell}$ obtained using the Theorem \ref{cont.lod.2} can be a non-integer. \cite{liu2017optimal} and \cite{Singh_eff_2019} provide efficient algorithms to transform this non-integer value to an integer value. For general $K$, obtaining analytic expression of the locally optimal design is extremely difficult, and numerical algorithms can be used to jointly optimize the NCP in the restricted design space $\Xi$. 
 
\begin{example} Consider a CRT with 5 treatment groups. Assume an exchangeable correlation structure of the working correlation matrix. Let the vector of ICCs $\boldsymbol{\rho}=(0.1,0.05,0.02,0.2,\\0.005)^{\rm T}$, vector of cluster sizes $\boldsymbol{n}=(30,25,20,25,23)^{\rm T}$, and the vector of treatment means $\boldsymbol{\theta}=(10,9,8,7,12)^{\rm T}$. It can be easily verified that, 
\begin{flalign*}
\underset{i\neq j}{\max}\left\{\frac{(\theta_{i}-\theta_{j})^2}{\left(\sqrt{1/r_i}+\sqrt{1/r_j}\right)^2}\right\} =\frac{(\theta_{2}-\theta_{4})^2}{\left(\sqrt{1/r_2}+\sqrt{1/r_4}\right)^2}.
\end{flalign*}
Therefore, using Theorem \ref{cont.lod.1}, the locally optimal design will be $\boldsymbol{\pi}_{\rm LOD}=(0,0,0.545,0,0.455)$.
\end{example}
\begin{example}{\label{example.2}}
Consider a CRT with four treatment arms. Suppose that the ICCs and the cluster sizes are same for all arms and let $\rho=0.05$ and n=25 be the common ICC and cluster size. Also, suppose that $\sigma^2=1$. Let the vector true response means is $\boldsymbol{\tilde{\theta}}=(5,5,4,6)$, then by the Corollary \ref{cont.lod.corollary.1}, the locally optimally design is $\boldsymbol{\tilde{\pi}}=(0,0,1/2,1/2).$ However, if the vector of true response means is $\boldsymbol{\theta^{*}}=(4,6,5,5)$ then the locally optimal design is $\boldsymbol{\pi^*}=(1/2,1/2,0,0)$. Observe that 

\begin{align*}
\phi(\boldsymbol{\tilde{\pi}},\boldsymbol{\tilde{\theta}},m,\boldsymbol{n},\boldsymbol{\rho})=\phi(\boldsymbol{\pi^{*}},\boldsymbol{\theta^{*}},m,\boldsymbol{n},\boldsymbol{\rho})= 11.36*m 
\end{align*} 
where m is the total number of clusters. It is trivial that the power of the test in both the cases increases to unity as the total number of clusters goes to infinity. Nonetheless, 
\begin{align*}
 \phi(\boldsymbol{\tilde{\pi}},\boldsymbol{\theta^{*}},m,\boldsymbol{n},\boldsymbol{\rho})=\phi(\boldsymbol{\pi^{*}},\boldsymbol{\tilde{\theta}},m,\boldsymbol{n},\boldsymbol{\rho})= 0 
\end{align*} 
irrespective of the total number of clusters $m$. From above equality, it can be concluded that the power of a test on design $\boldsymbol{\tilde{\pi}}$ reduces to the nominal significance level when the vector of true response means is $\boldsymbol{\theta^*}$ and likewise, the power of a test on the design $\boldsymbol{\pi^*}$ also reduces to the nominal significance level when the vector of true response means is $\boldsymbol{\tilde{\theta}}$. This demonstrates that locally optimal designs can perform very badly if the unknown parameters are not specified correctly. Furthermore, these response means are seldom known at the designing stage, therefore it is reasonable to find a design which performs sufficiently well without needing to specify the vector of response means $\boldsymbol{\theta}$. Next, we will discuss the Maximin approach for constructing the optimal designs for CRTs which do not require the prior specification of mean vector $\boldsymbol{\theta}$.
\end{example}
\subsection{Maximin designs}
The MM--design is obtained by first minimizing the objective function with respect to the
parameter $\boldsymbol{\theta}$ in the restricted space $\mathrm{\Delta}_{\delta}$ and then finding the design that will maximize this minimum value of the objective function. Following \cite{Singh_2019} and \cite{Singh_2020_EJS} the constrained space $\mathrm{\Delta}_{\delta}$ for the continuous responses is defined as
\begin{equation*}
 \mathrm{\Delta}_{\delta} = \{\boldsymbol{\theta}:\max_{i,j\in\mathcal{S}}|\theta_i-\theta_j|\geq \delta>0\}.
\end{equation*}

 The maximin problem defined in~\eqref{Eq.MM.design} can be formulated as a two-player zero-sum game where the experimenter chooses an allocation $\boldsymbol\pi\in\Pi$, which maximizes the NCP, while nature chooses a parameter configuration
$\boldsymbol\theta\in\mathbf{\Delta}_{\delta}$, which minimizes the NCP. For a detailed exploration of solving the maximin problems using game-theoretical approach, the reader may refer to \citet{singh2023some,ASO_2025_SJS,Singh_2020_EJS}. Here, we consider a mixed game in which nature assigns
probabilities called least favorable distribution (LFD) to the LFCs, while the experimenter chooses a
design that maximizes the resulting average NCP. The following theorem identifies the maximin design, the set of LFCs and the corresponding set of LFDs for solving the maximin problem given in equation~\eqref{Eq.MM.design} under fixed cluster sizes.
\begin{theorem}
\label{cont.mm.theorem.1.1} For $K\ge 1$, suppose that $m$, $\boldsymbol n$, $\boldsymbol{\rho}$ are fixed. Let $A=\sum\limits_{\substack{\ell=0}}^{K}r_\ell^{-1}$ and $A_\ell= A-r_{\ell}^{-1}$,\; $\ell \in \mathcal{S}$. Then the following statements hold.\\

\noindent\textbf{Case 1.} Suppose that $r_\ell^{-1}\leq A_{\ell},
\quad \forall \ell\in\mathcal S.$ Then the components of the maximin design $\boldsymbol{\pi}_{\rm MM}$ are given by \begin{equation}
\label{cont.mm.eqn.1.case1}
\pi_\ell
=r_{\ell}^{-1}/A,
\qquad \ell\in\mathcal S,
\end{equation} and the set of LFCs is given by $
\Theta_{\rm LFC}
=
\left\{
\boldsymbol\theta^{(ij)},\;-\boldsymbol\theta^{(ij)}
:\,i< j,\; i,j \in \mathcal{S}
\right\},
$ where the $\ell$th component of $\boldsymbol\theta^{(ij)}$ is
\begin{equation}
\label{cont.lfc.eqn.1.case1}
\theta_\ell^{(ij)}
=
\begin{cases}
\delta/2, & \ell=i,\\
-\delta/2, & \ell=j,\\
0, & \ell\notin\{i,j\},
\end{cases}
\end{equation}
with the set of LFDs given by 
\begin{equation}
\label{cont.lfd.eqn.1.case1}
\mathcal{Q}_{\rm LFD}
=
\left\{
\boldsymbol q=(q_{ij})_{i<j}:
\quad q_{ij}\geq0,\,
\displaystyle\sum_{0\leq i<j\leq K}q_{ij}=1,\,
\displaystyle
\sum_{i<\ell}q_{i\ell}
+
\sum_{j>\ell}q_{\ell j}
=2r_\ell^{-1}/A,\,
\quad \ell\in\mathcal S
\right\}.
\end{equation}
where $q_{ij}$ is the total probability of vector $\boldsymbol{\theta^{(ij)}}$ and its negative $-\boldsymbol{\theta^{(ij)}}$, and may be divided arbitrarily between $\boldsymbol\theta^{(ij)}$ and $-\boldsymbol{\theta}^{(ij)}$.\\

\noindent\textbf{Case 2.} Suppose that $r_s^{-1}>A_{s}$, for some $s \in \mathcal{S}$, then the components of the maximin design $\boldsymbol{\pi}_{\rm MM}$ are given by
\begin{equation}
\label{cont.mm.eqn.1.case2}
\pi_s
=
\frac{\sqrt{r_s^{-1}}}
{\displaystyle \sqrt{r_s^{-1}}+\sqrt{A_s}},
\qquad\text{and}\quad
\pi_\ell
=
\frac{r_{\ell}^{-1}}
{\displaystyle \sqrt{A_s}\left(\sqrt{r_s^{-1}}+\sqrt{A_s}\right)},
\quad \ell \in \mathcal{S}\setminus\{s\}.
\end{equation}
The set of LFCs is $\boldsymbol{\Theta}_{\rm LFC}
=
\left\{
\boldsymbol\theta^{(sj)},\,-\boldsymbol\theta^{(sj)}:\;
j\in \mathcal{S}\setminus\{s\}
\right\}$, where the $\ell$th component of $\boldsymbol\theta^{(sj)}$is
\begin{equation}
\label{cont.lfc.eqn.1.2}
\theta_\ell^{(sj)}
=
\begin{cases}
\frac \delta2,
& \ell=s,\\
-\frac \delta2,
& \ell=j,\\
\frac{\delta}{2}
\frac{\sqrt{A_s}-\sqrt{r_s^{-1}}}
{\sqrt{A_s}+\sqrt{r_s^{-1}}},
& \ell\notin\{s,j\},
\end{cases}
\end{equation}
and the set of LFDs is
\begin{equation}
\label{cont.lfd.eqn.1.case2}
\mathcal{Q}_{\rm LFD} =
\left\{
\boldsymbol q=(q_{sj})_{s\neq j}:
q_{sj}
=r_j^{-1}/A_s, j\in \mathcal{S}\setminus\{s\}
\right\}.
\end{equation}
where $q_{sj}$ is the total probability of vector $\boldsymbol{\theta}^{sj}$ and its negative -$\boldsymbol{\theta}^{sj}$, and may be divided arbitrarily between $\boldsymbol\theta^{(sj)}$ and $-\boldsymbol{\theta}^{(sj)}$.
\end{theorem}
 The optimal allocations in Case 1 of Theorem~\ref{cont.mm.theorem.1.1} are inversely proportional to $r_{\ell}$, and consequently the quantities $\pi_{\ell}r_{\ell}$ are equal for all treatment arms. Under exchangeable working structure of $R({\rho_{\ell}})$, $r_{\ell}=n_{\ell}/(1+(n_{\ell}-1)\rho_{\ell})$. Therefore, treatment arms with higher ICCs or smaller cluster sizes receive the large number of clusters. In Case 2, the $r_{s}^{-1}$ of $s$--th arm exceeds the sum of the corresponding quantities of the remaining arm and the optimal allocation allocates more than half of the clusters to this arm. The LFCs in this case, only involve pairs containing the $s$--th arm. In both cases, the optimal allocations and the LFDs do not depend on the choice of $\delta$, whereas the LFC set does depend on $\delta$.
\begin{corollary}
For two experimental groups (control versus treatment, $K = 1$), the MM--design follows from Case 2 of the above theorem. Consequently, the components of MM--design $\boldsymbol{\pi}_{\rm MM}$ are
\begin{equation*}
 \pi_{0}=\frac{\sqrt{r_{0}^{-1}}}{\sqrt{r_{0}^{-1}}+\sqrt{r_1^{-1}}},\;\mbox{and } \pi_{1}=1-\pi_{0}.
 \end{equation*} The set of LFCs is $\boldsymbol{\Theta_{\rm LFC}}=\{(\delta/2,-\delta/2)^{\rm T},(-\delta/2,\delta/2)^{\rm T}\}$ and the set of LFD is 
 \[\mathcal{Q}_{\rm LFD}=\left\{\left(\frac{r_{0}^{-1}}{r_{0}^{-1}+r_{1}^{-1}}, \frac{r_{1}^{-1}}{r_{0}^{-1}+r_{1}^{-1}}\right)\right\}
 \]
\end{corollary}

\begin{example}\label{cont.mm.example}
To illustrate Theorem~\ref{cont.mm.theorem.1.1}, consider a CRT with three arms and assuming an exchangeable working correlation structure. \\
\noindent\textbf{Case 1 illustration}: Let $\boldsymbol{n}=(28,25,30)^{\rm T}$, $\delta=0.1$ and $\rho = (0.04,0.05,0.03)$. Rounded to three decimal places, we obtain $\boldsymbol{r}=(13.462 ,11.364 ,16.043)^{\rm T}$, $(r_{0}^{-1},r_{1}^{-1},r_{2}^{-1})=(0.074,\\ 0.088 ,0.062)$, $(A,A_0,A_1,A_2)= (0.225,0.150, 0.137,0.162)$. Since $r_{\ell}^{-1}\leq A_{\ell}$,$\ell \in \{0,1,2\}$, the MM--design obtained using~\eqref{cont.mm.eqn.1.case1} is $\boldsymbol{\pi}_{\rm MM}=(0.331,0.392,0.277)^{\rm T}$. The set of $\boldsymbol{\Theta}_{\rm LFC}$ is set of permutations of vectors $(0.05,-0.05,0)$. 
To obtain the LFD, let $q_{ij}$ be the total probability
assigned to the LFC pair
$\{\boldsymbol\theta^{(ij)},-\boldsymbol\theta^{(ij)}\}$.
Case 1 conditions on LFDs require
\begin{equation}
q_{01}+q_{02}=\frac{2r_0^{-1}}{A},\qquad
q_{01}+q_{12}=\frac{2r_1^{-1}}{A},\qquad
q_{02}+q_{12}=\frac{2r_2^{-1}}{A}.\label{cont.example.3.incidence.eqn}
\end{equation}
On solving these equations, we get
\begin{eqnarray*}
q_{01}&=&
\frac{r_0^{-1}+r_1^{-1}-r_2^{-1}}{A}=0.445,\qquad
q_{02}=
\frac{r_0^{-1}+r_2^{-1}-r_1^{-1}}{A}=0.216,\\
q_{12}&=&
\frac{r_1^{-1}+r_2^{-1}-r_0^{-1}}{A}=0.339.
\end{eqnarray*}

\noindent\textbf{Case 2 illustration:} Now suppose that $\rho=(0.005,0.04,0.3)$, keeping $\boldsymbol{n}$ and $\delta$ unchanged. On calculations, we obtain $\boldsymbol{r}=(24.67,12.755,3.093)^{\rm T}$, $(r_{0}^{-1},r_{1}^{-1},r_{2}^{-1})=(0.041,0.078,0.323)$, $(A,A_0,A_1,A_2)=(0.442,0.402,0.364,0.119)$. Since $r_{2}^{-1}> A_{2}$, thus the condition of Case 2 apply with $s=2$. The MM--design obtained using~\eqref{cont.mm.eqn.1.case2} is $\boldsymbol{\pi}_{\rm MM}=(0.129,0.249,0.622)$. The set of LFCs, $\boldsymbol{\Theta}_{\rm LFC}=\{\theta^{(20)},-\theta^{(20)},\theta^{(21)},-\theta^{(21)}\}$, where $\theta^{(20)}=(-0.05,-0.012,0.05)^{\rm T}$ and $\theta^{(21)}=(-0.012,-0.05,0.05)^{\rm T}$. The set of LFDs is $\mathcal{Q}_{\rm LFD}=\{(q_{02},q_{12})\}$ with $q_{02}=r_{0}^{-1}/A_{2}=0.341,q_{12}=r_{1}^{-1}/A_{2}= 0.659$.
\end{example}
For $K\ge3$, the LFDs in Case 1 of theorem~\ref{cont.mm.theorem.1.1} may not be unique. In particular, any solution which satisfies~\eqref{cont.lfd.eqn.1.case1} is a LFD. The Algorithm~\ref{cont.lfd.construction} in Appendix outlines a method to construct one such LFD. 

Next we present the MM--design for an important special case in which all clusters have equal sizes and the ICCs are identical. This result follows immediately from Theorem \ref{cont.mm.theorem.1.1}.
\begin{corollary}
 \label{cont.mm.balanced.theorem.1.1}
Suppose that $r_{\ell}=r,\;\forall\; \ell\in \mathcal{S}$, then the MM--design is the balanced design given by $\boldsymbol{\pi}_{\rm MM}=(K+1)^{-1}\mathbf{1}_{K+1}$ and $LFC$ is any permutation of the vector $(-\delta/2,\delta/2,0,\hdots,0)^{\rm T}$.
\end{corollary}

 An alternate characterization of the maximin optimality under fixed cluster sizes is provided in the following equivalence theorem. The condition~\eqref{cont.equivalence.condition.1} is derived using the directional derivative approach as highlighted in \citet{muller1998applications}.

\begin{theorem}\label{cont.equivalence.theorem.1} For a candidate design $\boldsymbol\pi^*\in\Pi$, let $\Theta_{\rm LFC}(\boldsymbol\pi^*)=
\arg\,\min\limits_{\boldsymbol\theta\in\Delta_\delta}
\phi(\boldsymbol\pi^*,\boldsymbol\theta)$. Then $\boldsymbol\pi^*$ is a maximin design if and only if there exist finitely many vectors $\boldsymbol\theta^{(0)},\ldots,
\boldsymbol\theta^{(\mathcal G)}
\in\Theta_{\rm LFC}(\boldsymbol\pi^*)$ and a probability vector $\boldsymbol q=(q_0,\ldots,q_{\mathcal G})^{\rm T}$, with $q_h\geq0,\quad h=0,\ldots,\mathcal G,\quad \sum_{h=0}^{\mathcal G}q_h=1,$
such that 
\begin{eqnarray}
\sum_{h=0}^{ G}q_{h}\left(2\boldsymbol{\pi}_{s}^{\rm T}\mathbf{A}^{(h)}\boldsymbol{\pi}^{*}(\boldsymbol{r}^{\rm T}\boldsymbol{\pi}^{*})-\boldsymbol{\pi}^{*}\mathbf{A}^{(h)}\boldsymbol{\pi}^{*}(\boldsymbol{r}^{\rm T}\boldsymbol{\pi}^{*}+\boldsymbol{r}^{\rm T}\boldsymbol{\pi}_{s})\right)&\leq&0, \;\; \forall s\in \mathcal{S} \label{cont.equivalence.condition.1}
 \end{eqnarray}
where $\boldsymbol{\pi}_s$ is the vector of cluster proportion assigning unit weight to $s$--th treatment, $\mathbf{A}^{(h)}=\mathbf{A}(\boldsymbol \theta^{(h)})$and $\mathbf{A}(\boldsymbol \theta)=(a_{ij})_{i,j\in\mathcal{S}}$ is a symmetric matrix whose $(i,j)$--th entry is given by 
\begin{align*}
 a_{ij}=\begin{cases}
 0,&if\quad i=j,\\
 \frac{r_{i}r_{j}(\theta_{i}-\theta_{j})^2}{2},&if\quad i\neq j.
 \end{cases}
\end{align*} 
\end{theorem}

To demonstrate the utility of the above equivalence condition, we apply it to the balanced design established in Corollary \ref{cont.mm.balanced.theorem.1.1}. The resulting verification of the maximin property of the balanced design is provided in Result~\ref{balanced.design.verification.equivalence} of Appendix.

As in the locally optimal formulation, the cluster sizes
$\boldsymbol{n}$ may be optimized jointly with the cluster
allocation proportions $\boldsymbol{\pi}$. The following theorem identifies the joint MM-design for cluster allocations and the cluster sizes.

\begin{theorem}\label{cont.mm.theorem.joint}
For $K\ge 1$, define the matrix $\mathbf H=(h_{\ell j})$ by $
h_{\ell j} = \sqrt{\rho_\ell\rho_j} + \frac{\sqrt{(1-\rho_\ell)(1-\rho_j)}}{\eta},\;\\\ell,j\in\mathcal S$. Let $\Lambda$ be the largest eigenvalue of $\mathbf {H}$,
and $\boldsymbol e=(e_0,\ldots,e_K)^{\rm T}$ be the corresponding
positive normalized eigenvector such that $\sum_{\ell=0}^{K}e_{\ell}^{2}=1$. Let $u_{1\ell}=\sqrt{\rho_\ell}\ e_{\ell},\; u_{2\ell}=\sqrt{1-\rho_\ell}\ e_{\ell},\ \ U_{1}=\sum_{\ell=0}^{K}u_{1\ell} \mbox{ and } U_{2}=\sum_{\ell=0}^{K}u_{2\ell}$. Then the following statements hold.\\

\noindent\textbf{Case 1.}
Suppose that $
e_\ell^2\leq \frac12,\; \forall \ell\in\mathcal S$.
Then the components of the joint maximin design
$\boldsymbol\pi_{\rm MM}$ and
$\boldsymbol n_{\rm MM}$ are given by
\begin{equation}
\label{cont.mm.eqn.joint.case1}
\pi_\ell
=
\frac{u_{1\ell}}
{U_{1}},
\qquad
n_\ell
=
\frac{\eta\,U_{1}}
{U_{2}}\,
\sqrt{\frac{1-\rho_\ell}{\rho_\ell}}
,
\quad \ell\in\mathcal S.
\end{equation}
The set of LFCs is given by $\Theta_{\rm LFC}
=
\left\{
\boldsymbol\theta^{(ij)},\,
-\boldsymbol\theta^{(ij)}
:\ i<j,\ i,j\in\mathcal S
\right\},$
where the $\ell$th component of
$\boldsymbol\theta^{(ij)}$ is
\begin{equation}
\label{cont.lfc.eqn.joint.case1}
\theta_\ell^{(ij)}
=
\begin{cases}
\delta/2, & \ell=i,\\
-\delta/2, & \ell=j,\\
0, & \ell\notin\{i,j\}.
\end{cases}
\end{equation}
The set of LFDs is
\begin{equation}
\label{cont.mm.joint.lfd.case1}
\mathcal Q_{\rm LFD}
=
\left\{
\boldsymbol q=(q_{ij})_{i<j}:
\begin{array}{l}
q_{ij}\geq0,\qquad
\displaystyle\sum_{0\leq i<j\leq K}q_{ij}=1,\\[0.6em]
\displaystyle
\sum_{i<\ell}q_{i\ell}
+
\sum_{j>\ell}q_{\ell j}
=2e_\ell^2,
\quad \ell\in\mathcal S
\end{array}
\right\}.
\end{equation}
Here, $q_{ij}$ is the total probability assigned to the pair
$\{\boldsymbol\theta^{(ij)},-\boldsymbol\theta^{(ij)}\}$.\\

\noindent\textbf{Case 2.}
Suppose that $e_s^2>\frac12,\; \text{for some }s\in\mathcal S$. Define $\boldsymbol{h}_{s}=(h_{0s},\hdots,h_{(s-1)s},h_{(s+1)s},\hdots,h_{Ks})^{\rm T}$ and $\boldsymbol{H}_{ss}=(h_{\ell j})_{\ell,j\in \mathcal{S}\setminus\{s\}}$. Here, $\boldsymbol{h}_{s}$ is the $s$--th column of matrix $\boldsymbol{H}$ without the $s$--th element and $\boldsymbol{H}_{ss}$ is obtained from the matrix $\boldsymbol{H}$ after deleting the $s$-th row and $s$--th column. Let $z$ be the unique solution satisfying $
\left\|
(z\boldsymbol I-\mathbf H_{ss})^{-1}\boldsymbol h_{s}
\right\|_2=1,\quad z>\Lambda_s,\;$ with $\Lambda_s$ denoting the largest eigenvalue of $\boldsymbol{H}_{ss}$. Define $\boldsymbol t=(t_0,\hdots,t_{s-1},t_{s+1},\hdots, t_{K})^{\rm T}=(z\boldsymbol I-\mathbf H_{ss})^{-1}\boldsymbol h_{s}$ and construct $\boldsymbol v=(v_0,\ldots,v_K)^\top$ by setting $v_s=1,\quad v_\ell=t_\ell,\quad \ell\ne s$,
so that, $\sum_{\ell\ne s}v_\ell^2=1$. Let $u_{3\ell}=\sqrt{\rho_\ell}\ v_{\ell},\; u_{4\ell}=\sqrt{1-\rho_\ell}\ v_{\ell},\ \ U_{3}=\sum_{\ell=0}^{K}u_{3\ell} \mbox{ and } U_{4}=\sum_{\ell=0}^{K}u_{4\ell}$. Then components of the joint maximin design are
\begin{equation}
\label{cont.mm.eqn.joint.case2}
\pi_\ell
=
\frac{u_{3\ell}}
{U_{3}},
\qquad
n_\ell
=
\frac{\eta\,U_{3}}
{U_{4}}\,
\sqrt{\frac{1-\rho_\ell}{\rho_\ell}}
,
\quad \ell\in\mathcal S.
\end{equation}
The set of LFCs is $\Theta_{\rm LFC}=\left\{
\boldsymbol\theta^{(sj)},\,
-\boldsymbol\theta^{(sj)}
:\ j\in\mathcal S\setminus\{s\}
\right\}$, where the $\ell$th component of
$\boldsymbol\theta^{(sj)}$ is
\begin{equation}
\label{cont.mm.joint.lfc.case2}
\theta_\ell^{(sj)}
=
\begin{cases}
\delta/2, & \ell=s,\\
-\delta/2, & \ell=j,\\
\frac{\delta}{2}
\frac{z-\tau_s}{z+\tau_s},
& \ell\notin\{s,j\}, 
\end{cases}
\end{equation}
and $\tau_s=h_{ss}+\boldsymbol h_{s}^{\rm T}\ \boldsymbol t$.
Also, the set of LFDs is
\begin{equation}
\label{cont.mm.joint.lfd.case2}
\mathcal Q_{\rm LFD}
=\{
\boldsymbol q=(q_{sj})_{j\in\mathcal S\setminus\{s\}}:
q_{sj}\geq0,\quad \sum_{j\ne s}q_{sj}=1,\quad q_{sj}=v_j^2,
\quad j\in\mathcal S\setminus\{s\}\}.
\end{equation}
Here, $q_{sj}$ is the total probability assigned to
$\boldsymbol\theta^{(sj)}$ and
$-\boldsymbol\theta^{(sj)}$.
\end{theorem}
 The optimal allocations and the optimal cluster sizes in Case 1 of above theorem are determined by the positive principal eigenvector of $\mathbf H$ and LFCs vectors are permutations of the vector $(-\delta/2,\delta/2,0,0,\hdots,0)$. The resulting design equalizes the quantities $\pi_{\ell}r_{\ell}$ across all treatment arms and the optimal allocations are directly proportional to the quantities $u_{1\ell}$'s. In Case 2, when one squared component of the principal eigenvector exceeds 1/2, the optimal design calculations require solving a scalar equation. In this case, the LFCs and the associated LFDs involve only pairs containing that arm. In both cases, the optimal allocation proportions, cluster sizes, and LFDs are independent of the choice of $\delta$, whereas the LFCs do depend on $\delta$. To demonstrate the use of Theorem~\ref{cont.mm.theorem.joint}, an illustrative example (Example \ref{cont.mm.example.joint}) is provided in the Appendix. 

\begin{corollary}\label{cont.mm.corollary.joint.1}
For two experimental groups (control versus treatment, K = 1) and assuming an exchangeable working correlation matrix, the MM--design $(\boldsymbol{\pi}_{\rm MM},\boldsymbol{n}_{\rm MM})$ is same as the LOD in Theorem \ref{cont.lod.2} and the set of LFCs is $\boldsymbol{\Theta_{\rm LFC}}=\{(-\delta/2,\delta/2)^{\rm T},(\delta/2,-\delta/2)^{\rm T}\}$.
\end{corollary}
\begin{corollary}
 \label{cont.mm.balanced.corollary.joint}
Suppose that $\rho_{\ell}=\rho,\;\forall\; \ell\in \mathcal{S}$, then the MM--design $(\boldsymbol{\pi}_{\rm MM},\boldsymbol{n}_{\rm MM})$ is the balanced design given by $\boldsymbol{\pi}_{\rm MM}=(K+1)^{-1}\mathbf{1}_{K+1}$, $\boldsymbol{n}_{\rm MM}=\eta\mathbf{1}_{K+1}$ and $LFC$ is any permutation of the vector $(-\delta/2,\delta/2,0,0\hdots,0)^{\rm T}$.
\end{corollary}

\section{Designs for binary response}
The CRT with a binary response is widely analysed using the GEE approach \citep{LiangZeger1986}. The sample size calculations(i.e., number of clusters) along with the power analysis for CRTs has been extensively reported in the literature \citep{LiuLiang1997}, \cite{crespi2009using}. In the context of optimal cluster allocation, \cite{wu2017maximin} proposed locally optimal and MM--designs for two-arm CRTs with a binary outcome allowing both the unit cost and the ICC to differ between arms, and deriving results for the three most commonly used treatment effect measures --- risk difference(RR), relative risk(RD) and odds ratio(OR), under a GEE-based variance criterion. Building on this, \cite{liu2019optimal} extended the framework to three-level CRTs with a nested exchangeable correlation structure, again proposing locally optimal and MM--designs under a fixed sampling budget, jointly optimizing the cluster size, the proportion of clusters allocated to each arm, and the total number of clusters, though unlike \cite{wu2017maximin}, assuming a common ICC across arms. Furthermore, \cite{liu2023overview} proposed MM--designs for two-level and three-level CRTs, as well as for partially nested designs, based on GEE estimation with the variance of the treatment effect estimator for a binary response as the optimality criterion. Their designs jointly determined the optimal cluster size, the proportion of clusters allocated to each arm, and the total number of clusters, though for the two-level CRT they assumed a common ICC across both arms. Despite this substantial progress, all of the above designs remain confined to the two-arm control versus intervention settings. As in the continuous case, the problem of optimally allocating clusters and cluster sizes across different treatment arms in a multi-arm CRT allowing for varying ICCs when the design objective is hypothesis testing has not, to our knowledge, been addressed for binary outcomes. In this section, we provide the framework for optimally allocating the clusters and cluster sizes in a multi-arm CRT by optimizing the NCP of QST when the response is binary. 

For binary response employing the logit link function, the marginal model in equation \eqref{model.1} becomes
\begin{align}
\text{logit} (\mathbb{E}(y_{ij}|\boldsymbol{x}_{i})) =\lambda + \boldsymbol{x}^{\rm T}_{i}\boldsymbol{\psi},\qquad j=1,2,...,n_{i},\; i=1,2,...,m 
\end{align}

In context of binary response, $\theta_{\ell}$ represents the treatment specific response probabilities and we can directly model these response probabilities as
\begin{equation*}
 \theta_{0}=\frac{\exp (\lambda)}{1+\exp (\lambda)}, \quad \theta_{1}=\frac{\exp (\lambda+\psi_{1})}{1+\exp (\lambda+\psi_{1})}, \cdots, \theta_{K}=\frac{\exp (\lambda+\psi_{K})}{1+\exp (\lambda+\psi_{K})}. 
\end{equation*}
The NCP in this case is given by 
\begin{equation}\label{binary.ncp}
\phi(\boldsymbol{\pi},\boldsymbol{\theta},m,\boldsymbol{n},\boldsymbol{\rho})
= m\left\{\sum\limits_{l=0}^{K-1}\frac{\pi_{\ell}r_{\ell}}{\gamma_{\ell}}(\theta_{\ell}-\theta_K)^2-\frac{\left[\sum\limits_{l=0}^{K-1}\frac{\pi_{\ell}r_{\ell}}{\gamma_{\ell}}(\theta_{\ell}-\theta_{K})\right]^{2}}{\sum\limits_{l=0}^{K}\frac{\pi_{\ell}r_{\ell}}{\gamma_{\ell}}}\right\}
\end{equation}
where, $ \gamma_{\ell}=\theta_{\ell}(1-\theta_{\ell})$
\begin{remark}
The NCP in \eqref{binary.ncp} generalizes the NCP obtained by \cite{Tymofyeyev2007} and \cite{Baldi2021} for equal cluster sizes and ICCs. Furthermore, the property of concavity for the NCP (as a function of $\boldsymbol{\pi}$) follows directly from their work. 
\end{remark}

\subsection{Locally optimal designs}
For binary case the variance is a function of the mean parameter therefore optimal design depends on the variance also. However, LOD in this case is again a two point design similar to the normal response case. 
\begin{theorem}\label{binary.lod.1}
 Suppose that 
\begin{equation}\label{binary.lod.condition}
 \frac{(\theta_{s}-\theta_{t})^2}{\left(\sqrt{\gamma_s/r_s}+\sqrt{\gamma_t/r_t}\right)^2}=\underset{i\neq j}{\max}\left\{\frac{(\theta_{i}-\theta_{j})^2}{\left(\sqrt{\gamma_i/r_i}+\sqrt{\gamma_j/r_j}\right)^2}\right\}. 
\end{equation}
Then the LOD $\boldsymbol{\pi}_{\rm LOD}$ is given by
\begin{flalign}
\pi_{s}=\frac{\sqrt{\gamma_{s}/r_{s}}}{\sqrt{\gamma_{s}/r_{s}}+\sqrt{\gamma_{t}/r_t}},\; \pi_{t}=1-\pi_{s},
\;\text{and}\; \pi_{\ell}=0, \forall \quad \ell\in \mathcal{S}\setminus\{s,t\} \label{binary.lod.eq.1}
\end{flalign}
If more than one pair satisfies condition \eqref{binary.lod.condition}, then any convex combination of the corresponding two-point LODs is also an LOD. 
 \end{theorem}

As in the case of normal response, the proof of Theorem~\ref{binary.lod.1} follows that of Theorem~2.1 in \citet{Baldi2021}, with appropriate modifications to account for the $r_\ell$ factors. 
 \begin{corollary}\label{binary.lod.corollary.1}
 When $r_\ell=r$ for all $\ell\in \mathcal{S}$ and $\theta_{0}\leq\theta_{1}\leq... \leq\theta_{K}$, with at least one strict inequality, then $\boldsymbol{\pi}_{\rm LOD}$ is given by 
\begin{flalign*}
\pi_{0}=\frac{\sqrt{\gamma_{0}}}{\sqrt{\gamma_{0}}+\sqrt{\gamma_{K}}},\quad\pi_{K}=1-\pi_{0}, \;\text{and}\quad \pi_{\ell}=0, \forall \; \ell\in\mathcal{S}\setminus\{0,K\}.
\end{flalign*}
\end{corollary}

For a two-arm trial ($K=1$) with equal cluster sizes ($n_0=n_1$), the LOD in Theorem~\ref{binary.lod.1} coincides with the optimal allocation derived by \citet{wu2017maximin} for estimating the risk difference (RD) when per-cluster costs are equal across treatment arms. The following example illustrates the calculation of an LOD using Theorem~\ref{binary.lod.1}.

\begin{example} Consider a CRT with 5 treatment groups. Assume an exchangeable correlation structure of the working correlation matrix, vector of ICCs $\boldsymbol{\rho}=(0.05,0.002,0.04,0.03,0.04)^{\rm T}$ and $\boldsymbol{n}=(30,30,30,30,30)^{\rm T}$. Furthermore, let the vector of treatment means be $\boldsymbol{\theta}=(0.2,0.25,0.15,0.18,0.15)^{\rm T}$. It can be easily verified that, 
\begin{flalign*}
\underset{i\neq j}{\max}\left\{\frac{(\theta_{i}-\theta_{j})^2}{\left(\sqrt{\gamma_{i}/r_i}+\sqrt{\gamma_{j}/r_j}\right)^2}\right\} =\frac{(\theta_{1}-\theta_{2})^2}{\left(\sqrt{\gamma_{1}/r_1}+\sqrt{\gamma_{2}/r_{2}}\right)^2}=\frac{(\theta_{1}-\theta_{4})^2}{\left(\sqrt{\gamma_1/r_1}+\sqrt{\gamma_4/r_4}\right)^2}.
\end{flalign*}
Both pairs $\{1,2\}$ and $\{1,4\}$ satisfy the 
condition~\eqref{binary.lod.condition}. The corresponding locally
optimal designs, obtained from \eqref{binary.lod.eq.1}, are $
\boldsymbol{\pi}_{\rm LOD}^{(1)}=(0,0.459,0.541,0,0)^{\rm T}\mbox{ and } \boldsymbol{\pi}_{\rm LOD}^{(2)}=(0,0.459,0,0,\\0.541)^{\rm T}$. Therefore, by Theorem~\ref{binary.lod.1}, every convex combination $t\,\boldsymbol{\pi}_{\rm LOD}^{(1)}+(1-t)\,\boldsymbol{\pi}_{\rm LOD}^{(2)},\:\\ t\in[0,1],$ is also a locally optimal design.
\end{example}
Next theorem provides optimal cluster allocation proportions as well as optimal cluster sizes for two experimental groups. 

\begin{theorem}\label{binary.lod.2}
For two experimental groups (control versus treatment, K = 1), the components of LOD, $(\boldsymbol \pi_{LOD},\boldsymbol n_{LOD})=(\pi_0,\pi_1,n_0,n_1)$ are given by
\begin{eqnarray}
\pi_{0}&=&\frac{\sqrt{\gamma_{0}\rho_{0}}}{\sqrt{\gamma_{0}\rho_{0}}+\sqrt{\gamma_{1}\rho_1}},\; \pi_{1}=1-\pi_{0}, \\
n_{0}&=&\frac{\eta\sqrt{\gamma_{0}(1-\rho_{0})}}{\pi_{0}(\sqrt{\gamma_{0}(1-\rho_{0})}+\sqrt{\gamma_{1}(1-\rho_1)}\;)},\mbox{ and } n_{1}=\frac{\eta\sqrt{\gamma_{1}(1-\rho_{1})}}{\pi_{1}(\sqrt{\gamma_{0}(1-\rho_{0})}+\sqrt{\gamma_{1}(1-\rho_1)}\;)}.\label{binary.lod.eq.2}
\end{eqnarray}
 \end{theorem}
The proof of Theorem~\ref{binary.lod.2} follows in the similar way as the proof of Theorem~\ref{cont.lod.2} with adjustments for the arm-specific variances. Extension of the above theorem for analytical solution of LOD for general $K$ is difficult. However, numerical optimization techniques can be utilized to get LODs numerically. 

\subsection{Maximin designs}
For obtaining the MM--designs in case of binary response, the restricted space $\mathrm{\Delta}_{\delta}$ is defined as
\begin{equation*}
 \mathrm{\Delta}_{\delta} = \{\boldsymbol{\theta}:\max_{i,j\in\mathcal{S}}|\theta_i-\theta_j|\geq \delta,\quad \delta\in (0,1),\boldsymbol{\theta}\in (0,1)^{K+1}\}.
\end{equation*}
The following theorem characterizes the MM--design, the set of LFCs and associated LFDs when the clusters are held fixed. 

\begin{theorem}
\label{binary.mm.theorem.1.1}
 For $K\ge 1$, suppose that $m,\boldsymbol n, \boldsymbol \rho$ are fixed. Let $A=\sum\limits_{\substack{\ell=0}}^{K}r_\ell^{-1}$ and $A_\ell= A-r_{\ell}^{-1}$,\; $\ell \in \mathcal{S}$. Then the following statements hold.\\

\noindent\textbf{Case 1.} Suppose that $r_\ell^{-1}\leq A_{\ell},
\quad \forall \ell\in\mathcal S.$ Then the components of the maximin design $\boldsymbol{\pi}_{\rm MM}$ are given by \begin{equation}
\label{binary.mm.eqn.1.case1}
\pi_\ell
=r_{\ell}^{-1}/A,
\qquad \ell\in\mathcal S,
\end{equation} and the set of LFCs is given by $
\Theta_{\rm LFC}
=
\left\{
\boldsymbol\theta^{(ij)},\;\boldsymbol{1}-\boldsymbol\theta^{(ij)}
:\,i< j,\; i,j \in \mathcal{S}
\right\},
$ where the $\ell$th component of $\boldsymbol\theta^{(ij)}$ is
\begin{equation}
\label{binary.lfc.eqn.1.1}
\theta_\ell^{(ij)}
=
\begin{cases}
\frac{1+\delta}{2}, & \ell=i,\\
\frac{1-\delta}{2}, & \ell=j,\\
\frac12, & \ell\notin\{i,j\}.
\end{cases}
\end{equation}
and the set of LFDs is given by 
\begin{equation}
\label{eq:binary_mm_lfd_A}
\mathcal{Q}_{\rm LFD}
=
\left\{
\boldsymbol q=(q_{ij})_{i<j}:\quad
q_{ij}\geq0,\quad
\displaystyle\sum_{0\leq i<j\leq K}q_{ij}=1,\quad
\displaystyle
\sum_{i<\ell}q_{i\ell}
+
\sum_{j>\ell}q_{\ell j}
=2r_\ell^{-1}/A,
\quad \ell\in\mathcal S \right\}.
\end{equation}
where $q_{ij}$ is the total probability of vector $\boldsymbol{\theta^{(ij)}}$ and its componentwise complement $\boldsymbol{1}-\boldsymbol{\theta^{(ij)}}$, and may be divided arbitrarily between $\boldsymbol\theta^{(ij)}$ and $\boldsymbol{1}-\boldsymbol{\theta}^{(ij)}$.\\

\noindent\textbf{Case 2.} Suppose that $r_s^{-1}>A_{s}$, for some $s \in \mathcal{S}$, and let $z$ be the unique positive root of 
\begin{equation}\label{binary.mm.root.eqn.1.case2}
A_s(1+z)^2-r_{s}^{-1}(1+1/z)^2+4\delta^2(r_{s}^{-1}-A_s)=0
\end{equation}
then the components of the maximin design $\boldsymbol{\pi}_{\rm MM}$ are given by
\begin{equation}
\label{binary.mm.eqn.1.case2}
\pi_s
=
\frac{r_s^{-1}}
{\displaystyle r_s^{-1}+z\,A_s},
\qquad\text{and}\quad
\pi_\ell
=
\frac{z\,r_{\ell}^{-1}}
{\displaystyle r_s^{-1}+z\,A_s},
\quad \ell \in \mathcal{S}\setminus\{s\}.
\end{equation}
The set of LFCs is $\boldsymbol{\Theta}_{\rm LFC}
=
\left\{
\boldsymbol\theta^{(sj)},\boldsymbol{1}-\boldsymbol\theta^{(sj)},\;
j\in \mathcal{S}\setminus\{s\}
\right\}$, where the $\ell$th component of $\boldsymbol\theta^{(sj)}$is
\begin{equation}
\label{binary.lfc.eqn.1.2}
\theta_\ell^{(sj)}
=
\begin{cases}
\frac12+\frac {\delta}{1+z},
& \ell=s,\\
\frac12-\frac {\delta\,z}{1+z},
& \ell=j,\\
\frac12-\frac{\delta(z^2-1)}{(1+z)^2-4\delta^2\,z},
& \ell\notin\{s,j\},
\end{cases}
\end{equation}
and the set of LFDs is
\begin{equation}
\label{binary.lfd.eqn.1.case2}
\mathcal{Q}_{\rm LFD} =
\left\{
\boldsymbol q=(q_{sj})_{s\neq j}:
q_{sj}
=r_j^{-1}/A_s, j\in \mathcal{S}\setminus\{s\}
\right\}.
\end{equation}
where $q_{sj}$ is the total probability of vector $\boldsymbol{\theta}^{sj}$ and its componentwise complement $\boldsymbol{\theta}^{js}$.
\end{theorem}
 The optimal allocations and LFDs in Case 1 of Theorem~\ref{binary.mm.theorem.1.1} coincide with their continuous response counterparts, whereas the LFCs are the permutations of vector $((1+\delta)/2,(1-\delta)/2,1/2,1/2,\hdots,1/2)^{\rm T}$. In Case 2, contrary to the continuous response counterpart, the optimal allocations depend on $\delta$ through the scalar root of equation~\eqref{binary.mm.root.eqn.1.case2}, and only pairs involving the unique exceptional arm support the LFDs. The theorem therefore identifies conditions under which the optimal allocations are either independent or dependent on the type of response. The following corollary follows directly from Case 1 of the Theorem~\ref{binary.mm.theorem.1.1} and characterizes the MM--design when the cluster sizes and ICCs are identical across the treatment arms. 

\begin{corollary}\label{binary.mm.balanced.corrolary.1}
Suppose that $r_{\ell}=r,\;\forall\; \ell\in \mathcal{S}$, then the MM--design is the balanced design given by $\boldsymbol{\pi}_{\rm MM}=(K+1)^{-1}\boldsymbol{1}_{K+1}$ and $LFC$ is any permutation of the vector $((1-\delta)/2,(1+\delta)/2,1/2,\hdots,1/2)^{\rm T}$.
\end{corollary}

For fixed cluster sizes, the following equivalence theorem establishes the conditions under which a design is a MM--design.
\begin{theorem}\label{binary.equivalence.theorem}
For any design $\boldsymbol{\pi}^{*}\in \Pi$, the following two statements are equivalent
\begin{enumerate}
\item The design $\boldsymbol{\pi}^{*}$ is an MM--design , i.e. 
 \begin{align*}
 \boldsymbol{\pi}^{*}=\underset{\boldsymbol{\pi}\in \Pi}{\arg\max} \{\underset{\boldsymbol{\theta}\in \mathbf{\Delta}_{\delta}}{\min}\phi(\boldsymbol{\pi},\boldsymbol{\theta}
) \} \end{align*}
\item There exist finitely many configurations $\boldsymbol\theta^{(0)},\ldots,
\boldsymbol\theta^{(\mathcal G)}
\in\Theta_{\rm LFC}(\boldsymbol\pi^*)$ with $\Theta_{\rm LFC}(\boldsymbol\pi^*)=
\arg\,\min\limits_{\boldsymbol\theta\in\Delta_\delta}
\phi(\boldsymbol\pi^*,\boldsymbol\theta)$ and a probability vector $\boldsymbol q=(q_0,\ldots,q_{\mathcal G})^{\rm T}$, with $q_h\geq0,\quad h=0,\ldots,\mathcal G,\\\quad \sum_{h=0}^{\mathcal G}q_h=1,$ 
such that 
\begin{eqnarray}
\sum_{h=0}^{ G}\frac{q_{h}}{(\boldsymbol {b}^{(h)^{\rm T}}\pi^*)^2}\left(2\boldsymbol{\pi}_{s}^{\rm T}\mathbf{A}^{(h)}\boldsymbol{\pi}^{*}(\boldsymbol{b}^{(h)^{\rm T}}\boldsymbol{\pi}^{*})-\boldsymbol{\pi}^{*^{\rm T}}\mathbf{A}^{(h)}\boldsymbol{\pi}^{*}(\boldsymbol{b}^{(h)^{\rm T}}\boldsymbol{\pi}^{*}+\boldsymbol{b}^{(h)^{\rm T}}\boldsymbol{\pi}_{s})\right)&\leq&0, \;\; \forall s\in \mathcal{S} \label{binary.equivalence.condition}
 \end{eqnarray}
where $\boldsymbol{\pi}_s$ is the vector of cluster proportion assigning unit weight to $s$--th treatment, $\boldsymbol{b}^{(h)}=\boldsymbol{b}(\boldsymbol \theta^{(h)})$ ,$\boldsymbol{b}(\boldsymbol \theta)=(\frac{r_{0}}{\gamma_{0}},\frac{r_{1}}{\gamma_{1}},\hdots,\frac{r_{K}}{\gamma_{K}})^{\rm T}$, $\mathbf{A}^{(h)}=\mathbf{A}(\boldsymbol \theta^{(h)})$and $\mathbf{A}(\boldsymbol \theta)=(a_{ij})_{i,j\in\mathcal{S}}$ is a symmetric matrix whose $(i,j)$--th entry is given by
\begin{align*}
 a_{ij}=\begin{cases}
 0,&if\quad i=j,\\
 \frac{r_{i}r_{j}(\theta_{i}-\theta_{j})^2}{2\gamma_{i}\gamma_{j}},&if\quad i\neq j.
 \end{cases}
\end{align*}
\end{enumerate}
 \end{theorem}

The proof of Theorem~\ref{binary.equivalence.theorem} follows in the same fashion as the proof of its continuous counterpart Theorem~\ref{cont.equivalence.theorem.1} and is therefore omitted. Next, we provide the MM--design corresponding to the simultaneous maximization of $\boldsymbol{\pi}$ and $\boldsymbol{n}$ over the restricted region $\Xi$.

\begin{theorem}\label{binary.mm.theorem.joint}
For $K\geq1$, suppose that $m$ and $\boldsymbol \rho$ are fixed. Define the matrix $\mathbf H=(h_{\ell j})$ by $h_{\ell j}=\sqrt{\rho_\ell\rho_j}+\frac{\sqrt{(1-\rho_\ell)(1-\rho_j)}}{\eta},
\qquad \ell,j\in\mathcal S $.
Let $\Lambda$ denote the largest eigenvalue of $\mathbf H$, and let $\boldsymbol e=(e_0,\ldots,e_K)^{\rm T}$ be its
corresponding positive eigenvector, normalized so that $
\sum_{\ell=0}^{K}e_\ell^2=1$. Let $u_{1\ell}=\sqrt{\rho_\ell}\,e_\ell,\;
u_{2\ell}=\sqrt{1-\rho_\ell}\,e_\ell,\;
U_1=\sum_{\ell=0}^{K}u_{1\ell},\mbox{ and }
U_2=\sum_{\ell=0}^{K}u_{2\ell}$. Then the following statements hold.\\

\noindent\textbf{Case 1.}
Suppose that $e_\ell^2\leq\frac12,
\quad \ell\in\mathcal S $. Then the components of a joint maximin design
$(\boldsymbol\pi_{\rm MM},\boldsymbol n_{\rm MM})$
are given by
\begin{equation}
\label{binary.mm.eqn.joint.case1}
\pi_\ell=\frac{u_{1\ell}}{U_1},
\qquad
n_\ell= \frac{\eta U_1}{U_2}
\sqrt{\frac{1-\rho_\ell}{\rho_\ell}},
\qquad \ell\in\mathcal S.
\end{equation}
The set of LFCs is
\[
\Theta_{\rm LFC}=
\left\{
\boldsymbol\theta^{(ij)},
\boldsymbol1-\boldsymbol\theta^{(ij)}
:
i<j,\quad i,j\in\mathcal S
\right\},
\]
where the $\ell$th component of
$\boldsymbol\theta^{(ij)}$ is
\begin{equation}
\label{binary.lfc.eqn.joint.case1}
\theta_\ell^{(ij)}
=
\begin{cases}
(1+\delta)/2, & \ell=i,\\
(1-\delta)/2, & \ell=j,\\
1/2, & \ell\notin\{i,j\}.
\end{cases}
\end{equation}
The set of LFDs is
\begin{equation}
\label{binary.mm.joint.lfd.case1}
\mathcal Q_{\rm LFD}
=\{
\boldsymbol q=(q_{ij})_{i<j}:\;q_{ij}\geq0,\quad
\displaystyle\sum_{0\leq i<j\leq K}q_{ij}=1, \quad
\displaystyle
\sum_{i<\ell}q_{i\ell}
+
\sum_{j>\ell}q_{\ell j}
=
2e_\ell^2,
\quad \ell\in\mathcal S\}.
\end{equation}
Here, $q_{ij}$ is the total probability assigned to
$\boldsymbol\theta^{(ij)}$ and its component wise
complement $\boldsymbol1-\boldsymbol\theta^{(ij)}$.\\

\noindent\textbf{Case 2.}
Suppose that $e_s^2>\frac12
\quad\text{for some }s\in\mathcal S$. Define $\boldsymbol h_s=
(h_{0s},\ldots,h_{s-1,s},
 h_{s+1,s},\ldots,h_{Ks})^{\rm T},
\quad \\\mathbf H_{ss}=(h_{\ell j})_{\ell,j\in\mathcal S\setminus\{s\}}$. For $x>\Lambda_{s}$, define $\boldsymbol t(x)=(x\boldsymbol I_K-\mathbf H_{ss})^{-1}\boldsymbol h_s,
\quad
\tau_s(x)=h_{ss}+\boldsymbol h_s^{\rm T}\boldsymbol t(x)$, where $\Lambda_{s}$ denotes the
largest eigenvalue of $\mathbf H_{ss}$.
Let $z$ be the unique solution of
\begin{equation}
\label{binary.mm.joint.root}
\left\|\boldsymbol t(z)\right\|_2^2
=
\frac{
\{\tau_s(z)+z\}^2-4\delta^2\tau_s(z)^2
}{
\{\tau_s(z)+z\}^2-4\delta^2z^2
},
\qquad
\Lambda_{s}<z<\Lambda.
\end{equation}
Set $\boldsymbol t
=
(t_0,\ldots,t_{s-1},t_{s+1},\ldots,t_K)^{\rm T}
=
\boldsymbol t(z),
\qquad
\tau_s=\tau_s(z), $ and construct $\boldsymbol v=(v_0,\ldots,v_K)^{\rm T}$
by setting $v_s=1,
\quad
v_\ell=t_\ell,
\quad \ell\in\mathcal S\setminus\{s\}$. Let $u_{3\ell}=\sqrt{\rho_\ell}\,v_\ell,
\quad
u_{4\ell}=\sqrt{1-\rho_\ell}\,v_\ell,
\quad
U_3=\sum_{\ell=0}^{K}u_{3\ell},
\quad
U_4=\sum_{\ell=0}^{K}u_{4\ell}$. Then the components of a joint maximin design $(\boldsymbol\pi_{\rm MM},\boldsymbol n_{\rm MM})$ are
\begin{equation}
\label{binary.mm.eqn.joint.case2}
\pi_\ell
=
\frac{u_{3\ell}}{U_3},
\qquad
n_\ell
=
\frac{\eta U_3}{U_4}
\sqrt{\frac{1-\rho_\ell}{\rho_\ell}},
\qquad \ell\in\mathcal S.
\end{equation}
The set of LFCs is $\Theta_{\rm LFC}=\left\{
\boldsymbol\theta^{(sj)},
\boldsymbol1-\boldsymbol\theta^{(sj)}
:
j\in\mathcal S\setminus\{s\}
\right\}$,
where
\begin{equation}
\label{binary.mm.joint.lfc.case2}
\theta_\ell^{(sj)}
=
\begin{cases}
\displaystyle
\frac12+\frac{\delta z}{z+\tau_s},
& \ell=s,
\\
\displaystyle
\frac12-\frac{\delta\tau_s}{z+\tau_s},
& \ell=j,
\\
\displaystyle
\frac12-
\frac{\delta(\tau_s^2-z^2)}
{(z+\tau_s)^2-4\delta^2z\tau_s},
& \ell\notin\{s,j\}.
\end{cases}
\end{equation}
The set of LFDs is given by
\begin{equation}
\label{binary.mm.joint.lfd.case2}
\mathcal Q_{\rm LFD}
=
\left\{
\boldsymbol q=(q_{sj})_{j\in\mathcal S\setminus\{s\}}:\quad q_{sj}\geq0,\quad
\displaystyle\sum_{j\neq s}q_{sj}=1,\quad \displaystyle
q_{sj}=\frac{v_j^2}{\sum_{h\neq s}v_h^2},
\quad j\in\mathcal S\setminus\{s\}\right\}.
\end{equation} 
Here, $q_{sj}$ is the total probability assigned to
$\boldsymbol\theta^{(sj)}$ and its componentwise
complement. 
\end{theorem}
\begin{corollary}\label{binary.mm.balanced.corrolary.2}
Suppose that $\rho_{\ell}=\rho,\;\forall\; \ell\in \mathcal{S}$, then the MM--design is the balanced design given by $\boldsymbol{\pi}_{\rm MM}=(K+1)^{-1}\boldsymbol{1}_{K+1}$, $\boldsymbol{n}_{\rm MM}=\eta\boldsymbol{1}_{K+1}$ and $LFC$ is any permutation of the vector $((1-\delta)/2,(1+\delta)/2,1/2,\hdots,1/2)^{\rm T}$.
\end{corollary}
In Case 1 of the Theorem~\ref{binary.mm.theorem.joint}, the optimal allocations, optimal cluster sizes, and LFDs coincide with their continuous responses counterparts, although the LFCs in the binary case are permutations of the vector $((1-\delta)/2,(1+\delta)/2,1/2,1/2,\hdots,1/2)$. The optimal allocation proportions allocate larger number of clusters to treatment arms with higher values of $u_{1\ell}$'s. In Case 2, both the design and the LFDs generally depend on \(\delta\) through the scalar root of the equation~\eqref{binary.mm.joint.root}. Next we present the real life applications of these designs and compare these designs with the actual implemented design in the trial. 

\section{Numerical studies}\label{section.applications}
In this section, we illustrate the application of the
proposed optimal designs by re-examining two multi-arm
CRTs from a design perspective. We obtain the maximin optimal
designs for these studies and compare them with the designs
adopted in the original trials in terms of (1) the power achieved
under fixed sample-size constraints, and (2) the number
of clusters and observations required to achieve a prespecified
target power. We also compare the corresponding planning costs of the trial.
Before presenting the trial examples, we describe the power and
sample-size calculations and define the measures used for these
comparisons.

\subsection{Power and sample-size calculations}
\label{sample_size.section}

We describe the methodology for determining the required number
of clusters and the corresponding total number of observations
needed to achieve a prespecified statistical power in case of a normal response.
The NCP
$\phi(\boldsymbol{\pi},\boldsymbol{\theta},m,
 \boldsymbol{n},\boldsymbol{\rho})$
in equations \eqref{cont.ncp}
can be written as
$m\phi_{1}(\boldsymbol{\pi},\boldsymbol{\theta},
 \boldsymbol{n},\boldsymbol{\rho})$.
Thus, $\phi_{1}$ represents the NCP per cluster and can be
computed for any specified values of
$\boldsymbol{\pi},\boldsymbol{\theta},\boldsymbol{n}$, and
$\boldsymbol{\rho}$.
Under the alternative hypothesis, the QST statistic for testing
the hypothesis \eqref{hyp.1} follows a non-central chi-square distribution asymptotically, with
$K$ degrees of freedom. Therefore, for a significance level
$\alpha$, the asymptotic power is given by
\begin{equation}
\operatorname{Power}
=
1-
F_{\chi^2_K(\phi)}
\left(
\chi^2_{K,1-\alpha}
\right),
\label{power.eqn}
\end{equation}
where $F_{\chi^2_K(\phi)}(\cdot)$ denotes the cumulative distribution
function of a non-central chi-square distribution with $K$
degrees of freedom and non-centrality parameter $\phi$, and
$\chi^2_{K,1-\alpha}$ denotes the $1-\alpha$ quantile of the central
chi-square distribution with $K$ degrees of freedom.
For a target power $1-\beta>\alpha$, we first determine the
NCP $\phi^{*}$ for which the right-hand side of
\eqref{power.eqn} equals $1-\beta$. For fixed design and planning
values with $\phi_{1}>0$, the required number of clusters, before
adjustment for integer allocation to the treatment arms, is
\begin{equation*}
m=
\left\lceil
\frac{\phi^{*}}
{\phi_{1}(\boldsymbol{\pi},\boldsymbol{\theta},
 \boldsymbol{n},\boldsymbol{\rho})}
\right\rceil,
\end{equation*}
where $\lceil\cdot\rceil$ denotes the ceiling function.
This calculation gives the cluster requirement at the specified
treatment-mean vector. When the vector of cluster sizes $\boldsymbol{n}$ and the
vector of ICCs $\boldsymbol{\rho}$ are held fixed, the sample-size determination reduces to calculating the number of clusters $m$, once
the optimal allocation proportions $\boldsymbol{\pi}^{*}$ have
been obtained. The total number of observations is
then given by
$N=m\sum_{\ell=0}^{K}\pi_{\ell}^{*}n_{\ell}$.
The complete procedure is presented in
Algorithm~\ref{algorithm.1} of the Appendix, together with an
illustrative example (Example~\ref{algorithm.cont.sample.size.example}).
However, when both $\boldsymbol{\pi}$ and $\boldsymbol{n}$
are optimized, we additionally need to specify the average cluster
size $\eta$.
Starting with an initial value of $\eta$, the optimal design is obtained
over the restricted space $\Xi$, subject to
$\sum_{\ell=0}^{K}\pi_{\ell}n_{\ell}=\eta$.
The required number of clusters is then calculated using the
resulting optimal allocation and cluster sizes, and the total
number of observations is $N(\eta)=m\eta$.
The complete procedure is presented in
Algorithm~\ref{algorithm.2} of the Appendix. The sample size and power calculations in case of binary response follow similarly by replacing the normal response NCP with its counterpart binary response NCP in the above calculations.

\subsubsection{Relative gain in power}

For power comparison of two designs, the relative gain (RG) in power
of $\boldsymbol{\pi}_{1}$ compared with
$\boldsymbol{\pi}_{2}$ is defined as
\begin{equation*}
\operatorname{RG}
(\boldsymbol{\pi}_{1},\boldsymbol{\pi}_{2})
=
\frac{
\mathcal{P}(\boldsymbol{\pi}_{1})
-
\mathcal{P}(\boldsymbol{\pi}_{2})
}{
\mathcal{P}(\boldsymbol{\pi}_{2})
}
\times 100,
\end{equation*}
where $\mathcal{P}(\boldsymbol{\pi}_{\ell})$ denotes the
power calculated from \eqref{power.eqn} for design
$\boldsymbol{\pi}_{\ell}$. Both $\mathcal{P}(\boldsymbol{\pi}_{1})$ and $\mathcal{P}(\boldsymbol{\pi}_{2})$ are calculated under the same sample-size constraints and
planning assumptions specified for the comparison.

\subsection{Cost comparison}
To assess the resource implications of the different designs,
we compare the numbers of clusters, total numbers of
observations, and total costs required to achieve a prespecified target power under these designs.
Following \citet{wu2017maximin,liu2023overview}, we define
the total cost function (CF) as
\begin{equation*}
\mathrm{CF}
=
\sum_{\ell=0}^{K}
m\pi_{\ell}(c_{1\ell}+n_{\ell}c_{2\ell})
,
\end{equation*}
where $c_{1\ell}$ denotes the cost per cluster in the
$\ell$--th treatment arm and $c_{2\ell}$ denotes the cost
per individual in the same arm. We assume that the costs are
the same across different treatment arms, i.e.,
$c_{1\ell}=c_{1}$ and $c_{2\ell}=c_{2}$.

\subsection{Applications to real multi-arm trials}
\vspace{0.5cm}
\subsubsection{WASH Benefits Bangladesh trial} The WASH Benefits CRT enrolled pregnant women across 720 clusters in Bangladesh to evaluate the effects of different hygiene and nutrition interventions on early child development, with follow-up at one and two years \citep{Luby2018}. The seven treatment arms comprised control; water; sanitation; handwashing; combined water, sanitation, and handwashing; nutrition; and all four interventions combined. The control arm received twice as many clusters as each intervention arm, giving the trial design $\boldsymbol{\pi}_{\rm A}=(0.25,0.125,\ldots,0.125)^{\rm T}$. For the continuous response child length-for-age Z-score at the final follow-up, the observed arm-specific means, variance estimate, GEE-estimated exchangeable ICCs, and average cluster sizes were
\[
\begin{aligned}
\boldsymbol{\theta}_{\rm obs}
&=(-1.787,-1.860,-1.795,-1.854,-1.759,-1.532,-1.673)^{\rm T},\;\quad \hat{\sigma}^{2}=1.05,\\
\widehat{\boldsymbol{\rho}}
&=(0.037,0.016,0.093,0.068,0.006,0.062,0.027)^{\rm T},\;
\boldsymbol{n}_{\rm av}
=(6,7,6,6,6,6,7)^{\rm T}.
\end{aligned}
\]
Following \cite{wu2017maximin}, we set $\boldsymbol{n}=\boldsymbol{n}_{\rm av}$, yielding $\boldsymbol{r}=(5.063,6.387,4.096,\\4.478,5.825,4.580,6.024)^{\rm T}$. For this retrospective design comparison, we treated $\widehat{\boldsymbol{\rho}}$ as fixed planning values of $\boldsymbol{\rho}$ and set $\delta=\max_{\ell\ne\ell'}|\theta_{{\rm obs},\ell}-\theta_{{\rm obs},\ell'}|=0.328$. With fixed cluster sizes and using the results of Theorem~\ref{cont.mm.theorem.1.1}, the MM--design was obtained as
\[
\boldsymbol{\pi}_{\rm MM}
=(0.143,0.114,0.177,0.162,0.125,0.158,0.121)^{\rm T}.
\]
Joint optimization of cluster allocations and cluster sizes over $\Xi$, subject to the average cluster-size constraint, yielded
\[
\begin{aligned}
\boldsymbol{\pi}_{\rm MM}
&=(0.136,0.084,0.236,0.196,0.049,0.185,0.113)^{\rm T},\\
\boldsymbol{n}_{\rm MM}
&=(6.58,10.11,4.03,4.77,16.59,5.01,7.74)^{\rm T}.
\end{aligned}
\]
The above design was obtained on applying the results of Theorem~\ref{cont.mm.theorem.joint}. 
All power calculations used $m=240$ clusters, one-third of the original trial size, at the type I error rate of $\alpha=0.05$. Results are summarized in Table~\ref{wash_trial}. When evaluated at $\boldsymbol{\theta}_{\rm obs}$, the fixed cluster size and jointly optimized MM--designs increased power relative to the actual trial allocation by approximately $10.53\%$ and $12.08\%$, respectively. The fourth column reports the number of clusters required to achieve the target power of $0.712$, obtained using Algorithm \ref{algorithm.1} (see Appendix) for the first two designs and Algorithm \ref{algorithm.2} (see Appendix) for the jointly optimized design. The latter requires the fewest clusters ($m=202$), compared with $m=240$ for the actual allocation. The last column reports the corresponding total costs, assuming common per-cluster and per-individual costs of $c_1=5$ and $c_2=1$, respectively. \textcolor{blue}Relative to the actual allocation, the MM--designs with fixed and jointly optimized cluster sizes reduce modeled costs by $13.87\%$ and $15.83\%$, respectively.

\begin{table}[H]
\centering
\caption{Comparison of actual trial design with MM--designs for WASH trial}
\label{wash_trial}
\fontsize{6}{9}\selectfont
\renewcommand{\arraystretch}{2.5}
\setlength{\tabcolsep}{0.5pt}
\begin{tabularx}{\textwidth}{
 >{\raggedright\arraybackslash}p{0.20\textwidth}
 >{\raggedright\arraybackslash}p{0.40\textwidth}
 >{\raggedright\arraybackslash}p{0.11\textwidth}
 >{\raggedright\arraybackslash}p{0.11\textwidth}
 >{\centering\arraybackslash}p{0.20\textwidth}
}
 \toprule
 \hline
 Design&Cluster allocation proportions ($\boldsymbol{\pi}$)\newline and cluster sizes ($\boldsymbol{n}$)&$Power$&No. of clusters to achieve 71.2\% power & Total cost\newline $(c_{1}, c_{2})=(5,1)$\\
 \hline
Actual trial $\boldsymbol{\pi}$ and $\boldsymbol{n}$&
$\boldsymbol{\pi}_{\rm A} = (0.25, 0.125,\ldots, 0.125)^{\rm T}$\newline
$\boldsymbol{n}=(6.00,7.00,6.00,6.00,6.00,6.00,7.00)^{\rm T}$&
0.712&240&2700.0 \\

Maximin over $\boldsymbol{\pi}$ and trial $\boldsymbol{n}$&
$\pi_{\rm MM} = (0.143,0.114,0.177,0.162,0.125,0.158,0.121)^{\rm T}$\newline
$\boldsymbol{n}=(6.00,7.00,6.00,6.00,6.00,6.00,7.00)^{\rm T}$&
0.787&207&2325.5 \\

Maximin over both $\boldsymbol{\pi}$ and $\boldsymbol{n}$&
$\boldsymbol{\pi}_{\rm MM} = (0.136,0.084,0.236,0.196,0.049,0.185,0.113)^{\rm T}$\newline
$\boldsymbol{n}_{\rm MM}=(6.58,10.11,4.03,4.77,16.59,5.01,7.74)^{\rm T}$&
0.798&202&2272.5 \\

 \bottomrule
\end{tabularx}
\end{table}
\vspace{0.5cm}
\subsubsection{SCORE C\^ote d'Ivoire trial}

The SCORE trial in C\^ote d'Ivoire compared four strategies for
controlling \emph{Schistosoma haematobium} transmission over three
years, with 64 villages allocated equally across treatment arms
\citep{Ouattara2022}. The interventions were annual mass drug
administration (MDA) before peak transmission, annual MDA after
peak transmission, biannual MDA, and standard MDA combined with
snail control. The primary objective of the trial was to compare infection prevalence
and intensity across treatment arms among children aged 9--12 years. Infection status was recorded as a binary outcome, and we use the corresponding data from the final follow-up survey in the present analysis. The observed arm-specific infection prevalence proportions, GEE-estimated exchangeable ICCs, and average cluster sizes were
\[
\begin{aligned}
\boldsymbol{\theta}_{\rm obs}
&=(0.097,0.035,0.006,0.052)^{\rm T},
\widehat{\boldsymbol{\rho}}
=(0.259,0.132,-0.002,0.201)^{\rm T},\\
\boldsymbol{n}_{\rm av}=&(92,90,89,82)^{\rm T}.
\end{aligned}
\]
For the planning calculations, the small negative ICC estimate
in the third arm was replaced by a small positive value of $0.0001$
\citep{crespi2009using,donner1994cluster}, giving
$\boldsymbol{\rho}=(0.259,0.132,0.0001,0.201)^{\rm T}$.
As in the WASH example, we treated these ICCs as fixed,
set $\boldsymbol{n}=\boldsymbol{n}_{\rm av}$, and specified
$\delta$ as the largest observed difference in prevalence proportions,
yielding $\delta=0.091$.
With fixed cluster sizes, the MM--design was obtained as
\[
\boldsymbol{\pi}_{\rm MM}
= (0.423,0.225,0.018,0.334)^{\rm T}.
\]
Also, the joint maximin design using the results of Theorem~\ref{binary.mm.theorem.joint} subject to the average
cluster-size constraint of $\eta=88.25$, yielded
\[
\begin{aligned}
\boldsymbol{\pi}_{\rm MM}
&= (0.434,0.226,0.001,0.339)^{\rm T},\\
\boldsymbol{n}_{\rm MM}
&=(73.04,110.73,4318.02,86.10)^{\rm T}.
\end{aligned}
\]
Table~\ref{score_trial} summarizes the comparisons at
$\boldsymbol{\theta}_{\rm obs}$ and $\alpha=0.05$.
With $m=64$ clusters, the calculated powers were $0.823$,
$0.869$, and $0.875$ for the actual allocation, fixed-size
MM--design, and joint MM--design, respectively. The latter
two correspond to relative gains in power of approximately
$5.59\%$ and $6.32\%$. For a target power of $0.823$,
the sample-size procedures in Algorithms~\ref{algorithm.1}
and \ref{algorithm.2} gave requirements of 57 and 56 clusters
for the fixed-size and joint MM--designs, respectively,
compared with 64 for the actual allocation. Under the same
illustrative cost structure as in the WASH example
($c_1=5$ and $c_2=1$), the corresponding reductions in
modeled cost were $11.03\%$ and $12.5\%$.

The joint MM--design assigns a cluster proportion
of only $0.001$ to the third arm, together with a cluster
size of $4318.02$. This extreme allocation is driven in
part by the near-zero planning ICC and is not directly
implementable. Practical recruitment requirements could
be incorporated by re-optimizing subject to minimum
numbers of clusters per arm and feasible bounds on cluster
sizes, followed by integer allocation and reassessment
of power. The unconstrained joint design should therefore
be interpreted as a theoretical efficiency benchmark
rather than a recommended trial allocation.
\begin{table}[H]
\centering
\caption{Comparison of actual trial design with MM--designs for SCORE trial}
\label{score_trial}
\fontsize{6}{9}\selectfont
\renewcommand{\arraystretch}{2.5}
\setlength{\tabcolsep}{0.5pt}
\begin{tabularx}{\textwidth}{
 >{\raggedright\arraybackslash}p{0.20\textwidth}
 >{\raggedright\arraybackslash}p{0.40\textwidth}
 >{\raggedright\arraybackslash}p{0.11\textwidth}
 >{\raggedright\arraybackslash}p{0.11\textwidth}
 >{\centering\arraybackslash}p{0.20\textwidth}
}
 \toprule
 \hline
 Design&Cluster allocation proportions ($\boldsymbol{\pi}$)\newline and cluster sizes ($\boldsymbol{n}$)&Power&No. of clusters to achieve 82.3\% power & Total cost\newline $(c_{1}, c_{2})=(5,1)$\\
 \hline
Actual trial $\boldsymbol{\pi}$ and $\boldsymbol{n}$&
$\boldsymbol{\pi}_{\rm A} = (0.250,0.250,0.250,0.250)^{\rm T}$\newline
$\boldsymbol{n}=(92.00,90.00,89.00,82.00)^{\rm T}$&
0.823&64&5968.0 \\

Maximin over $\boldsymbol{\pi}$ and trial $\boldsymbol{n}$&
$\pi_{\rm MM} = (0.423,0.225,0.018,0.334)^{\rm T}$\newline
$\boldsymbol{n}=(92.00,90.00,89.00,82.00)^{\rm T}$&
0.869&57&5309.9 \\

Maximin over both $\boldsymbol{\pi}$ and $\boldsymbol{n}$&
$\boldsymbol{\pi}_{\rm MM} = (0.434,0.226,0.001,0.339)^{\rm T}$\newline
$\boldsymbol{n}_{\rm MM}=(73.04,110.73,4318.02,86.10)^{\rm T}$&
0.875&56&5222.0 \\

 \bottomrule
\end{tabularx}
\end{table}

\section{Summary and Discussion}
In this article, we propose optimal allocations of clusters and cluster sizes in multi-arm cluster randomized trials when the objective of the study is to test the hypothesis of equality of treatment means. To test this hypothesis, we employ the QST statistic developed by \citet{LiuLiang1997} and the QST follows a non-central chi-square distribution asymptotically under the alternative hypothesis. The expression for the non-centrality parameter (NCP) is derived in closed form for both continuous and binary responses, allowing the intracluster correlation coefficient (ICC) and the cluster size to vary by treatment arms. The NCP provides a common basis for maximizing the asymptotic power of the test and is therefore used as design criterion throughout this article.

We first derive the locally optimal designs, which maximize the NCP with respect to the cluster sizes and cluster allocation proportions for specified planning values of other parameters, subject to the design constraints. Assuming that cluster sizes are held fixed in advance along with other parameters, the closed-form locally optimal designs are obtained for the cluster allocations in the general $(K+1)$-arm setting. For the joint optimization of cluster allocations and cluster sizes, closed-form of the locally optimal designs are obtained for the simple control--versus--intervention case ($K=1$), while numerical methods are used in the general case. Locally optimal designs depend on the model parameters, which are seldom known at the design stage of a trial. As demonstrated in Example \ref{example.2}, a locally optimal design may perform well for one set of parameters but can perform very poorly for other values of parameters. Therefore, we also propose robust maximin optimal designs that avoid specifying a treatment-mean vector. The MM--design maximizes, over $\boldsymbol\pi$ or $(\boldsymbol\pi,\boldsymbol n)$, the minimum NCP over a prespecified set $\Delta_{\delta}$ of alternative treatment--mean vectors. For both continuous and binary responses, the analytical forms of the MM-designs and of the corresponding LFCs are derived explicitly. Under the conditions stated in the corresponding results, the MM--designs turn out to be the balanced designs that allocate equal proportions of clusters and equal cluster sizes to each treatment arm.

The locally optimal designs obtained here are consistent, in the two-arm special case, with the optimal cluster-size and cluster-allocation results of, \citet{van2018efficient}, and \citet{copas2021optimal}: higher ICCs can favour allocating a larger share of clusters to an arm, while increasing cluster size yields diminishing returns because the information multiplier $n_\ell/\{1+(n_\ell-1)\rho_\ell\}$ approaches $1/\rho_\ell$ as $n_\ell\to\infty$ for fixed $0<\rho_\ell<1$, a pattern well documented in the CRT design literature \citep{Donner_2000,crespi2009using}. Allowing $\rho_\ell$ and $n_\ell$ to differ by arm, rather than assuming a common ICC and cluster size, shows that this trade-off between the number of clusters and cluster size operates arm-by-arm, so that heterogeneity in $\boldsymbol\rho$ across arms can contribute to unequal optimal allocations, together with cluster sizes, marginal variances, and the treatment-mean configurations over which power is optimized.

Additionally, the sample-size calculation procedure discussed in Section~\ref{sample_size.section} connects the optimal allocation schemes to the numbers of clusters and observations required to attain a target power. The applications based on the WASH Benefits Bangladesh trial and the SCORE trial in C\^ote d'Ivoire illustrate the procedure for continuous and binary outcomes, respectively, wherein alongside the calculation of number of clusters, their allocation, and, when optimized jointly, cluster sizes are optimally obtained. In the reported comparisons ( see Tables~\ref{wash_trial} and~\ref{score_trial}), the MM--design with joint optimization of cluster proportions and cluster sizes achieves higher calculated power and requires fewer clusters to achieve a prespecified power than the designs adopted in the original trials, under the stated planning assumptions. These examples show how trial-informed planning values can be used to examine alternative allocations and recruitment targets when designing future multi-arm CRTs. The choice of an implementable design should also reflect cluster availability and recruitment capacity, in addition to its statistical properties.

This robust planning approach is related to earlier designs that
address uncertainty in the ICC
\citep{van2021maximin,liu2017optimal}. In the present
formulation, however, the MM--design protects against uncertainty
in the treatment effects, while the arm-specific ICCs
$\boldsymbol{\rho}$ are treated as specified planning inputs.
When reliable point values are unavailable, plausible ranges can
instead be specified using pilot data or previously published
estimates \citep{wu2017maximin}. The maximin criterion can be extended to accommodate uncertainty in both
treatment effects and ICCs by minimizing the NCP over both
$\boldsymbol{\theta}$ and $\boldsymbol{\rho}$. Suppose that the ICC
uncertainty region is
$\mathcal{R}_{\rho}
=\prod_{\ell=0}^{K}[\underline{\rho}_{\ell},\bar{\rho}_{\ell}]$,
with bounds specified independently of the treatment means and
design variables, and with all combinations admissible under the
planning model. Let
$\bar{\boldsymbol{\rho}}
=(\bar{\rho}_{0},\ldots,\bar{\rho}_{K})^{\rm T}$.
Under the exchangeable correlation model, the NCP is nonincreasing
in each $\rho_{\ell}$ for fixed cluster allocations, cluster sizes
$n_{\ell}\geq 1$, treatment means, and marginal variances.
The minimum over $\boldsymbol{\rho}$ is therefore attained at
$\bar{\boldsymbol{\rho}}$. Consequently,
\begin{equation}
\label{icc_uncertainty_reduction}
\min_{\boldsymbol{\theta}\in\Delta_{\delta}}
\min_{\boldsymbol{\rho}\in\mathcal{R}_{\rho}}
\phi(\boldsymbol{\pi},\boldsymbol{n},
 \boldsymbol{\theta},\boldsymbol{\rho})
=
\min_{\boldsymbol{\theta}\in\Delta_{\delta}}
\phi(\boldsymbol{\pi},\boldsymbol{n},
 \boldsymbol{\theta},\bar{\boldsymbol{\rho}}).
\end{equation}
Thus, the MM--design accounting for both sources of uncertainty
can be obtained using the existing procedures, with each arm-specific
ICC fixed at its upper bound.

In the SCORE trial example, unconstrained joint optimization
of cluster allocations and cluster sizes under the maximin
criterion produced an extreme allocation: an arm with a
low ICC was assigned a cluster allocation proportion close
to zero together with a disproportionately large cluster
size. Such a design may be impractical to implement
and, more importantly, can compromise the accuracy of asymptotic
GEE-based inference when too few independent clusters are
allocated to an arm. This implementation issue
could be addressed by re-optimizing the criterion subject
to a minimum number of clusters per arm and, where relevant, a
maximum feasible cluster size. This would allow practical
recruitment requirements to be incorporated into the optimization,
although the effects of these constraints on power and the
required sample size would need to be evaluated.

One of the limitations of the present work is that it only considers the hypothesis of equality of treatment means, whereas in some situations, the pairwise comparisons may also be of interest. Under different hypotheses, the functional form of the NCP for the QST has to be derived separately, and a similar approach may then be followed to obtain the optimal designs. Simultaneous comparisons additionally require a power criterion aligned with the testing procedure and appropriate multiplicity control. In addition to this, optimal allocation of clusters and cluster sizes in multi-arm longitudinal cluster randomized trials for hypothesis testing is a direction for further research that we are currently pursuing. Furthermore, a natural extension to the present work concerns trials with multiple responses or mixed response types. Sample size calculations and power analyses have been carried out by \citet{li2020power} and \citet{yang2023power} for multiple binary co-primary endpoints and multiple continuous co-primary endpoints, respectively. Extending the quasi-score test based optimality criterion developed here to a vector-valued outcome, and to the more demanding case of mixed continuous and binary co-primary endpoints, would provide a basis for constructing locally optimal and MM--designs for multiple-outcome CRTs. Overall, the proposed framework links cluster allocation, within-cluster recruitment, and sample-size determination to the testing objective while allowing treatment-specific heterogeneity and parameter uncertainty to be addressed at the design stage.

\section*{Acknowledgments:}
The work of Satya Prakash Singh is supported by Indian Institute of Technology Kanpur, India grant no. IITK/MATH/2021277. The authors acknowledge the use of ChatGPT (OpenAI) for assistance with language editing, improving clarity, and refining the presentation of the manuscript.

\section*{Conflict of Interest:} On behalf of all authors, the corresponding author states that there is no conflict of interest.

\section*{Data Availability Statement:} The WASH benefits and SCORE trial data used in Section~\ref{section.applications} is publicly available at \url{https://clinepidb.org/ce/app}.
\newpage
\bibliographystyle{apalike}
\bibliography{references}

@article{LiuLiang1997,
  author = {Liu, G. and Liang, K.-Y.},
  title = {Sample Size Calculations for Studies with Correlated Observations},
  journal = {Biometrics},
  volume = {53},
  number = {3},
  pages = {937--947},
  year = {1997}
}

@article{LiangZeger1986,
  author = {Liang, K.-Y. and Zeger, S. L.},
  title = {Longitudinal data analysis using generalized linear models},
  journal = {Biometrika},
  volume = {73},
  number = {1},
  pages = {13--22},
  year = {1986}
}

@article{SelfMauritsen1988,
  author = {Self, Steven and Mauritsen, Robert},
  title = {Power/sample size calculations for generalized linear models},
  journal = {Biometrics},
  volume = {44},
  number = {1},
  pages = {79--86},
  year = {1988}
}

@book{CoxHinkley1974,
  author = {Cox, D. R. and Hinkley, D. V.},
  title = {Theoretical Statistics},
  publisher = {Chapman and Hall},
  address = {London},
  year = {1974}
}

@article{Tymofyeyev2007,
  author = {Tymofyeyev, Yulia and Rosenberger, William F. and Hu, Feifang},
  title = {Implementing optimal allocation in sequential binary response experiments},
  journal = {Journal of the American Statistical Association},
  volume = {102},
  number = {477},
  pages = {224--234},
  year = {2007}
}

@article{Ouattara2022,
  author = {Ouattara, M. and Bassa, F. K. and Diakité, N. R. and Hattendorf, J. and Coulibaly, J. T. and Yao, P. K. and N'Goran, E. K.},
  title = {Effectiveness of four different interventions against Schistosoma haematobium in a seasonal transmission setting of Côte d’Ivoire: a cluster randomized trial},
  journal = {Clinical Infectious Diseases},
  volume = {74},
  number = {12},
  pages = {2181--2190},
  year = {2022}
}

@article{Luby2018,
  author = {Luby, Stephen P. and Rahman, M. and Arnold, Benjamin F. and Unicomb, L. and Ashraf, S. and Winch, P. J. and Colford, J. M.},
  title = {Effects of water quality, sanitation, handwashing, and nutritional interventions on diarrhoea and child growth in rural Bangladesh: a cluster randomised controlled trial},
  journal = {The Lancet Global Health},
  volume = {6},
  number = {3},
  pages = {e302--e315},
  year = {2018}
}

@article{Baldi2021,
  author = {Baldi Antognini, A. and Frieri, R. and Novelli, M. and Zagoraiou, M.},
  title = {Optimal designs for testing the efficacy of heterogeneous experimental groups},
  journal = {Electronic Journal of Statistics},
  volume = {15},
  number = {1},
  pages = {3217--3248},
  year = {2021}
}

@book{hayes2017cluster,
  title={Cluster randomised trials},
  author={Hayes, Richard J and Moulton, Lawrence H},
  year={2017},
  publisher={Chapman and Hall/CRC}
}

@article{copas2021optimal,
  title={Optimal design of cluster randomized trials allowing unequal allocation of clusters and unequal cluster size between arms},
  author={Copas, Andrew J and Hooper, Richard},
  journal={Statistics in medicine},
  volume={40},
  number={25},
  pages={5474--5486},
  year={2021},
  publisher={Wiley Online Library}
}

@article{wu2017maximin,
  title={Maximin optimal designs for cluster randomized trials},
  author={Wu, Sheng and Wong, Weng Kee and Crespi, Catherine M},
  journal={Biometrics},
  volume={73},
  number={3},
  pages={916--926},
  year={2017},
  publisher={Oxford University Press}
}

@article{van2021maximin,
  title={Maximin design of cluster randomized trials with heterogeneous costs and variances},
  author={van Breukelen, G. J. P. and Candel, M. J. J. M.},
  journal={Biometrical Journal},
  volume={63},
  number={7},
  pages={1444--1463},
  year={2021},
  publisher={Wiley Online Library}
}

@article{kennedy2022power,
  title={Power and sample size calculations for cluster randomized trials with binary outcomes when intracluster correlation coefficients vary by treatment arm},
  author={Kennedy-Shaffer, Lee and Hughes, Michael D},
  journal={Clinical Trials},
  volume={19},
  number={1},
  pages={42--51},
  year={2022},
  publisher={SAGE Publications Sage UK: London, England}
}

@article{van2018efficient,
  title={Efficient design of cluster randomized trials with treatment--dependent costs and treatment--dependent unknown variances},
  author={van {B}reukelen, G. J. P. and Candel, M. J. J. M.},
  journal={Statistics in medicine},
  volume={37},
  number={21},
  pages={3027--3046},
  year={2018},
  publisher={Wiley Online Library}
}

@article{singh2023some,
  title={Some design considerations for cluster randomized trials with binary responses},
  author={Singh, Arpan and Pal, Meghendar and Singh, Satya Prakash},
  journal={Communications in Statistics-Theory and Methods},
  volume={52},
  number={12},
  pages={4310--4328},
  year={2023},
  publisher={Taylor \& Francis}
}

@article{omar2000analysis,
  title={Analysis of a cluster randomized trial with binary outcome data using a multi-level model},
  author={Omar, Rumana Z and Thompson, Simon G},
  journal={Statistics in Medicine},
  volume={19},
  number={19},
  pages={2675--2688},
  year={2000},
  publisher={Wiley Online Library}
}

@article{crespi2009using,
  title={Using second-order generalized estimating equations to model heterogeneous intraclass correlation in cluster-randomized trials},
  author={Crespi, Catherine M and Wong, Weng Kee and Mishra, Shiraz I},
  journal={Statistics in Medicine},
  volume={28},
  number={5},
  pages={814--827},
  year={2009},
  publisher={Wiley Online Library}
}

@article{singh_JSPI_2026,
  title={Max--min experimental designs for comparing pairs of treatments
with binary outcomes},
  author={Singh, S. P. and Davidov, O.},
  journal={Journal of Statistical Planning and Inference},
  volume={243},
  number={},
  pages={106389},
  year={2026}
}

@article{ASO_2025_SJS,
  title={Optimal designs for testing pairwise differences:
A graph-based game theoretic approach},
  author={Singh, A. and Singh, S. P. and Davidov, O.},
  journal={Scandinavian Journal of Statistics},
  volume={52},
  number={2},
  pages={533--571},
  year={2025}
}

@article{Singh_2019,
title	= {On the design of experiments with ordered treatments},
author	= {Singh, S. P. and Davidov, O.},
journal	= {Journal of Royal Statistical Society: Series B},
volume	= {85},
number	= {5},
pages	= {881-900},
year	= {2019},
}

@Article{Singh_2020_EJS,
	author = {Singh, S. P. and Davidov, O.},
	title = {On {B}ayes and {N}ash experimental
designs for hypothesis testing problems},
	journal = {Electronic Journal of Statistics},
	volume = {14},
	number = {},
	pages = {3976-4003},
	year = {2020},
}

@article{liu2003statistical,
  title={Statistical power and optimum sample allocation ratio for treatment and control having unequal costs per unit of randomization},
  author={Liu, X.},
  journal={Journal of Educational and Behavioral Statistics},
  volume={28},
  number={3},
  pages={231--248},
  year={2003},
  publisher={Sage Publications Sage CA: Los Angeles, CA}
}

@article{candel2015sample,
  title={Sample size calculation for treatment effects in randomized trials with fixed cluster sizes and heterogeneous intraclass correlations and variances},
  author={Candel, M. J. J. M. and {V}an {B}reukelen, G. J. P.},
  journal={Statistical Methods in Medical Research},
  volume={24},
  number={5},
  pages={557--573},
  year={2015},
  publisher={SAGE Publications Sage UK: London, England}
}

@article{Terpstra_2003,
  title={A new nonparametric test for the ordered alternative problem},
  author={Terpstra, J. T. and Magel, R. C.},
  journal={Journal of Nonparametric Statistics},
  volume={15},
  number={3},
  pages={289--301},
  year={2003},
  publisher={}
}

@article{liu2023overview,
  title={An overview of optimal designs under a given budget in cluster randomized trials with a binary outcome},
  author={Liu, Jingxia and Liu, Lei and James, Aimee S and Colditz, Graham A},
  journal={Statistical methods in medical research},
  volume={32},
  number={7},
  pages={1420--1441},
  year={2023},
  publisher={SAGE Publications Sage UK: London, England}
}

@article{muller1998applications,
  title={Applications of necessary and sufficient conditions for maximin efficient designs},
  author={M{\"u}ller, Christine H and P{\'a}zman, Andrej},
  journal={Metrika},
  volume={48},
  number={1},
  pages={1--19},
  year={1998},
  publisher={Springer}
}

@article{liu2019optimal,
  title={Optimal designs in three-level cluster randomized trials with a binary outcome},
  author={Liu, Jingxia and Liu, Lei and Colditz, Graham A},
  journal={Statistics in medicine},
  volume={38},
  number={20},
  pages={3733--3746},
  year={2019},
  publisher={Wiley Online Library}
}

@article{phillips2020evaluating,
  title={Evaluating the impact of biannual school-based and community-wide treatment on urogenital schistosomiasis in Niger},
  author={Phillips, Anna E and Tohon, Zilahatou and Dhanani, Neerav A and Sofo, Boubacar and Gnandou, Issa and Sidikou, Boubacar and Noma, Adamou Garba and Madougou, Bassirou and Alto, Oumarou and Sebangou, Hannatou and others},
  journal={Parasites \& Vectors},
  volume={13},
  number={1},
  pages={557},
  year={2020},
  publisher={Springer}
}

@article{phillips2017assessing,
  title={Assessing the benefits of five years of different approaches to treatment of urogenital schistosomiasis: A SCORE project in Northern Mozambique},
  author={Phillips, Anna E and Gazzinelli-Guimaraes, Pedro H and Aurelio, Herminio O and Ferro, Josefo and Nala, Rassul and Clements, Michelle and King, Charles H and Fenwick, Alan and Fleming, Fiona M and Dhanani, Neerav},
  journal={PLoS neglected tropical diseases},
  volume={11},
  number={12},
  pages={e0006061},
  year={2017},
  publisher={Public Library of Science San Francisco, CA USA}
}

@article{team2015sanitation,
  title={The sanitation hygiene infant nutrition efficacy (SHINE) trial: Rationale, design, and methods},
  author={Team, Trial},
  journal={Clinical Infectious Diseases: An Official Publication of the Infectious Diseases Society of America},
  volume={61},
  number={Suppl 7},
  pages={S685},
  year={2015}
}

@article{humphrey2019independent,
  title={Independent and combined effects of improved water, sanitation, and hygiene, and improved complementary feeding, on child stunting and anaemia in rural Zimbabwe: a cluster-randomised trial},
  author={Humphrey, Jean H and Mbuya, Mduduzi NN and Ntozini, Robert and Moulton, Lawrence H and Stoltzfus, Rebecca J and Tavengwa, Naume V and Mutasa, Kuda and Majo, Florence and Mutasa, Batsirai and Mangwadu, Goldberg and others},
  journal={The Lancet Global Health},
  volume={7},
  number={1},
  pages={e132--e147},
  year={2019},
  publisher={Elsevier}
}

@article{null2018effects,
  title={Effects of water quality, sanitation, handwashing, and nutritional interventions on diarrhoea and child growth in rural Kenya: a cluster-randomised controlled trial},
  author={Null, Clair and Stewart, Christine P and Pickering, Amy J and Dentz, Holly N and Arnold, Benjamin F and Arnold, Charles D and Benjamin-Chung, Jade and Clasen, Thomas and Dewey, Kathryn G and Fernald, Lia CH and others},
  journal={The Lancet Global Health},
  volume={6},
  number={3},
  pages={e316--e329},
  year={2018},
  publisher={Elsevier}
}

@article{liu2017optimal,
  title={Optimal design of longitudinal data analysis using generalized estimating equation models},
  author={Liu, Jingxia and Colditz, Graham A},
  journal={Biometrical Journal},
  volume={59},
  number={2},
  pages={315--330},
  year={2017},
  publisher={Wiley Online Library}
}

@BOOK{Donner_2000,
author = "Donner, A and Klar, N",
title= "Design and Analysis of Cluster Randomised Trials in Health Research",
publisher = "Arnold",
edition = {},
year = {2000},
volume = {},
address = "London",
}

@article{Breukelen_2015,
title	= {Efficient design of cluster randomized and multicentre trials with unknown intraclass correlation},
author	= {Van {B}reukelen, G. J. P. and Candel, M. J. J. M.},
journal	= {Statistical Methods in Medical Research},
volume	= {24},
number	= {5},
pages	= {540-1556},
year	= {2015},
}

@article{Singh_2015,
title	= {Comparison of three-level cluster randomized trials using quantile dispersion graphs},
author	= {Singh, S P and Mukhopadhyay, S and Roy, A},
journal	= {Journal of Applied Statistics},
volume	= {42},
number	= {8},
pages	= {1792-1812},
year	= {2015},
}

@article{Singh_2016,
title	= {Bayesian optimal cluster designs},
author	= {Singh, S P and Mukhopadhyay, S},
journal	= {Statistical Methodology},
volume	= {32},
number	= {},
pages	= {36-52},
year	= {2016},
}

@Article{Lewis_2021,
	author = {Lewis, J. and Julious, S. A.},
	title = {Sample sizes for cluster-randomised trials with continuous outcomes: accounting for uncertainty in a single intra-cluster correlation estimate},
	journal = {Stat Methods Med Res},
	volume = {30},
	number = {11},
	pages = {2459-2470},
	year = {2021},
}

@article{donner1994cluster,
  title={Cluster randomization trials in epidemiology: theory and application},
  author={Donner, Allan and Klar, Neil},
  journal={Journal of Statistical Planning and Inference},
  volume={42},
  number={1-2},
  pages={37--56},
  year={1994},
  publisher={Elsevier}
}

@article{Singh_eff_2019,
title	= {On efficient exact experimental designs for ordered treatments},
author	= {Singh, S. P. and Davidov, O.},
journal	= {Computational Statistics and Data Analysis},
volume	= {164},
number	= {107305},
pages	= {},
year	= {2021},
}

@article{li2020power,
  title={Power analysis for cluster randomized trials with multiple binary co-primary endpoints},
  author={Li, Dateng and Cao, Jing and Zhang, Song},
  journal={Biometrics},
  volume={76},
  number={4},
  pages={1064--1074},
  year={2020},
  publisher={Oxford University Press}
}

@article{yang2023power,
  title={Power analysis for cluster randomized trials with continuous coprimary endpoints},
  author={Yang, Siyun and Moerbeek, Mirjam and Taljaard, Monica and Li, Fan},
  journal={Biometrics},
  volume={79},
  number={2},
  pages={1293--1305},
  year={2023},
  publisher={Wiley Online Library}
}

\section*{Appendix}
\section*{Proof of Theorem \ref{cont.lod.2}} 
 \begin{proof}
For $K=1$, the NCP given in equation~\eqref{cont.ncp} simplifies to
\begin{equation}
\phi(\boldsymbol\pi,\boldsymbol\theta,m,
 \boldsymbol n,\boldsymbol\rho)
=\left(\frac{m}{\sigma^2}\right)\frac{(\theta_1-\theta_0)^2}
{1/(\pi_0r_0)+1/(\pi_1r_1)}.
\label{cont.two_arm_ncp}
\end{equation}
Under exchangeable correlation,
\begin{equation}
r_{\ell}=\frac{n_{\ell}}{1+(n_{\ell}-1)\rho_{\ell}}\implies \frac{1}{r_\ell}
=\rho_\ell+\frac{1-\rho_\ell}{n_\ell}.\nonumber
\end{equation}
Since $m(\theta_1-\theta_0)^2/\sigma^2$ is fixed and positive, maximizing
\eqref{cont.two_arm_ncp} is equivalent to minimizing
\begin{equation}
g(\boldsymbol\pi,\boldsymbol n)
=\frac{1}{\pi_0r_0}+\frac{1}{\pi_1r_1}=\frac{\rho_{0}}{\pi_0}+\frac{\rho_1}{\pi_1}+\frac{1-\rho_0}{\pi_0n_0}+\frac{1-\rho_1}{\pi_1n_1},
\label{cont.lod_objective}\nonumber
\end{equation}
The Lagrangian of this minimization problem over the space $\Xi=\{(\pi_0,\pi_1,n_0,n_1):\pi_0+\pi_1=1; \pi_0n_0+\pi_1n_1=\eta;\pi_{0},\pi_{1}>0\}$ is given as 
\begin{equation}
\begin{aligned}
\mathcal L(\boldsymbol\pi,\boldsymbol n,\kappa,\nu)
&=\frac{\rho_{0}}{\pi_0}+\frac{\rho_1}{\pi_1}+\frac{1-\rho_0}{\pi_0n_0}+\frac{1-\rho_1}{\pi_1n_1}
+\kappa(\pi_0+\pi_1-1)
+\nu(\pi_0n_0+\pi_1n_1-\eta),
\end{aligned}\nonumber
\end{equation}
where $\kappa$ and $\nu$ are the Lagrange multipliers.
For $\ell=0,1$, the first-order conditions are
\begin{eqnarray}
\frac{\partial\mathcal L}{\partial n_\ell}
&=&-\frac{1-\rho_\ell}{\pi_\ell n_\ell^2}+\nu\pi_\ell=0,
\label{cont.lod_foc_n}\\
\frac{\partial\mathcal L}{\partial\pi_\ell}
&=&-\frac{\rho_\ell}{\pi_\ell^2}
-\frac{1-\rho_\ell}{\pi_\ell^2n_\ell}
+\kappa+\nu n_\ell=0.
\label{cont.lod_foc_pi}
\end{eqnarray}

\noindent Using the constraints $\pi_0+\pi_1=1$, $\pi_0n_0+\pi_1n_1=\eta$ and solving equations \eqref{cont.lod_foc_n}, \eqref{cont.lod_foc_pi}, we obtain
\begin{equation}
\pi_\ell=\frac{\sqrt{\rho_\ell}}
{\sqrt{\rho_0}+\sqrt{\rho_1}},\quad n_\ell
=\frac{\eta\sqrt{1-\rho_\ell}}
{\pi_\ell(\sqrt{1-\rho_0}+\sqrt{1-\rho_1})},
\qquad \ell=0,1 \label{cont.sol}
\end{equation}
which proves the expressions in \eqref{cont.lod.eq.2} of Theorem \ref{cont.lod.2} and the value of the function $g$ at this solution is
\begin{equation}
g_{\min}
=\left(\sqrt{\rho_0}
 +\sqrt{\rho_1}\right)^2
 +\frac{1}{\eta}
 \left(\sqrt{1-\rho_0}
 +\sqrt{1-\rho_1}\right)^2.\nonumber
\end{equation}
Now the Cauchy--Schwarz inequality gives,
\begin{align}
g(\boldsymbol\pi,\boldsymbol n)
&=\frac{\rho_{0}}{\pi_0}+\frac{\rho_1}{\pi_1}+\frac{1-\rho_0}{\pi_0n_0}+\frac{1-\rho_1}{\pi_1n_1}\nonumber\\
&\geq
\frac{(\sqrt{\rho_0}+\sqrt{\rho_1})^2}{\pi_0+\pi_1}
+\frac{(\sqrt{1-\rho_0}+\sqrt{1-\rho_1})^2}{t_0+t_1},\quad t_0=\pi_0 n_0, t_1=\pi_1 n_1\nonumber\\
&=(\sqrt{\rho_0}+\sqrt{\rho_1})^2
+\frac{(\sqrt{1-\rho_0}+\sqrt{1-\rho_1})^2}{\eta}=g_{min}.\nonumber
\label{cont.lod_global_bound}
\end{align}
Thus the stationary point in equation \eqref{cont.sol} is the unique global minimizer of $g$ and,
equivalently, the unique maximizer of the NCP in equation~\eqref{cont.two_arm_ncp}.
\end{proof}

\section*{Proof of Theorem \ref{cont.mm.theorem.1.1}}
The proof of Theorem \ref{cont.mm.theorem.1.1} requires the following lemma.
\begin{lemma}\label{cont.lfc.lemma}
Under the restriction, $\theta_{s}-\theta_{t}=\delta$, the minimizer of the ncp in \eqref{cont.ncp} as a function of $\boldsymbol{\theta}$ is $\boldsymbol{\theta^*}=(\theta_{0}^*,\theta_{1}^*,\hdots,\theta_{K}^*)$ with $\theta_{s}^*=a+\delta,\quad \theta_{t}^*=a, \quad \theta_{\ell}^*=a+\frac{r_{s}\pi_{s}}{r_{s}\pi_{s}+r_{t}\pi_{t}}\delta, \qquad\forall \ell\in \{0,1,2,\hdots,K\}/\{s,t\}$, where $a\in \mathbb{R}$ is some arbitrary constant. 
\end{lemma}
\begin{proof}
Without loss of generality, we assume $0<s<t<K$. The Lagrangian function is given by 
\begin{equation*}
 \mathcal L(\boldsymbol{\theta},\mu) =\phi(\boldsymbol{\pi},\boldsymbol{\theta},\boldsymbol{\rho},\boldsymbol{n},m)+\mu(\theta_{s}-\theta_{t}-\delta),
\end{equation*} 
where $\mu\in\mathbb{R}$. Ignoring the multiplicative constant $m/\sigma^2$ in NCP, the optimal value must satisfy the following set of equations
\begin{eqnarray*}
 \frac{\partial \mathcal L}{\partial \theta_{\ell}}&=& 2r_{\ell}\pi_{\ell}(\theta_{\ell}-\theta_{K})-\frac{2r_{\ell}\pi_{\ell}}{\boldsymbol{r}^{\rm T}\boldsymbol{\pi}}\sum\limits_{\ell=0}^{K-1}r_{\ell}\pi_{\ell}(\theta_{\ell}-\theta_{K})=0,\; \forall \ell\in \mathcal{S}/\{s,t,K\}\\
 \frac{\partial \mathcal L}{\partial \theta_{s}}&=& 2r_{s}\pi_{s}(\theta_{s}-\theta_{K})-\frac{2r_{s}\pi_{s}}{\boldsymbol{r}^{\rm T}\boldsymbol{\pi}}\sum\limits_{\ell=0}^{K-1}r_{\ell}\pi_{\ell}(\theta_{\ell}-\theta_{K})+\mu=0\\
 \frac{\partial \mathcal L}{\partial \theta_{t}}&=& 2r_{t}\pi_{t}(\theta_{t}-\theta_{K})-\frac{2r_{t}\pi_{t}}{\boldsymbol{r}^{\rm T}\boldsymbol{\pi}}\sum\limits_{\ell=0}^{K-1}r_{\ell}\pi_{\ell}(\theta_{\ell}-\theta_{K})-\mu=0\\
 \frac{\partial \mathcal L}{\partial \theta_{K}}&=&-2\sum\limits_{\ell=0}^{K-1}r_{\ell}\pi_{\ell}(\theta_{\ell}-\theta_{K})+\frac{2}{\boldsymbol{r}^{\rm T}\boldsymbol{\pi}}\sum\limits_{\ell=0}^{K-1}r_{\ell}\pi_{\ell}\sum\limits_{\ell=0}^{K-1}r_{\ell}\pi_{\ell}(\theta_{\ell}-\theta_{K})=0\\
 \frac{\partial \mathcal L}{\partial \mu}&=&\theta_{s}-\theta_{t}-\delta=0
\end{eqnarray*}
It can be easily verified that $\boldsymbol{\theta} = \boldsymbol{\theta}^*$ with $\mu=(2r_{s}r_{t}\pi_{s}\pi_{t})/(r_{s}\pi_{s}+r_{t}\pi_{t})$ satisfy the above set of equations. 
Next we shall prove that $\boldsymbol{\theta}^*$ is a global minimizer. The NCP (as a function of $\boldsymbol{\theta}$) can be written as 
\begin{equation*}
 \phi(\boldsymbol{\theta})=\frac{m}{\sigma^{2}}\boldsymbol{v}^{\rm T}({\rm D}-\frac{\boldsymbol{w}\boldsymbol{w}^{\rm T}}{\boldsymbol{r}^{\rm T}\boldsymbol{\pi}})\boldsymbol{v},
\end{equation*} where $\boldsymbol{v}=(\theta_0-\theta_K,\theta_1-\theta_K,\hdots,\theta_{K-1}-\theta_K)^{\rm T}$ , $\boldsymbol{w}=(r_{0}\pi_{0},r_{1}\pi_{1},\hdots,r_{K-1}\pi_{K-1})^{\rm T}$ and $\rm D=diag(\boldsymbol{w})$. Using the Cauchy--Schwarz inequality it can be shown that 
$({\rm D}-\boldsymbol{ww}^{\rm T}/(\boldsymbol{r}^{\rm T}\boldsymbol{\pi})$ is a positive definite matrix. Now the convexity of $\phi(\boldsymbol{\theta})$ follows since the convexity is preserved under linear transformation. Further, the constraint $\theta_{s}-\theta_{t}=\delta$ is a convex set. Thus $\boldsymbol{\theta} = \boldsymbol{\theta}^*$ is the global minimizer of the NCP. This completes the proof.
\end{proof}

Now we will continue with the proof of Theorem \ref{cont.mm.theorem.1.1}.

\begin{proof}
We will prove the theorem in two parts: \textbf{Part I} and \textbf{Part II}. In \textbf{Part I}, we  take the proposed LFCs and an LFD in $\mathcal Q_{\rm LFD}$ as candidates, and show that the average NCP is maximized at the stated design $\boldsymbol{\pi}_{\rm MM}$. \textbf{Part II}, fixes
design and we show that $\boldsymbol{\Theta}_{\rm LFC}$ is the set of LFC and $\mathcal Q_{\rm LFD}$ is the corresponding set of LFD. The maximin property is then established using the fact that the  Nash equilibrium is the solution of this maximin problem. We follow the notations defined in the statement of theorem. 
\\

\noindent\textbf{Part I:} The NCP in equation~\eqref{cont.ncp} can be written as 
\begin{equation*}
\phi(\boldsymbol{\pi},\boldsymbol\theta)
=\frac{m}{\sigma^2}\sum_{\ell=0}^{K}w_\ell(\theta_\ell-\bar\theta)^2= \frac{m}{\sigma^2}\min\limits_{c}\sum_{\ell=0}^{K}w_\ell(\theta_\ell-c)^2,
\end{equation*}
where, $w_{\ell}=r_{\ell}\pi_{\ell}$ and $\bar{\theta}=\frac{\sum _{\ell=0}^{K}w_{\ell}\theta_{\ell}}{\sum_{\ell=0}^{K}w_{\ell}}$. Therefore, for any fixed c, we have
\begin{equation}
\phi(\boldsymbol{\pi},\boldsymbol\theta)
\leq\frac {m}{\sigma^2}\sum_{\ell=0}^{K}w_\ell(\theta_\ell-c)^2,\label{cont.ncp.upperbound}
\end{equation}
\textbf{Case 1.} The average NCP function with respect to $\boldsymbol{\Theta}_{\rm LFC}$ and any $\boldsymbol{q} ^{*}\in \mathcal Q_{\rm LFD}$ is defined as 
\begin{eqnarray*}
 \Psi(\boldsymbol{\pi}, \boldsymbol{q}^{*})= \sum\limits_{0\leq i<j \leq K} q_{ij}^{*}\phi(\boldsymbol{\pi},\boldsymbol{\theta}^{(ij)})
\end{eqnarray*}
Since $\ell$th component of $\boldsymbol\theta^{(ij)}$ is
\begin{equation}
\theta_\ell^{(ij)}
=
\begin{cases}
\delta/2, & \ell=i,\\
-\delta/2, & \ell=j,\\
0, & \ell\notin\{i,j\}.
\end{cases}
\end{equation}
Taking $c=0$ in \eqref{cont.ncp.upperbound}  gives $\phi(\boldsymbol\pi,\boldsymbol\theta^{(ij)})
\leq
\frac{m\delta^2}{4\sigma^2}
(\pi_i r_i+\pi_j r_j)$, and 
\begin{align}
\Psi(\boldsymbol{\pi}, \boldsymbol{q}^{*})
&\leq\frac{m\delta^2}{4\sigma^2}\sum\limits_{0\leq i<j \leq K} q_{ij}^{*}(\pi_i r_i+\pi_j r_j)=
\frac{m\delta^2}{4\sigma^2}
\sum_{\ell=0}^{K}
\pi_\ell r_\ell
\left(
\sum_{i<\ell}q_{i\ell}^{*}
+
\sum_{j>\ell}q_{\ell j}^{*}
\right)\notag\\
&=
\frac{m\delta^2}{4\sigma^2}
\sum_{\ell=0}^{K}
\pi_\ell r_\ell
\frac{2r_\ell^{-1}}{\sum_{h=0}^{K}r_h^{-1}}=
\frac{m\delta^2}{4\sigma^2}\frac2 A= \frac{m\delta^2}{2\sigma^2 A}
\label{cont.average.ncp.upperbound.eqn}
\end{align}
Note that, $\phi(\boldsymbol{\pi}_{\rm MM},\boldsymbol{\theta}^{(ij)})=\frac{m\delta^2}{4\sigma^2}\frac{2}{A}=\frac{m\delta^2}{2\sigma^2 A}$, thus 
\begin{eqnarray*}
 \Psi(\boldsymbol{\pi}_{\rm MM}, \boldsymbol{q}^{*})= \sum\limits_{0\leq i<j \leq K} q_{ij}^{*}\phi(\boldsymbol{\pi}_{\rm MM},\boldsymbol{\theta}^{(ij)})=\frac{m\delta^2}{2\sigma^2 A}
\end{eqnarray*}
Hence, $\boldsymbol{\pi}_{\rm MM}$ attains the upper bound~\eqref{cont.average.ncp.upperbound.eqn} and is therefore the global maximizer of the average NCP. 

\noindent\textbf{Case 2.} Now suppose that for some $s\in \mathcal{S}$, $r_{s}^{-1}>A_{s}$. Since the $\ell$th component of $\boldsymbol\theta^{(sj)}$is
\begin{equation}
\theta_\ell^{(sj)}
=
\begin{cases}
\frac \delta2,
& \ell=s,\\
-\frac \delta2,
& \ell=j,\\
\frac{\delta}{2}
\frac{\sqrt{A_s}-\sqrt{r_s^{-1}}}
{\sqrt{A_s}+\sqrt{r_s^{-1}}},
& \ell\notin\{s,j\},
\end{cases}
\end{equation}
Taking $
c=\frac{\delta}{2}
\frac{\sqrt{A_s}-\sqrt{r_s^{-1}}}
{\sqrt{A_s}+\sqrt{r_s^{-1}}}$, 
and since, $\theta_s^{(sj)}-\theta_j^{(sj)}=\delta$, direct subtraction gives
\[
\theta_s^{(sj)}-c=
\frac{\delta\,\sqrt{r_s^{-1}}}
{\sqrt{A_s}+\sqrt{r_s^{-1}}},
\qquad \theta_j^{(sj)}-c=
-\ \frac{\delta\,\sqrt{A_{s}}}
{\sqrt{A_s}+\sqrt{r_s^{-1}}}.
\]
Also, $\theta_{\ell}=c, \forall \ell \in \mathcal{S}\setminus\{s,j\}$, thus from~\eqref{cont.ncp.upperbound} we obtain
\[
\phi(\boldsymbol\pi,\boldsymbol\theta^{(sj)})
\leq\frac{m\delta^2}{\sigma^2(\sqrt{A_s}+\sqrt{r_s^{-1}})^{2}}
\left(r_s r_s^{-1}\pi_s+r_jA_{s}\pi_j
\right)=\frac{m\delta^2}{\sigma^2(\sqrt{A_s}+\sqrt{r_s^{-1}})^{2}}
\left(\pi_s+r_jA_{s}\pi_j
\right).
\]
Let $\boldsymbol{q}^{*}\in \mathcal Q_{\rm LFD}$, defined in~\eqref{cont.lfd.eqn.1.case2}. Its $j$--th element is given by $q_{sj}^{*}=\frac{r_j^{-1}}{A_{s}},\;\text{and}\;\sum_{j\in \mathcal{S}\setminus\{s\}}q_{sj}^{*}=1$.
Therefore , the average NCP is given by
\begin{align}
\Psi(\boldsymbol\pi,\boldsymbol q^{*})=
\sum_{j\neq s}^{K}
q_{sj}\phi(\boldsymbol\pi,\boldsymbol\theta^{(sj)})
\notag
&\leq \frac{m\delta^2}{\sigma^2(\sqrt{A_s}+\sqrt{r_s^{-1}})^{2}}\sum_{j\neq s}^{K}
q_{sj}^{*}
\left(\pi_s+r_jA_{s}\pi_j\right)
\notag\\
&=\frac{m\delta^2}{\sigma^2(\sqrt{A_s}+\sqrt{r_s^{-1}})^{2}}
\left(\pi_s+\sum_{j\neq s}^{K}
q_{sj}^{*}r_jA_{s}\pi_j\right)
\notag\\
&=\frac{m\delta^2}{\sigma^2(\sqrt{A_s}+\sqrt{r_s^{-1}})^{2}}
\left(\pi_s+\sum_{j\neq s}^{K}\frac{r_{j}^{-1}}{A_{s}}r_jA_{s}\pi_j\right)
\notag\\
&=\frac{m\delta^2}{\sigma^2(\sqrt{A_s}+\sqrt{r_s^{-1}})^{2}}
\left(\sum_{\ell=0}^{K}\pi_\ell\right)=\frac{m\delta^2}{\sigma^2(\sqrt{A_s}+\sqrt{r_s^{-1}})^{2}}
\label{cont.average.ncp.upperbound.case2}
\end{align}

It can be easily verified that the equality holds in~\eqref{cont.average.ncp.upperbound.case2} for the $\boldsymbol{\pi}_{\rm MM}$ given in \eqref{cont.mm.eqn.1.case2}. Thus, the design $\boldsymbol{\pi}_{\rm MM}$ is a global maximizer of the average NCP.\\

\noindent\textbf{Part II:} Fixing the design $\boldsymbol{\pi}$ to be $\boldsymbol{\pi}_{\rm MM}$, we will show that the set of LFC is $\Theta_{\rm LFC}$ defined in the statement of therorem with the associated LFD $\mathcal Q_{\rm LFD}$. Let $\boldsymbol{\theta}^{(ij)}$ be the minimizer of NCP given by Lemma~\ref{cont.lfc.lemma} when the maximum separation of $\delta$ is in pair \{i,j\} with i<j . Since the NCP is a non-decreasing function of the absolute differences $|\theta_i-\theta_j|$. Thus, the minimum over $\Delta_\delta$ is found among
configurations having range exactly $\delta$, that 
\begin{eqnarray*}
 \arg\min\limits_{\boldsymbol{\theta}\in \Delta_{\delta}}\phi(\boldsymbol \pi_{\rm MM},\boldsymbol{\theta})\in\{\boldsymbol{\theta}^{ij},\ \boldsymbol{\theta}^{ji}:i<j, i,j\in \mathcal{S}\}, 
\end{eqnarray*}
where the $\ell$--th component of $\boldsymbol \theta^{(ij)}$ is given by 
\begin{eqnarray}\label{cont.lfc.eqn.from.lemma}
 \theta_{\ell}^{(ij)}=\begin{cases}
 & \delta/2, \qquad \ell=i\\
 & -\delta/2, \qquad \ell =j,\\
 & \frac\delta2\frac{r_{i}\pi_{i}-r_{j}\pi_{j}}{r_{i}\pi_{i}+r_{j}\pi_{j}}, \qquad \ell\in \mathcal{S}\setminus\{i,j\}.
 \end{cases}
\end{eqnarray}
Here $ \theta_{\ell}^{(ij)}$ is obtained by taking $a=-\delta/2$ in the Lemma~\ref{cont.lfc.lemma}.\\

\noindent\textbf{Case 1.}
Since $\boldsymbol{\pi}_{\rm MM}$ in this case is such that all $\pi_\ell r_\ell$ are equal, therefore from the equation~\eqref{cont.lfc.eqn.from.lemma}, we get 
\begin{eqnarray}
 \theta_{\ell}^{(ij)}=\begin{cases}
 & \delta/2, \qquad \ell=i\\
 & -\delta/2, \qquad \ell =j,\\
 & 0, \qquad \ell\in \mathcal{S}\setminus\{i,j\}.
 \end{cases}\nonumber
\end{eqnarray}
which is same as that of the set given in~\eqref{cont.lfc.eqn.1.case1} and also $\boldsymbol{\theta}^{(ji)}=-\boldsymbol{\theta}^{(ij)}$. Thus the set $\boldsymbol{\Theta}_{\rm LFC}=\{\boldsymbol{\theta}^{ij},\ -\boldsymbol{\theta}^{ij}:i<j, i,j\in \mathcal{S}\}$ is the set of LFCs. Also, 
\[
\phi(\boldsymbol\pi_{\rm MM},\boldsymbol{\theta}^{(ij)})
=
\frac{m\delta^2}
{2\sigma^2A},\quad \forall \;\boldsymbol{\theta}^{(ij)}\in \Theta_{\rm LFC} .
\]
and the average NCP for any probability distribution $\boldsymbol{q}$ on the set $\Theta_{\rm LFC}$ is 
\begin{eqnarray*}
 \psi(\boldsymbol{\pi}_{\rm MM}, \boldsymbol{q})= \sum\limits_{0\leq i<j \leq K} q_{ij}\phi(\boldsymbol{\pi}_{\rm MM},\boldsymbol{\theta}^{(ij)})=\frac{m\delta^2}
{2\sigma^2A}.
\end{eqnarray*}
This is a constant, therefore, every proposed
$\boldsymbol q\in\mathcal Q_{\rm LFD}$ is a minimizing response.\\

\noindent\textbf{Case 2.} Since $\boldsymbol{\pi}_{\rm MM}$ in this case is such that all $\pi_s r_s=\sqrt{r_{s}}\sqrt{A_{s}} \pi_{\ell}r_{\ell}, \forall \ell \in \mathcal{S}\setminus\{s\}$ and $ \pi_{i}r_{i}=\pi_{j}r_{j}, \forall i,j \in \mathcal{S}\setminus\{s\}$. Therefore, for $i\neq j$, the $\ell$--th element of $\boldsymbol{\theta}^{(ij)}$ using equation~\eqref{cont.lfc.eqn.from.lemma}, is given 
\begin{eqnarray*}
\theta_\ell^{(ij)}
&=&
\begin{cases}
\delta/2,
& \ell=i,\\[0.3em]
-\delta/2,
& \ell=j,\\[0.3em]
0,
& \ell\notin\{i,j\},\quad s\notin\{i,j\},\\[0.6em]
\displaystyle
\frac{\delta}{2}
\frac{\sqrt{A_s}-\sqrt{r_s^{-1}}}
{\sqrt{A_s}+\sqrt{r_s^{-1}}},
& \ell\notin\{i,j\},\quad i=s,\\[1em]
\displaystyle
-\frac{\delta}{2}
\frac{\sqrt{A_s}-\sqrt{r_s^{-1}}}
{\sqrt{A_s}+\sqrt{r_s^{-1}}},
& \ell\notin\{i,j\},\quad j=s.
\end{cases}
\end{eqnarray*}
In particular, $\boldsymbol\theta^{(js)}=-\boldsymbol\theta^{(sj)}$. Also, from \eqref{cont.lfc.eqn.from.lemma}
\[
\bar\theta^{(ij)}
=\frac{\sum_{\ell=0}^{K}r_{\ell}\pi_{\ell}\theta_{\ell}^{(ij)}}{\sum_{\ell=0}^{K}r_{\ell}\pi_{\ell}}=
\frac{\pi_i r_i(\delta/2)+\pi_j r_j(-\delta/2)}
{\pi_i r_i+\pi_j r_j}.
\]
Moreover, $
\theta_i^{(ij)}-\bar\theta^{(ij)}
=
\frac{\delta\,\pi_j r_j}{\pi_i r_i+\pi_j r_j},
\qquad
\theta_j^{(ij)}-\bar\theta^{(ij)}
=
-\frac{\delta\,\pi_i r_i}{\pi_i r_i+\pi_j r_j}$ and $\theta_{\ell}^{(ij)}-\bar{\theta}^{(ij)}=0, \ell\notin \{i,j\}$. Thus,
\begin{eqnarray*}
\begin{aligned}
\phi(\boldsymbol\pi_{\rm MM},\boldsymbol\theta^{(ij)})
&=
\frac{m}{\sigma^2}
\left\{
\pi_i r_i
\left(
\frac{\delta\,\pi_j r_j}{\pi_i r_i+\pi_j r_j}
\right)^2
+
\pi_j r_j
\left(
\frac{\delta\,\pi_i r_i}{\pi_i r_i+\pi_j r_j}
\right)^2
\right\}=
\frac{m\delta^2}{\sigma^2}
\frac{
\pi_i r_i\pi_j r_j
(\pi_i r_i+\pi_j r_j)
}{
(\pi_i r_i+\pi_j r_j)^2
}\\
&=
\frac{m\delta^2}{\sigma^2}
\frac{\pi_i r_i\pi_j r_j}
{\pi_i r_i+\pi_j r_j}=
\frac{m\delta^2}{\sigma^2}
\left(
\frac1{\pi_i r_i}+\frac1{\pi_j r_j}
\right)^{-1}.
\end{aligned}
\end{eqnarray*}
For a pair $(s,j)$ with $j\neq s$, substitution gives
\begin{align*}
\phi(\boldsymbol\pi_{\rm MM},\boldsymbol\theta^{(sj)})
&=
\frac{m\delta^2}{\sigma^2}
\left\{
\sqrt{r_s^{-1}}
\left(\sqrt{r_s^{-1}}+\sqrt{A_s}\right)
+
\sqrt{A_s}
\left(\sqrt{r_s^{-1}}+\sqrt{A_s}\right)
\right\}^{-1}=
\frac{m\delta^2}
{\sigma^2
 \left(\sqrt{r_s^{-1}}+\sqrt{A_s}\right)^2}.
\end{align*}
This value is the same for every $j\neq s$ and for the opposite orientation $-\boldsymbol\theta^{(sj)}$.
For distinct $i,j\in\mathcal S\setminus\{s\}$,
\begin{align*}
\phi(\boldsymbol\pi_{\rm MM},\boldsymbol\theta^{(ij)})=
\frac{m\delta^2}{\sigma^2}
\left\{
2\sqrt{A_s}
\left(\sqrt{r_s^{-1}}+\sqrt{A_s}\right)
\right\}^{-1}=
\frac{m\delta^2}
{2\sigma^2\sqrt{A_s}
 \left(\sqrt{r_s^{-1}}+\sqrt{A_s}\right)}.
\end{align*}
Since $r_s^{-1}>A_s$, we have $\sqrt{r_s^{-1}}+\sqrt{A_s}>2\sqrt{A_s}$. Therefore, for any $h\neq s$ and distinct $i,j\neq s$,
\[
\frac{
\phi(\boldsymbol\pi_{\rm MM},\boldsymbol\theta^{(sh)})
}{
\phi(\boldsymbol\pi_{\rm MM},\boldsymbol\theta^{(ij)})
}
=
\frac{2\sqrt{A_s}}
{\sqrt{r_s^{-1}}+\sqrt{A_s}}
<1.
\]
Thus, among all $\theta^{(ij)}$ with the maximum separation between the $i$--th and $j$--th components being $\delta$, only the vectors where the maximum separated pair involves arm $s$ attain
the smallest NCP. Consequently, the set of LFC is $\Theta_{\rm LFC}=\left\{
\boldsymbol\theta^{(sj)},
-\boldsymbol\theta^{(sj)}
:
j\in\mathcal S\setminus\{s\}
\right\}$. Also, the average NCP for any probability distribution $\boldsymbol{q}$ on the set $\Theta_{\rm LFC}$ is 
\begin{eqnarray*}
 \Psi(\boldsymbol{\pi}_{\rm MM}, \boldsymbol{q})= \sum\limits_{j\neq s} q_{sj}\phi(\boldsymbol{\pi}_{\rm MM},\boldsymbol{\theta}^{(sj)})=\frac{m\delta^2}
{\sigma^2
 \left(\sqrt{r_s^{-1}}+\sqrt{A_s}\right)^2}.
\end{eqnarray*}
Since this is a constant, therefore every proposed $\boldsymbol q^{*}\in\mathcal Q_{\rm LFD}$ is a minimizing distribution.\\
\medskip
\noindent
In either case, we established that, 
\[
\max_{\boldsymbol\pi\in\Pi}
\min_{\boldsymbol q \in \mathcal Q}
\Psi(\boldsymbol\pi,\boldsymbol q)
=
\min_{\boldsymbol q \in \mathcal Q}
\max_{\boldsymbol\pi\in\Pi}
\Psi(\boldsymbol\pi,\boldsymbol q)
=\Psi(\boldsymbol\pi_{\rm MM},\boldsymbol q^{*})
\]
where $\boldsymbol q^{*}\in\mathcal Q_{\rm LFD}$and $\mathcal Q$ is any probability distribution defined on $\Theta_{\rm LFC}$. Thus $(\boldsymbol\pi_{\rm MM},\boldsymbol q^{*})$ is the saddle point and therefore, $\boldsymbol\pi_{\rm MM}$ is the required maximin design.  This completes the proof. 
\end{proof}

\section*{Proof of Theorem \ref{cont.equivalence.theorem.1}}
\begin{proof}
Using the notations established in the statement of the Theorem \ref{cont.equivalence.theorem.1} and let $\mathbf{A}=\mathbf{A(\theta)}$. The NCP in equation \eqref{cont.ncp} can be written as
\begin{equation*}
 \phi(\boldsymbol{\pi},\boldsymbol{\theta}, m,\boldsymbol n,\boldsymbol\rho)=\frac{m\boldsymbol{\pi}^{\rm T}{\rm A}\boldsymbol{\pi}}{\sigma^{2}\boldsymbol{r^{\rm T}\boldsymbol{\pi}}},
\end{equation*}

The directional derivative of the NCP in the point $\boldsymbol{\pi}^{*}$ and in the direction $\boldsymbol{\pi} $ is given by 
\begin{eqnarray} \nabla(\boldsymbol{\pi}^{*},\boldsymbol{\pi} ,\boldsymbol{\theta}, m,\boldsymbol n,\boldsymbol\rho)&=&\lim_{\varepsilon \to 0^+} \frac{1}{\varepsilon}\left\{\phi(\varepsilon\boldsymbol{\pi} +(1-\varepsilon)\boldsymbol{\pi}^{*},\boldsymbol{\theta}, m,\boldsymbol n,\boldsymbol\rho)-\phi(\boldsymbol{\pi}^{*},\boldsymbol{\theta}, m,\boldsymbol n,\boldsymbol\rho)\right\}\nonumber\\
 &=&\frac{m}{\sigma^{2}} \lim_{\varepsilon \to 0^+} \frac{1}{\varepsilon}\left(\frac{(\varepsilon\boldsymbol{\pi} +(1-\varepsilon)\boldsymbol{\pi}^{*})^{\rm T}{\rm A} (\varepsilon\boldsymbol{\pi} +(1-\varepsilon)\boldsymbol{\pi}^{*})}{\boldsymbol{r}^{\rm T}(\varepsilon\boldsymbol{\pi} +(1-\varepsilon)\boldsymbol{\pi}^{*})}-\frac{\boldsymbol{\pi}^{*^{\rm T}}{\rm A}\boldsymbol{\pi}^{*}}{\boldsymbol{r^{\rm T}\boldsymbol{\pi}^{*}}}\right)\nonumber\\
 &=&\frac{m}{\sigma^{2}}\left\{\frac{2\boldsymbol{\pi}^{\rm T}\mathbf{A}\boldsymbol{\pi}^{*}(\boldsymbol{r}^{\rm T}\boldsymbol{\pi}^{*})-\boldsymbol{\pi}^{*}\mathbf{A}\boldsymbol{\pi}^{*}(\boldsymbol{r}^{\rm T}\boldsymbol{\pi}^{*}+\boldsymbol{r}^{\rm T}\boldsymbol{\pi})}{(\boldsymbol{r}^{\rm T}\boldsymbol{\pi}^{*})^{2}}\right\}\nonumber
\end{eqnarray}

In context of minimax designs, \cite{muller1998applications} established an equivalence theorem that provides the necessary and sufficient condition for any design to be a minimax design. Adopting their results to the present setting with the necessary modifications, the necessary and sufficient condition for any design $\boldsymbol{\pi}^{*}$ to be a maximin optimal design is 
\begin{eqnarray}
\sum\limits_{h=0}^{\mathcal{G}}q_{h}\nabla(\boldsymbol{\pi}^{*},\boldsymbol{\pi} ,\boldsymbol{\theta}^{(h)}, m,\boldsymbol n,\boldsymbol\rho)&\leq& 0, \quad \forall \boldsymbol{\pi} \in \Pi \nonumber\\
\implies \sum\limits_{h=0}^{\mathcal{G}}q_{h}\left(2\boldsymbol{\pi}^{\rm T}\mathbf{A}^{(h)}\boldsymbol{\pi}^{*}(\boldsymbol{r}^{\rm T}\boldsymbol{\pi}^{*})-\boldsymbol{\pi}^{*^{\rm T}}\mathbf{A}^{(h)}\boldsymbol{\pi}^{*}(\boldsymbol{r}^{\rm T}\boldsymbol{\pi}^{*}+\boldsymbol{r}^{\rm T}\boldsymbol{\pi})\right) &\leq& 0, \quad \forall \boldsymbol{\pi} \in \Pi \nonumber.
\end{eqnarray}
Since the design space $\Pi$ is convex and compact, the above inequality reduces to 
\begin{eqnarray}
\sum_{h=0}^{ G}q_{h}\left(2\boldsymbol{\pi}_{s}^{\rm T}\mathbf{A}^{(h)}\boldsymbol{\pi}^{*}(\boldsymbol{r}^{\rm T}\boldsymbol{\pi}^{*})-\boldsymbol{\pi}^{*}\mathbf{A}^{(h)}\boldsymbol{\pi}^{*}(\boldsymbol{r}^{\rm T}\boldsymbol{\pi}^{*}+\boldsymbol{r}^{\rm T}\boldsymbol{\pi}_{s})\right)&\leq&0, \;\; \forall s\in \mathcal{S},
 \end{eqnarray}
where $\boldsymbol{\pi}_{s}\in\Pi$ assigns unit weight to $s$--th treatment arm. This completes the proof.\\\\
\end{proof}
The following result illustrates the utility of above equivalence condition. 

\begin{result}\label{balanced.design.verification.equivalence} 
As an example, we verify the maximin optimality of balanced design given in Corollary \ref{cont.mm.balanced.theorem.1.1} using above equivalence condition. Therefore, $\boldsymbol{\pi}^{*}=(K+1)^{-1}\mathbf{1}_{K+1}$ and the set of $LFCs$ is $\boldsymbol{\Theta}_{\rm LFC}=(\boldsymbol{\theta}^{ij}; i\neq j,\quad i,j\in \mathcal{S})$, where the $\ell$-th component of $\boldsymbol{\theta}^{ij}$ is given by 
\begin{equation} 
\theta^{ij}_{\ell}=\begin{cases}
\delta/2, &\mbox{ if }\ell=i,\\
-\delta/2,& \mbox{ if }\ell=j,\\
0,& \mbox{ if } \ell\in \mathcal{S}/\{i,j\}.
\end{cases}
\end{equation}
 Thus we must have, 
\begin{eqnarray*}
 \frac{1}{K(K+1)}\sum\limits_{i,j\in \mathcal{S}}(2\boldsymbol{\pi}^{\rm T}\mathbf{A}^{ij}\boldsymbol{\pi}^{*}(\boldsymbol{r}^{\rm T}\boldsymbol{\pi}^{*})-\boldsymbol{\pi}^{*^{\rm T}}\mathbf{A}^{ij}\boldsymbol{\pi}^{*}(\boldsymbol{r}^{\rm T}\boldsymbol{\pi}^{*}+\boldsymbol{r}^{\rm T}\boldsymbol{\pi})&\leq& 0, \quad \forall \boldsymbol{\pi} \in \Pi 
\end{eqnarray*}
where $\mathbf{A}^{ij}$ is the matrix $\mathbf{A}$ with $\boldsymbol{\theta}=\boldsymbol{\theta}^{ij}$. Since $r_{\ell}=r,\quad \forall \ell \in \mathcal{S}$, therefore, $\boldsymbol{r}^{\rm T}\boldsymbol{\pi}=r, \quad\forall \pi \in \Pi$

\[
 \boldsymbol\pi^{*\rm T}\mathbf{A}^{ij}\boldsymbol\pi^*
 = \frac{r^2\delta^2}{2(K+1)}, \qquad
 \boldsymbol\pi^{\rm T}\mathbf{A}^{ij}\boldsymbol\pi^*
 = \frac{r^2\delta^2}{K+1}\left(\frac14+\frac{(K+1)(\pi_{i}+\pi_{j})}{8}\right).
\]

\[
2\boldsymbol\pi^{\rm T}\mathbf{A}^{ij}\boldsymbol\pi^*
 (\boldsymbol r^{\rm T}\boldsymbol\pi^*)
 -\boldsymbol\pi^{*\rm T}\mathbf{A}^{ij}\boldsymbol\pi^*
 (\boldsymbol r^{\rm T}\boldsymbol\pi^*+\boldsymbol r^{\rm T}\boldsymbol\pi)
 = \frac{r^3\delta^2}{4}\left(\pi_i+\pi_j-\frac{2}{K+1}\right)
\]

\smallskip
Using
$\sum_{i}\sum_{j\ne i}(\pi_i+\pi_j)=2K$ , we get
\begin{align*}
 \sum_{\substack{i,j\in\mathcal S\\ i\ne j}} (2\boldsymbol\pi^{\rm T}\mathbf{A}^{ij}\boldsymbol\pi^*
 (\boldsymbol r^{\rm T}\boldsymbol\pi^*)
 -\boldsymbol\pi^{*\rm T}\mathbf{A}^{ij}\boldsymbol\pi^*
 (\boldsymbol r^{\rm T}\boldsymbol\pi^*+\boldsymbol r^{\rm T}\boldsymbol\pi)) =\frac{r^3\delta^2}{4}\left[2K-K(K+1)\cdot\frac{2}{K+1}\right]=0.
\end{align*}
which establishes the required equivalence condition, confirming that the
balanced design $\boldsymbol\pi^{*}=(K+1)^{-1}\mathbf 1_{K+1}$ is the maximin
design when $r_{\ell}=r,\quad \forall \ell \in \mathcal{S}$.
\end{result}

\section*{Proof of Theorem~\ref{cont.mm.theorem.joint}.} 
The proof of Theorem~\ref{cont.mm.theorem.joint} requires the following Lemma.
\begin{lemma}\label{cont.mm.root.lemma}
Following the notations established in the statement of 
Theorem~\ref{cont.mm.theorem.joint},
suppose that $e_s^2>1/2$ for some $s\in\mathcal S$.
For $x>\Lambda_s$, define $\boldsymbol t(x)=(x\boldsymbol I_K-\mathbf H_{ss})^{-1}\boldsymbol h_s,
\quad\tau_s(x)
=h_{ss}+\boldsymbol h_s^{\rm T}\boldsymbol t(x).$ Then the equation
$\|\boldsymbol t(x)\|_2^2=1$ has exactly one solution $z$ in $(\Lambda_s,\Lambda)$
and no additional solutions for $x\geq\Lambda$. Moreover, $\tau_s(z)>z>0.$
\end{lemma}

\begin{proof}
The proof follows the same argument as that of
Lemma~\ref{binary.mm.root.lemma}, with the right-hand
side of the root equation replaced by $1$.
\end{proof}

We now proceed with the proof of Theorem~\ref{cont.mm.theorem.joint}
\begin{proof} 
As in the proof of Theorem~\ref{cont.mm.theorem.1.1}, the proof of Theorem~\ref{cont.mm.theorem.joint} follows in two parts: \textbf{Part I} and \textbf{Part II}. In Part I, we hold the proposed LFC set and LFDs set fixed and show that the
average NCP is maximized at the proposed joint design $(\boldsymbol{\pi}_{\rm MM},\boldsymbol{n}_{\rm MM})$.
In Part II, we fix the joint design at $(\boldsymbol{\pi}_{\rm MM},\boldsymbol{n}_{\rm MM})$, and show that the LFC
set is $\Theta_{\rm LFC}$ with associated set of LFDs $\mathcal Q_{\rm LFD}$. We follow the notations defined in the statement of theorem. 

\noindent\textbf{Part I:} \\
\noindent\textbf{Case 1.}
For any $\boldsymbol{\theta}^{(ij)}\in \Theta_{\rm LFC}$, and $\boldsymbol{q}^{*}\in \mathcal Q_{\rm LFD}$, the average NCP is bounded by
\begin{eqnarray}\label{cont.average.ncp.upperbound.joint.case1}
\begin{aligned}
\Psi(\boldsymbol{\pi},\boldsymbol{n}, \boldsymbol{q}^{*})
\leq&\frac{m\delta^2}{4\sigma^2}\sum\limits_{0\leq i<j \leq K} q_{ij}^{*}(\pi_i r_i+\pi_j r_j)\\
=&\frac{m\delta^2}{4\sigma^2}
\sum_{\ell=0}^{K}
\pi_\ell r_\ell
\left(
\sum_{i<\ell}q_{i\ell}^{*}
+
\sum_{j>\ell}q_{\ell j}^{*}
\right)=\frac{m\delta^2}{2\sigma^2}
\sum_{\ell=0}^{K}
\pi_\ell r_\ell e_{\ell}^{2}
\end{aligned}
\end{eqnarray}
Since, $\mathbf H\boldsymbol e=\Lambda\boldsymbol e, \mbox{ with } \sum_{\ell=0}^{K}e_\ell^2=1$, the $\ell$--th component of this equation is $\Lambda e_\ell=\sqrt{\rho_\ell}\,U_1+\frac{\sqrt{1-\rho_\ell}}{\eta}\,U_2$.
Using Cauchy--Schwarz inequality, we get
\begin{eqnarray*}
\begin{aligned}
\Lambda^2e_\ell^2
&=
\left(
\sqrt{\rho_\ell}\,U_1
+
\sqrt{\frac{1-\rho_\ell}{n_\ell}}\,
\frac{\sqrt{n_\ell}\,U_2}{\eta}
\right)^2\\
&\leq
\left(
\rho_\ell+\frac{1-\rho_\ell}{n_\ell}
\right)
\left(
U_1^2+\frac{n_\ell U_2^2}{\eta^2}
\right)=\frac1{r_\ell}
\left(
U_1^2+\frac{n_\ell U_2^2}{\eta^2}
\right).
\end{aligned}
\end{eqnarray*}
Multiplying by $\pi_\ell r_\ell$ and taking the summation on both sides gives
\begin{eqnarray*}
\Lambda^2\sum_{\ell=0}^{K}e_\ell^2\pi_\ell r_\ell
\leq U_1^2\sum_{\ell=0}^{K}\pi_\ell+\frac{U_2^2}{\eta^2}\sum_{\ell=0}^{K}\pi_\ell n_\ell=U_1^2+\frac{U_2^2}{\eta}=\boldsymbol e^{\rm T}\mathbf H\boldsymbol e
=\Lambda\boldsymbol e^{\rm T}\boldsymbol e =\Lambda
\end{eqnarray*}
Combining this with equation \eqref{cont.average.ncp.upperbound.joint.case1} gives the upper bound on the average NCP as
\begin{equation}
\Psi(\boldsymbol\pi,\boldsymbol n,\boldsymbol q^*)
\leq
\frac{m\delta^2}{2\sigma^2 \Lambda}.
\end{equation}
The joint maximin designs $\boldsymbol{\pi}_{\rm MM}$ and $\boldsymbol{n}_{\rm MM}$ under Case 1 are given by 
\[
\pi_\ell
=
\frac{\sqrt{\rho_\ell}\,e_\ell}{U_1},
\qquad
n_\ell
=
\frac{\eta U_1}{U_2}
\sqrt{\frac{1-\rho_\ell}{\rho_\ell}}.
 \]
 Consequently,
\begin{align*}
\frac1{\pi_\ell r_\ell}=
\frac{\rho_\ell}{\pi_\ell}
+
\frac{1-\rho_\ell}{\pi_\ell n_\ell}=
\frac{
\sqrt{\rho_\ell}\,U_1+
\eta^{-1}\sqrt{1-\rho_\ell}\,U_2
}{e_\ell}=\Lambda.
\end{align*}
Thus, for every $\boldsymbol{\theta}^{(ij)} \in \Theta_{\rm LFC}$ , we have 

\[
\phi(\boldsymbol\pi_{\rm MM},\boldsymbol\theta^{(ij)},
 m,\boldsymbol n_{\rm MM},\boldsymbol\rho)=
\frac{m\delta^2}{2\sigma^2\Lambda}
\]
Consequently,
\[
\Psi(\boldsymbol\pi_{\rm MM},\boldsymbol n_{\rm MM},\boldsymbol q^*)=
\frac{m\delta^2}{2\sigma^2\Lambda}.
\]
Therefore, the proposed joint design attains the upper bound on the average NCP for $\boldsymbol q^*\in\mathcal Q_{\rm LFD}$.

\noindent\textbf{Case 2.} In this case we have, $e_{s}^{2}>1/2$. For any $\boldsymbol{\theta}^{(sj)}\in \Theta_{\rm LFC}$, its $\ell$--th component is given by 
\begin{equation*}
\theta_\ell^{(sj)}
=
\begin{cases}
\delta/2, & \ell=s,\\
-\delta/2, & \ell=j,\\
\frac{\delta}{2}
\frac{z-\tau_s}{z+\tau_s},
& \ell\notin\{s,j\}, 
\end{cases}
\end{equation*}
Using $c=\frac{\delta}{2}
\frac{z-\tau_s}{z+\tau_s}$, in the equation~\eqref{cont.ncp.upperbound}, we get, 
\begin{eqnarray*}
\begin{aligned}
\phi(\boldsymbol{\pi},\boldsymbol{n},\boldsymbol{\theta}^{(sj)})&\leq
\frac{m\delta^2}{\sigma^2(\tau_s+z)^2}
\left\{
\tau_s^2\pi_sr_s+z^2\pi_jr_j
-
\frac{
(\tau_s\pi_sr_s-z\pi_jr_j)^2
}{
\sum_{\ell=0}^{K}\pi_\ell r_\ell
}
\right\}\\
&\leq\frac{m\delta^2}{\sigma^2(\tau_s+z)^2}
(\tau_s^2\pi_sr_s+z^2\pi_jr_j)
\end{aligned}
\end{eqnarray*}
For any $\boldsymbol{q}^{*}\in \mathcal Q_{\rm LFD}$, the average NCP is 
\begin{eqnarray}
\begin{aligned}
\Psi(\boldsymbol\pi,\boldsymbol n,\boldsymbol q^*)
&=
\sum_{j\neq s}^{K}q_{sj}^*
\phi(\boldsymbol\pi,\boldsymbol\theta^{(sj)},
 m,\boldsymbol n,\boldsymbol\rho)
\notag\\
&\leq
\frac{m\delta^2}{\sigma^2(\tau_s+z)^2}
\sum_{j\neq s}^{K}q_{sj}^*
\left(
\tau_s^2\pi_sr_s+z^2\pi_jr_j
\right)
\notag\\
&=
\frac{m\delta^2}{\sigma^2(\tau_s+z)^2}
\left\{
\tau_s^2\pi_sr_s
+
z^2\sum_{j\neq s}^{K}v_j^2\pi_jr_j
\right\}.
\label{cont.average.ncp.upperbound.joint.case2}
\end{aligned}
\end{eqnarray}
From the statement of theorem, we have $\boldsymbol t=
(z\boldsymbol I_K-\mathbf H_{ss})^{-1}
\boldsymbol h_{s},\;
\boldsymbol v=(t_0,\hdots,t_{s-1},1,t_{s+1},\hdots,t_{K})^{\rm T},\\
\sum_{j\neq s}^{K}v_j^2=1,
\tau_s=h_{ss}+\boldsymbol h_{s}^{\rm T}\boldsymbol t>z$. Since $\boldsymbol h_{s}+\mathbf H_{ss}\boldsymbol t
=z\boldsymbol t$, therefore
\begin{equation}
\label{cont.joint.case2.Hv}
\mathbf H\boldsymbol v
=\left(
zv_0,
\hdots,
zv_{s-1},
\tau_s,
zv_{s+1},\hdots,
zv_{K}
\right)^{\rm T}.
\end{equation}
From the definition of $\mathbf H$,
\begin{equation}
\label{cont.joint.case2.Hv.components}
(\mathbf H\boldsymbol v)_\ell
=
\sqrt{\rho_\ell}\,U_3
+
\frac{\sqrt{1-\rho_\ell}}{\eta}\,U_4
=
\begin{cases}
\tau_s, & \ell=s,\\
zv_\ell, & \ell\in \mathcal{S}\setminus\{s\}.
\end{cases}
\end{equation}
Using the Cauchy--Schwarz inequality, we get
\begin{align*}
\bigl\{(\mathbf H\boldsymbol v)_\ell\bigr\}^2
&=
\left(
\sqrt{\rho_\ell}\,U_3
+
\sqrt{\frac{1-\rho_\ell}{n_\ell}}\,
\frac{\sqrt{n_\ell}\,U_4}{\eta}
\right)^2\\
&\leq
\left(
\rho_\ell+\frac{1-\rho_\ell}{n_\ell}
\right)
\left(
U_3^2+\frac{n_\ell U_4^2}{\eta^2}
\right)=
\frac1{r_\ell}
\left(
U_3^2+\frac{n_\ell U_4^2}{\eta^2}
\right).
\end{align*}
Multiplying by $\pi_\ell r_\ell$ gives
\[
\pi_\ell r_\ell
\bigl\{(\mathbf H\boldsymbol v)_\ell\bigr\}^2
\leq
U_3^2\pi_\ell
+
\frac{U_4^2}{\eta^2}\pi_\ell n_\ell.
\]
Summing and using
\eqref{cont.joint.case2.Hv.components},
\begin{align*}
\tau_s^2\pi_sr_s
+
z^2\sum_{j\neq s}^{K}v_j^2\pi_jr_j
&\leq
U_3^2\sum_{\ell=0}^{K}\pi_\ell
+
\frac{U_4^2}{\eta^2}
\sum_{\ell=0}^{K}\pi_\ell n_\ell=
U_3^2+\frac{U_4^2}{\eta}.
\end{align*}
From the definition of $\mathbf H$ and \eqref{cont.joint.case2.Hv}, we have
\begin{equation*}
U_3^2+\frac{U_4^2}{\eta}=\boldsymbol v^{\rm T}\mathbf H\boldsymbol v=v_s\tau_s+z\sum_{j\neq s}^{K}v_j^2=\tau_s+z.
\end{equation*}
Therefore, $
\tau_s^2\pi_sr_s
+
z^2\sum_{j\neq s}^{K}v_j^2\pi_jr_j
\leq\tau_s+z$.
and combining this with \eqref{cont.average.ncp.upperbound.joint.case2},we obtain
\begin{equation}
\label{cont.average.ncp.upperbound.joint.case2.2}
\Psi(\boldsymbol\pi,\boldsymbol n,\boldsymbol q^*)
\leq
\frac{m\delta^2}{\sigma^2(\tau_s+z)}.
\end{equation}
This upper bound holds for every feasible design.
Under Case 2, the joint design is given by
\[
\pi_\ell
=
\frac{\sqrt{\rho_\ell}\,v_\ell}{U_3},
\qquad
n_\ell
=
\frac{\eta U_3}{U_4}
\sqrt{\frac{1-\rho_\ell}{\rho_\ell}},
\qquad \ell\in\mathcal S.
\]
Consequently, we have 
\begin{align*}
\frac1{\pi_\ell r_\ell}
&=
\frac{\rho_\ell}{\pi_\ell}
+
\frac{1-\rho_\ell}{\pi_\ell n_\ell}=
\frac{
\sqrt{\rho_\ell}\,U_3
+
\eta^{-1}\sqrt{1-\rho_\ell}\,U_4
}{v_\ell}=
\frac{(\mathbf H\boldsymbol v)_\ell}{v_\ell}.
\end{align*}
Since $v_s=1$, equation
\eqref{cont.joint.case2.Hv.components} gives
\[
\frac1{\pi_s r_s}=\tau_s,
\quad\text{and}\quad
\frac1{\pi_j r_j}=z, 
\quad\implies\quad
\pi_s r_s=\frac1{\tau_s},
\quad \text{and} \quad
\pi_j r_j=\frac1z,
\quad j \in \mathcal{S}\setminus\{s\}.
\]
Thus for every pair $(s,j)$,
\[
\tau_s\pi_s r_s-z\pi_j r_j
=1-1=0.
\]
Using these results, for any $\boldsymbol{\theta}^{(sj)}\in \Theta_{\rm LFC}$, we obtain
\begin{equation*}
\phi(\boldsymbol\pi_{\rm MM},\boldsymbol\theta^{(sj)},
 m,\boldsymbol n_{\rm MM},\boldsymbol\rho)=
\frac{m\delta^2}{\sigma^2(\tau_s+z)^2}
\left(
\tau_s^2\frac1{\tau_s}
+z^2\frac1z
\right)=
\frac{m\delta^2}{\sigma^2(\tau_s+z)}.
\end{equation*}
Since $\sum_{j\neq s}^{K}q_{sj}^*=1$,
\[
\Psi(\boldsymbol\pi_{\rm MM},\boldsymbol n_{\rm MM},\boldsymbol q^*)
=
\frac{m\delta^2}{\sigma^2(\tau_s+z)}.
\]
Thus, from~\eqref{cont.average.ncp.upperbound.joint.case2.2}, the proposed joint design attains the upper bound and therefore, globally maximizes the
average NCP under the fixed candidate LFC and LFD.

\noindent\textbf{Part II:} In this part we fix the joint design to be $\boldsymbol\pi_{\rm MM}$ and $\boldsymbol n_{\rm MM}$ and show that the set of LFC is $\Theta_{\rm LFC}$ with corresponding set of LFD as $\mathcal Q_{\rm LFD}$.\\

\noindent\textbf{Case 1.} Since $\boldsymbol{\pi}_{\rm MM}$ and $\boldsymbol{n}_{\rm MM}$ in this case are such that all $\pi_\ell r_\ell$ are equal to $\Lambda^{-1}$, therefore from the equation~\eqref{cont.lfc.eqn.from.lemma}, we get 
\begin{eqnarray}
 \theta_{\ell}^{(ij)}=\begin{cases}
 & \delta/2, \qquad \ell=i\\
 & -\delta/2, \qquad \ell =j,\\
 & 0, \qquad \ell\in \mathcal{S}\setminus\{i,j\}.
 \end{cases}\nonumber
\end{eqnarray}
which is same as that of the set given in~\eqref{cont.lfc.eqn.joint.case1} and also $\boldsymbol{\theta}^{(ji)}=-\boldsymbol{\theta}^{(ij)}$. Thus the set $\boldsymbol{\Theta}_{\rm LFC}=\{\boldsymbol{\theta}^{ij},\ -\boldsymbol{\theta}^{ij}:i<j, i,j\in \mathcal{S}\}$ is the set of LFCs. Also, 
\[
\phi(\boldsymbol\pi_{\rm MM},\boldsymbol{\theta}^{(ij)},m,\boldsymbol{n}_{\rm MM},\boldsymbol \rho)
=
\frac{m\delta^2}
{2\sigma^2\Lambda},\quad \forall \boldsymbol{\theta}^{(ij)}\in \Theta_{\rm LFC} .
\]
and the average NCP for any probability distribution $\boldsymbol{q}$ on the set $\Theta_{\rm LFC}$ is 
\begin{eqnarray*}
 \Psi(\boldsymbol{\pi}_{\rm MM},\boldsymbol{n}_{\rm MM},\boldsymbol{q})= \sum\limits_{0\leq i<j \leq K} q_{ij}\phi(\boldsymbol{\pi}_{\rm MM},\boldsymbol{\theta}^{(ij)},m,\boldsymbol{n}_{\rm MM},\boldsymbol \rho)=\frac{m\delta^2}
{2\sigma^2 \Lambda}.
\end{eqnarray*}
Since this is a constant, therefore, every proposed
$\boldsymbol q\in\mathcal Q_{\rm LFD}$ is a minimizing response.\\

\noindent\textbf{Case 2.}
Fixing the design at
$(\boldsymbol\pi_{\rm MM},\boldsymbol n_{\rm MM})$, and its components are given by 
\[\pi_\ell=\frac{\sqrt{\rho_\ell}\,v_\ell}{U_3},\qquad n_\ell = \frac{\eta U_3}{U_4}\sqrt{\frac{1-\rho_\ell}{\rho_\ell}},
\qquad \ell\in\mathcal S,
\]
Now we have, 
\begin{eqnarray*}
\frac1{\pi_\ell r_\ell}=
\frac{\rho_\ell}{\pi_\ell}
+
\frac{1-\rho_\ell}{\pi_\ell n_\ell}=
\frac{
\sqrt{\rho_\ell}\,U_3
+
\eta^{-1}\sqrt{1-\rho_\ell}\,U_4
}{v_\ell}=
\frac{(\mathbf H\boldsymbol v)_\ell}{v_\ell}=\begin{cases}
\tau_s, & \ell=s,\\
z, & \ell\neq s.
\end{cases}
\end{eqnarray*}

Therefore,
\begin{eqnarray*}
\begin{aligned}
\pi_s r_s&=\frac1{\tau_s},
\quad \text{and} \quad
\pi_\ell r_\ell=\frac1z,
\quad \ell\in\mathcal S\setminus\{s\},\\ \implies\quad \pi_s r_s&=\frac{z}{\tau_s}\pi_\ell r_\ell<\pi_\ell r_\ell,\quad\ell\neq s,\quad \text{and}\quad\pi_i r_i=\pi_j r_j, \quad i,j\in\mathcal S\setminus\{s\}.
\end{aligned}
\end{eqnarray*}
 
For any distinct $i,j\in\mathcal S$, choose the orientation
$\theta_i-\theta_j=\delta$ and fix the endpoints at
$\theta_i=\delta/2$ and $\theta_j=-\delta/2$.
By equation~\eqref{cont.lfc.eqn.from.lemma}, the remaining
components of the pairwise minimizing configuration are
\[
\theta_\ell^{(ij)}
=
\frac{\pi_i r_i(\delta/2)+\pi_j r_j(-\delta/2)}
{\pi_i r_i+\pi_j r_j}
=
\frac{\delta}{2}
\frac{\pi_i r_i-\pi_j r_j}
{\pi_i r_i+\pi_j r_j},
\qquad \ell\notin\{i,j\}.
\]
Substituting the information contributions above yields
\begin{eqnarray*}
\theta_\ell^{(ij)}
&=&
\begin{cases}
\delta/2,
& \ell=i,\\[0.3em]
-\delta/2,
& \ell=j,\\[0.3em]
0,
& \ell\notin\{i,j\},\quad s\notin\{i,j\},\\[0.6em]
\displaystyle
\frac{\delta}{2}\frac{z-\tau_s}{z+\tau_s},
& \ell\notin\{i,j\},\quad i=s,\\[0.9em]
\displaystyle
-\frac{\delta}{2}\frac{z-\tau_s}{z+\tau_s},
& \ell\notin\{i,j\},\quad j=s.
\end{cases}
\end{eqnarray*}
In particular,
$\boldsymbol\theta^{(js)}=-\boldsymbol\theta^{(sj)}$.
Let $\bar\theta^{(ij)}$ denote the information-weighted
mean of this minimizing configuration. Since every
remaining component equals the weighted mean of the
two endpoints, we have
\[
\begin{aligned}
\bar\theta^{(ij)}
&=
\frac{
\sum_{\ell=0}^{K}
\pi_\ell r_\ell\theta_\ell^{(ij)}
}{
\sum_{\ell=0}^{K}\pi_\ell r_\ell
}=
\frac{
\pi_i r_i(\delta/2)+\pi_j r_j(-\delta/2)
}{
\pi_i r_i+\pi_j r_j
}.
\end{aligned}
\]
Thus,
\[
\theta_i^{(ij)}-\bar\theta^{(ij)}
=\frac{\delta\,\pi_j r_j}{\pi_i r_i+\pi_j r_j},
\qquad
\theta_j^{(ij)}-\bar\theta^{(ij)}=
-\frac{\delta\,\pi_i r_i}{\pi_i r_i+\pi_j r_j}, \quad \mbox{ and }\quad
\theta_\ell^{(ij)}-\bar\theta^{(ij)}=0,
\qquad \ell\notin\{i,j\}.
\]

Substitution into the weighted-sum-of-squares expression
for the original NCP gives
\begin{eqnarray*}
\begin{aligned}
\phi(\boldsymbol\pi_{\rm MM},\boldsymbol\theta^{(ij)},
 m,\boldsymbol n_{\rm MM},\boldsymbol\rho)&=
\frac{m}{\sigma^2}
\left\{
\pi_i r_i
\left(
\frac{\delta\,\pi_j r_j}{\pi_i r_i+\pi_j r_j}
\right)^2
+
\pi_j r_j
\left(
\frac{\delta\,\pi_i r_i}{\pi_i r_i+\pi_j r_j}
\right)^2
\right\}\\[0.3cm]
&\quad=
\frac{m\delta^2}{\sigma^2}
\frac{
\pi_i r_i\pi_j r_j(\pi_i r_i+\pi_j r_j)
}{
(\pi_i r_i+\pi_j r_j)^2
}=
\frac{m\delta^2}{\sigma^2}
\frac{\pi_i r_i\pi_j r_j}{\pi_i r_i+\pi_j r_j}=
\frac{m\delta^2}{\sigma^2}
\left(
\frac1{\pi_i r_i}+\frac1{\pi_j r_j}
\right)^{-1}.
\end{aligned}
\end{eqnarray*}
For a pair $(s,j)$ with $j\neq s$, this becomes
\begin{align*}
&\phi(\boldsymbol\pi_{\rm MM},\boldsymbol\theta^{(sj)},
 m,\boldsymbol n_{\rm MM},\boldsymbol\rho)=
\frac{m\delta^2}{\sigma^2}
\frac{(1/\tau_s)(1/z)}{1/\tau_s+1/z}=
\frac{m\delta^2}{\sigma^2(\tau_s+z)}.
\end{align*}
This value is the same for every $j\neq s$ and for the
opposite orientation $-\boldsymbol\theta^{(sj)}$.

For distinct $i,j\in\mathcal S\setminus\{s\}$, whenever
such a pair exists,
\begin{equation*}
\phi(\boldsymbol\pi_{\rm MM},\boldsymbol\theta^{(ij)},
 m,\boldsymbol n_{\rm MM},\boldsymbol\rho)=
\frac{m\delta^2}{\sigma^2}
\frac{(1/z)(1/z)}{1/z+1/z}=
\frac{m\delta^2}{2\sigma^2z}.
\end{equation*}

Since $\tau_s>z$, we have $\tau_s+z>2z$.
Therefore, for any $h\neq s$ and distinct $i,j\neq s$,
\[
\frac{
\phi(\boldsymbol\pi_{\rm MM},\boldsymbol\theta^{(sh)},
 m,\boldsymbol n_{\rm MM},\boldsymbol\rho)
}{
\phi(\boldsymbol\pi_{\rm MM},\boldsymbol\theta^{(ij)},
 m,\boldsymbol n_{\rm MM},\boldsymbol\rho)
}=\frac{2z}{\tau_s+z}<1.
\]
Thus, among the pairwise minimizing configurations with
separation $\delta$, only pairs involving arm $s$ attain
the smallest NCP. 
Thus the LFC set is $\Theta_{\rm LFC}
=
\left\{
\boldsymbol\theta^{(sj)},
-\boldsymbol\theta^{(sj)}
:
j\in\mathcal S\setminus\{s\}
\right\}.$
Let $q$ be any probability distribution defined on this LFC set, then the average NCP is given by 
\begin{equation*}
\Psi(\boldsymbol\pi_{\rm MM},\boldsymbol n_{\rm MM},\boldsymbol q)=
\sum_{j\neq s}q_{sj}
\phi(\boldsymbol\pi_{\rm MM},\boldsymbol\theta^{(sj)},
 m,\boldsymbol n_{\rm MM},\boldsymbol\rho)=
\sum_{j\neq s}q_{sj}
\frac{m\delta^2}{\sigma^2(\tau_s+z)}=
\frac{m\delta^2}{\sigma^2(\tau_s+z)}.
\end{equation*}
Hence, every probability distribution supported on this LFC set is a
minimizing distribution at the fixed design
$(\boldsymbol\pi_{\rm MM},\boldsymbol n_{\rm MM})$.
In particular, this holds for $\mathcal Q_{\rm LFD}=\{q*=(q_{sj})_{\ell\neq s}: q_{sj}^*=v_j^2,\quad j\neq s,
\quad
\sum_{j\neq s}v_j^2=1\}$.\\

\noindent In both the cases, we showed that 
\[
\max_{(\boldsymbol\pi,\boldsymbol n)\in\Xi}
\min_{\boldsymbol q \in Q}
\Psi(\boldsymbol\pi,\boldsymbol n,\boldsymbol q)
=
\min_{\boldsymbol q \in Q}
\max_{(\boldsymbol\pi,\boldsymbol n)\in\Xi}
\Psi(\boldsymbol\pi,\boldsymbol n,\boldsymbol q)
=\Psi(\boldsymbol\pi_{\rm MM},\boldsymbol n_{\rm MM},\boldsymbol q^{*})
\]
where $\boldsymbol q^{*}\in\mathcal Q_{\rm LFD}$ and $\mathcal Q$ is any probability distribution defined on $\Theta_{\rm LFC}$. Hence, $(\pi_{\rm MM},\boldsymbol{n}_{\rm MM}, \boldsymbol q^{*})$ is a saddle point and therefore $(\pi_{\rm MM},\boldsymbol{n}_{\rm MM})$ is the maximin design under joint optimization of cluster allocations and cluster sizes. This completes the proof. 
\end{proof}

\vspace{1cm}

We now present the derivation of the expression for the NCP given in equation~\eqref{binary.ncp} when the response is a binary response. The derivation of NCP in the case of a normal response follows from using similar calculations. 
\section*{Derivation of NCP in case of the binary response.}
\begin{proof}
Let $\boldsymbol{y}_{i}=\left(y_{i1},\ldots,y_{in_i}\right)^{T}$
be the vector of responses from the $i$th cluster, and
$\boldsymbol{x}_i$ be the row vector denoting the treatment covariate.
\begin{eqnarray*}
E(y_{ij})=\mu_{ij}=E(y_{ij}^{2})
&=&
\frac{\exp(\lambda+\boldsymbol{x}_{i}\boldsymbol{\psi})}
{1+\exp(\lambda+\boldsymbol{x}_{i}\boldsymbol{\psi})},\\[0.4em]
V(y_{ij})
&=&
\frac{\exp(\lambda+\boldsymbol{x}_{i}\boldsymbol{\psi})}
{1+\exp(\lambda+\boldsymbol{x}_{i}\boldsymbol{\psi})}
\left(1-
\frac{\exp(\lambda+\boldsymbol{x}_{i}\boldsymbol{\psi})}
{1+\exp(\lambda+\boldsymbol{x}_{i}\boldsymbol{\psi})}
\right).
\end{eqnarray*}

Since $\operatorname{Var}(y_{ij})
=\operatorname{Var}(y_{ij'})$ for
$j,j'=1,2,\ldots,n_i$, we have

\[
\begin{aligned}
V_i
&=\operatorname{cov}(\boldsymbol{y}_i)
=\Delta_i^{1/2}R_i\Delta_i^{1/2}\\
&=\frac{\exp(\lambda+\boldsymbol{x}_{i}\boldsymbol{\psi})}
{1+\exp(\lambda+\boldsymbol{x}_{i}\boldsymbol{\psi})}
\left(1-
\frac{\exp(\lambda+\boldsymbol{x}_{i}\boldsymbol{\psi})}
{1+\exp(\lambda+\boldsymbol{x}_{i}\boldsymbol{\psi})}
\right)R_i,
\end{aligned}
\]
where $R_i$ is the working correlation matrix of the $i$th
cluster, and
$\Delta_i$ is the diagonal matrix of marginal variances.
Also, the mean vector of the $i$th cluster is given by
\[
\boldsymbol{\mu}_i
=(\mu_{i1},\mu_{i2},\ldots,\mu_{in_i})^{\rm T}
=\frac{\exp(\lambda+\boldsymbol{x}_i\boldsymbol{\psi})}
{1+\exp(\lambda+\boldsymbol{x}_i\boldsymbol{\psi})}
\mathbf{1}_{n_i}.
\]
For testing the hypothesis
$H_0:\boldsymbol{\psi}=\boldsymbol{\psi}^{(0)}=\boldsymbol{0}$
against
$H_1:\boldsymbol{\psi}=\boldsymbol{\psi}^{(1)}\neq\boldsymbol{0}$,
the noncentrality parameter is given by
\[
\phi=\boldsymbol{\zeta}^{T}\mathbf{\Sigma}_1^{-1}\boldsymbol{\zeta},
\]
where $\boldsymbol{\zeta}$ and $\mathbf{\Sigma}_1$ are the expectation and covariance
of the null-constrained score under $H_1$, which are approximated by:
\[
\begin{aligned}
\boldsymbol{\zeta}=E_{H_1}\!\left[
S_{\boldsymbol{\psi}}
\left(\boldsymbol{\psi}^{(0)},\hat{\lambda}_0\right)
\right]
&\approx
\sum_{i=1}^{m}P_i^{*}V_i^{-1}
\left(\boldsymbol{\mu}_i^{1}-\boldsymbol{\mu}_i^{*}\right),\\[0.4em]
\mathbf{\Sigma}_1=\operatorname{cov}_{H_1}\!\left[
S_{\boldsymbol{\psi}}
\left(\boldsymbol{\psi}^{(0)},\hat{\lambda}_0\right)
\right]
&\approx
\sum_{i=1}^{m}P_i^{*}V_i^{-1}
\operatorname{cov}_{H_1}(\mathbf{y}_i)V_i^{-1}
(P_i^{*})^{T}.
\end{aligned}
\]
\[
P_i^{*}
=\left(
\frac{\partial\boldsymbol{\mu}_i}
{\partial\boldsymbol{\psi}}
\right)^{\prime}
-I_{\boldsymbol{\psi}\lambda}^{*}
(I_{\lambda\lambda}^{*})^{-1}
\left(\frac{\partial\boldsymbol{\mu}_i}{\partial\lambda}\right)^{\prime}
,
\]
\[
\begin{aligned}
I_{\boldsymbol{\psi}\lambda}^{*}
&=\sum_{i=1}^{m}
\left(
\frac{\partial\boldsymbol{\mu}_i}
{\partial\boldsymbol{\psi}}
\right)^{\rm T}
V_i^{-1}
\left(\frac{\partial\boldsymbol{\mu}_i}{\partial\lambda}\right),\\
I_{\lambda\lambda}^{*}
&=\sum_{i=1}^{m}
\left(\frac{\partial\boldsymbol{\mu}_i}{\partial\lambda}\right)^{\rm T}
V_i^{-1}
\left(\frac{\partial\boldsymbol{\mu}_i}{\partial\lambda}\right).
\end{aligned}
\]
where, $\boldsymbol{\mu}_i^{1}$ and $\operatorname{cov}_{H_1}(\boldsymbol{y}_i)$ are mean and covariance of $\boldsymbol{y}_{i}$ under the alternative hypothesis, and each of above expressions are evaluated at $\mathbf{\psi}^{(0)}$ and $\lambda^{*}_{0}$. Here, $\lambda_{0}^{*}$ is the limiting value of $\hat{\lambda}_{0}$ under $H_{1}$ as $m \rightarrow \infty$, which is the
solution of the following equation,
\[
\lim_{m\rightarrow\infty}m^{-1}
E_{H_1}\!\left[
S_{\lambda}(\boldsymbol{\psi}^{(0)},\lambda_0^{*});
\boldsymbol{\psi}^{(1)},\lambda_1
\right]=0.
\]
Now, we have
\[
\begin{aligned}
\frac{\partial\boldsymbol{\mu}_i}
{\partial\boldsymbol{\psi}}
&=
\frac{\exp(\lambda+\boldsymbol{x}_i\boldsymbol{\psi})}
{1+\exp(\lambda+\boldsymbol{x}_i\boldsymbol{\psi})}
\left(1-
\frac{\exp(\lambda+\boldsymbol{x}_i\boldsymbol{\psi})}
{1+\exp(\lambda+\boldsymbol{x}_i\boldsymbol{\psi})}
\right)
\mathbf{1}_{n_i}\boldsymbol{x}_i\\
&=w_i\mathbf{1}_{n_i}\boldsymbol{x}_i,\\
\frac{\partial\boldsymbol{\mu}_i}{\partial\lambda}
&=w_i\mathbf{1}_{n_i},
\end{aligned}
\]
where the first derivative is an $n_i\times K$ matrix and
\[
w_i=
\frac{\exp(\lambda+\boldsymbol{x}_i\boldsymbol{\psi})}
{1+\exp(\lambda+\boldsymbol{x}_i\boldsymbol{\psi})}
\left(1-
\frac{\exp(\lambda+\boldsymbol{x}_i\boldsymbol{\psi})}
{1+\exp(\lambda+\boldsymbol{x}_i\boldsymbol{\psi})}
\right).
\]
\begin{eqnarray*}
S_{\boldsymbol{\psi}}
(\boldsymbol{\psi}^{(0)},\lambda_0^{*})
&=&\sum_{i=1}^{m}
\left(
\frac{\partial\boldsymbol{\mu}_i}
{\partial\boldsymbol{\psi}}
\right)^{\rm T}
V_i^{-1}(\boldsymbol{y}_i-\boldsymbol{\mu}_i)\\
&=&\sum_{i=1}^{m}
\boldsymbol{x}_i^{T}\mathbf{1}_{n_i}^{T}R_i^{-1}
\left(\boldsymbol{y}_i-
\frac{\exp(\lambda_0^{*})}{1+\exp(\lambda_0^{*})}
\mathbf{1}_{n_i}\right).
\end{eqnarray*}
Similarly,
\[
S_{\lambda}(\boldsymbol{\psi}^{(0)},\lambda_0^{*})
=\sum_{i=1}^{m}\mathbf{1}_{n_i}^{T}R_i^{-1}
\left(\boldsymbol{y}_i-
\frac{\exp(\lambda_0^{*})}{1+\exp(\lambda_0^{*})}
\mathbf{1}_{n_i}\right).
\]
Let $\boldsymbol{\pi}=(\pi_0,\pi_1,\ldots,\pi_K)^{\rm T}$ be the
vector whose $\ell$th component denotes the proportion of
clusters assigned to the $\ell$th treatment arm, with
$\ell=0$ denoting the control arm Thus stating in terms of probabilities:
\[
P(\boldsymbol{x}_i=\boldsymbol{u}_\ell)=\pi_\ell,
\qquad \ell=0,1,\ldots,K,
\]
where $\boldsymbol{u}_0=\boldsymbol{0}^{\rm T}$ and $\{\boldsymbol{u}_1^{\rm T},\boldsymbol{u}_2^{\rm T},\ldots,
\boldsymbol{u}_K^{\rm T}\}$ is the usual basis of $\mathbb{R}^K$.
We also assume that the correlation matrix and cluster size
are the same for all clusters within a treatment arm; equivalently,
\[
P\left(R_i=R_\ell,n_i=n_\ell
\mid\boldsymbol{x}_i=\boldsymbol{u}_\ell\right)=1,
\qquad \ell=0,1,\ldots,K.
\]

With these assumptions,
\[
\begin{aligned}
&\lim_{m\rightarrow\infty}m^{-1}
E_{H_1}\!\left[S_\lambda(\boldsymbol{\psi}^{(0)},\lambda_0^{*});
\boldsymbol{\psi}^{(1)},\lambda_1\right]=0\\
&\Longrightarrow
\lim_{m\rightarrow\infty}m^{-1}E_{H_1}\!\left[
\sum_{i=1}^{m}\mathbf{1}_{n_i}^{T}R_i^{-1}
\left(\boldsymbol{y}_i-
\frac{\exp(\lambda_0^{*})}{1+\exp(\lambda_0^{*})}
\mathbf{1}_{n_i}\right)\right]=0\\
&\Longrightarrow
\lim_{m\rightarrow\infty}m^{-1}\left[
\sum_{i=1}^{m}\sum_{\ell=0}^{K}
\mathbf{1}_{n_\ell}^{T}R_\ell^{-1}
\left\{
\frac{\exp(\lambda_1+\boldsymbol{u}_\ell\boldsymbol{\psi}^{(1)})}
{1+\exp(\lambda_1+\boldsymbol{u}_\ell\boldsymbol{\psi}^{(1)})}-\frac{\exp(\lambda_0^{*})}{1+\exp(\lambda_0^{*})}
\right\}\mathbf{1}_{n_\ell}\pi_\ell\right]=0\\
&\Longrightarrow
\frac{\exp(\lambda_0^{*})}{1+\exp(\lambda_0^{*})}
=\frac{\sum_{\ell=0}^{K}r_\ell\pi_\ell\theta_\ell}
{\sum_{\ell=0}^{K}r_\ell\pi_\ell}=:\theta^{*},\\
&\Longrightarrow\quad
\lambda_0^{*}
=\operatorname{logit}\!\left(
\frac{\sum_{\ell=0}^{K}r_\ell\pi_\ell\theta_\ell}
{\sum_{\ell=0}^{K}r_\ell\pi_\ell}\right),
\end{aligned}
\]
where
\[
r_\ell=\mathbf{1}_{n_\ell}^{T}R_\ell^{-1}\mathbf{1}_{n_\ell},
\qquad
\theta_\ell=
\frac{\exp(\lambda_1+\boldsymbol{u}_\ell\boldsymbol{\psi}^{(1)})}
{1+\exp(\lambda_1+\boldsymbol{u}_\ell\boldsymbol{\psi}^{(1)})}.
\]
Also,

\begin{eqnarray*}
I_{\boldsymbol{\psi}\lambda}^{*}
&=&\sum_{i=1}^{m}
\left(\frac{\partial\boldsymbol{\mu}_i}
{\partial\boldsymbol{\psi}}\right)^{\rm T}
V_i^{-1}
\left(\frac{\partial\boldsymbol{\mu}_i}{\partial\lambda}\right)
=\sum_{i=1}^{m}
(w_0^{*}\mathbf{1}_{n_i}\boldsymbol{x}_i)^{T}
(w_0^{*}R_i)^{-1}(w_0^{*}\mathbf{1}_{n_i})
=\sum_{i=1}^{m}w_0^{*}\boldsymbol{x}_i^{T}
\mathbf{1}_{n_i}^{T}R_i^{-1}\mathbf{1}_{n_i}\\
&=&m\theta^{*}(1-\theta^{*})
\sum_{\ell=0}^{K}\boldsymbol{u}_\ell^{T}r_\ell\pi_\ell
=m\theta^{*}(1-\theta^{*})
\begin{bmatrix}r_1\pi_1\\r_2\pi_2\\\vdots\\r_K\pi_K\end{bmatrix},
\end{eqnarray*}
where
$w_0^{*}=w_i(\boldsymbol{\psi}^{(0)},\lambda_0^{*})
=\theta^{*}(1-\theta^{*})$.
\begin{eqnarray*}
\begin{aligned}
I_{\lambda\lambda}^{*}
&=\sum_{i=1}^{m}
\left(\frac{\partial\boldsymbol{\mu}_i}{\partial\lambda}\right)^{\rm T}
V_i^{-1}
\left(\frac{\partial\boldsymbol{\mu}_i}{\partial\lambda}\right)\\
&=\sum_{i=1}^{m}
(w_0^{*}\mathbf{1}_{n_i})^{T}
(w_0^{*}R_i)^{-1}(w_0^{*}\mathbf{1}_{n_i})
=m\theta^{*}(1-\theta^{*})\sum_{\ell=0}^{K}r_\ell\pi_\ell,\\[0.5em]
I_{\boldsymbol{\psi}\lambda}^{*}(I_{\lambda\lambda}^{*})^{-1}
&=\frac{1}{\sum_{\ell=0}^{K}r_\ell\pi_\ell}
\begin{bmatrix}r_1\pi_1\\r_2\pi_2\\\vdots\\r_K\pi_K\end{bmatrix}.
\end{aligned}
\end{eqnarray*}
\begin{eqnarray*}
P_i^{*}
&=&\left(\frac{\partial\boldsymbol{\mu}_i}
{\partial\boldsymbol{\psi}}\right)^{\rm T}
-I_{\boldsymbol{\psi}\lambda}^{*}
(I_{\lambda\lambda}^{*})^{-1}
\left(\frac{\partial\boldsymbol{\mu}_i}{\partial\lambda}\right)^{\rm T}=(w_0^{*}\mathbf{1}_{n_i}\boldsymbol{x}_i)^{T}
-\frac{1}{\sum_{\ell=0}^{K}r_\ell\pi_\ell}
\begin{bmatrix}r_1\pi_1\\r_2\pi_2\\\vdots\\r_K\pi_K\end{bmatrix}
(w_0^{*}\mathbf{1}_{n_i})^{T}\\
&=&\theta^{*}(1-\theta^{*})\left\{
(\mathbf{1}_{n_i}\boldsymbol{x}_i)^{T}
-\frac{1}{S}
\begin{bmatrix}r_1\pi_1\\r_2\pi_2\\\vdots\\r_K\pi_K\end{bmatrix}
(\mathbf{1}_{n_i})^{T}\right\},
\end{eqnarray*}
where $S=\sum_{\ell=0}^{K}r_\ell\pi_\ell$.
Now, $\boldsymbol{\zeta}$ becomes
\begin{eqnarray*}
\boldsymbol{\zeta}
&=&\sum_{i=1}^{m}P_i^{*}V_i^{-1}
\left(\boldsymbol{\mu}_i^{1}-\boldsymbol{\mu}_i^{*}\right)=m\sum_{\ell=0}^{K}P_\ell^{*}V_\ell^{-1}
(\boldsymbol{\mu}_\ell^{1}-\boldsymbol{\mu}_\ell^{*})\pi_\ell,
\\[0.5em]
P_\ell^{*}
&=&\theta^{*}(1-\theta^{*})\left\{
(\mathbf{1}_{n_\ell}\boldsymbol{u}_\ell)^{T}
-\frac{1}{S}
\begin{bmatrix}r_1\pi_1\\r_2\pi_2\\\vdots\\r_K\pi_K\end{bmatrix}
(\mathbf{1}_{n_\ell})^{T}\right\},\quad
V_\ell
=w_0^{*}R_\ell=\theta^{*}(1-\theta^{*})R_\ell,\\[0.5em]
P_\ell^{*}V_\ell^{-1}
&=&\left\{
(\mathbf{1}_{n_\ell}\boldsymbol{u}_\ell)^{T}
-\frac{1}{S}
\begin{bmatrix}r_1\pi_1\\r_2\pi_2\\\vdots\\r_K\pi_K\end{bmatrix}
(\mathbf{1}_{n_\ell})^{T}\right\}R_\ell^{-1}=\left\{\boldsymbol{u}_\ell^{T}
-\frac{1}{S}
\begin{bmatrix}r_1\pi_1\\r_2\pi_2\\\vdots\\r_K\pi_K\end{bmatrix}
\right\}\mathbf{1}_{n_\ell}^{T}R_\ell^{-1}.
\end{eqnarray*}
\begin{eqnarray*}
\boldsymbol{\zeta}
&=&m\sum_{\ell=0}^{K}\pi_\ell
\left\{\boldsymbol{u}_\ell^{T}
-\frac{1}{S}
\begin{bmatrix}r_1\pi_1\\r_2\pi_2\\\vdots\\r_K\pi_K\end{bmatrix}
\right\}\mathbf{1}_{n_\ell}^{T}R_\ell^{-1}\left(\theta_{\ell}-\theta^{*}\right)\mathbf{1}_{n_\ell}\\[0.5em]
&=&m
\begin{bmatrix}
-\dfrac{r_1\pi_1}{S} & 1-\dfrac{r_1\pi_1}{S}
 & -\dfrac{r_1\pi_1}{S} &\cdots & -\dfrac{r_1\pi_1}{S}\\[0.4em]
-\dfrac{r_2\pi_2}{S} & -\dfrac{r_2\pi_2}{S}
 & 1-\dfrac{r_2\pi_2}{S} &\cdots & -\dfrac{r_2\pi_2}{S}\\[0.4em]
\vdots & \vdots & \vdots & \ddots & \vdots\\[0.4em]
-\dfrac{r_K\pi_K}{S} & -\dfrac{r_K\pi_K}{S}
 & -\dfrac{r_K\pi_K}{S} & \cdots & 1-\dfrac{r_K\pi_K}{S}
\end{bmatrix}
\begin{bmatrix}
r_0\pi_0(\theta_0-\theta^{*})\\
r_1\pi_1(\theta_1-\theta^{*})\\
r_2\pi_2(\theta_2-\theta^{*})\\
\vdots\\
r_K\pi_K(\theta_K-\theta^{*})
\end{bmatrix}\\[0.5em]
&=&mAv\\
&=&m\sum_{\ell=1}^{K}r_\ell\pi_\ell(\theta_\ell-\theta^{*})
\boldsymbol{u}_\ell^{T}
=m\begin{bmatrix}
r_1\pi_1(\theta_1-\theta^{*})\\
r_2\pi_2(\theta_2-\theta^{*})\\
\vdots\\
r_K\pi_K(\theta_K-\theta^{*})
\end{bmatrix}=:m\mathbf{h}.
\end{eqnarray*}
Here, $A$ and $v$ denote the matrix and vector displayed above. Also,
\begin{eqnarray*}
\mathbf{\Sigma}_1
&=m\sum_{\ell=0}^{K}P_\ell^{*}V_\ell^{-1}
\operatorname{cov}_{H_1}(\boldsymbol{y}_i
\mid\boldsymbol{x}_i=\boldsymbol{u}_\ell)
V_\ell^{-1}(P_\ell^{*})^{T}\pi_\ell,\\[0.4em]
\end{eqnarray*}
where $\operatorname{cov}_{H_1}(\boldsymbol{y}_i
\mid\boldsymbol{x}_i=\boldsymbol{u}_\ell)
=w_\ell^{(1)}R_\ell, $ and $w_\ell^{(1)}=\theta_\ell(1-\theta_\ell).$
\begin{eqnarray*}
\mathbf{\Sigma}_1&=m\sum_{\ell=0}^{K}\pi_\ell P_\ell^{*}V_\ell^{-1}
\frac{\exp(\lambda_1+\boldsymbol{u}_\ell\boldsymbol{\psi}^{(1)})}
{1+\exp(\lambda_1+\boldsymbol{u}_\ell\boldsymbol{\psi}^{(1)})}
\left(1-
\frac{\exp(\lambda_1+\boldsymbol{u}_\ell\boldsymbol{\psi}^{(1)})}
{1+\exp(\lambda_1+\boldsymbol{u}_\ell\boldsymbol{\psi}^{(1)})}
\right)R_\ell V_\ell^{-1}(P_\ell^{*})^{T}=mADA^{T},
\end{eqnarray*}
where $D=\operatorname{diag}\!\left(
 r_0\pi_0\theta_0(1-\theta_0),\,
 r_1\pi_1\theta_1(1-\theta_1),\,\ldots,\,
 r_K\pi_K\theta_K(1-\theta_K)
\right).$
Let $
\mathbf{s}_1=
\left(\frac{r_1\pi_1}{S},\frac{r_2\pi_2}{S},\ldots,
\frac{r_K\pi_K}{S}\right)^{T}
$ and $\mathbf{d}_1=
\left(r_1\pi_1\theta_1(1-\theta_1),r_2\pi_2\theta_2(1-\theta_2),\ldots,
r_K\pi_K\theta_K(1-\theta_K)\right)^{T}$.
\newpage
\noindent Note that, $A=\left[-\mathbf{s}_1,\,
\boldsymbol{u}_1^{T}-\mathbf{s}_1,\,\ldots,\,
\boldsymbol{u}_K^{T}-\mathbf{s}_1\right]$
and $ADA^{\rm T}$ can be written as $D_1+LML^{T}$, with $
D_1=\operatorname{diag}\!\left(
r_1\pi_1\theta_1(1-\theta_1),\,
r_2\pi_2\theta_2(1-\theta_2),\,\ldots,\,
r_K\pi_K\theta_K(1-\theta_K)\right)
$ and
\begin{align*}
L&=\begin{bmatrix}\mathbf{s}_1&\mathbf{d}_1\end{bmatrix},
\qquad
M=\begin{bmatrix}c&-1\\-1&0\end{bmatrix}, 
\qquad\text{where}\qquad
c=\sum_{\ell=0}^{K}r_\ell\pi_\ell \theta_\ell(1-\theta_\ell).
\end{align*}
Now, using the Woodbury matrix identity, we obtain $(ADA^{\rm T})^{-1}=D_1^{-1}-D_1^{-1}LJ^{-1}L^{T}D_1^{-1}$, where $J=M^{-1}+L^{T}D_1^{-1}L$. 
Therefore, the non-centrality parameter is given by
\begin{align*}
\phi=\boldsymbol{\zeta}^{T}\mathbf{\Sigma}_1^{-1}\boldsymbol{\zeta}=m\mathbf{h}^{T}(ADA^{\rm T})^{-1}\mathbf{h}&=m\mathbf{h}^{T}(D_1^{-1}-D_1^{-1}LJ^{-1}L^{T}D_1^{-1})\mathbf{h}\\
&=m\left(\mathbf{h}^{T}D_1^{-1}\mathbf{h}
-\mathbf{h}^{T}D_1^{-1}LJ^{-1}L^{T}D_1^{-1}\mathbf{h}\right).
\end{align*}
Here, 
\begin{align*}
L^{T}D_1^{-1}\mathbf{h}
&=\begin{bmatrix}
\dfrac{1}{S}\displaystyle\sum_{\ell=1}^{K}
\dfrac{r_\ell\pi_\ell}{\theta_\ell(1-\theta_\ell)}(\theta_\ell-\theta^{*})\\[0.8em]
\displaystyle\sum_{\ell=1}^{K}r_\ell\pi_\ell(\theta_\ell-\theta^{*})
\end{bmatrix}
=:\begin{bmatrix}f_1\\f_2\end{bmatrix},\quad \quad
\mathbf{h}^{T}D_1^{-1}L
=\begin{bmatrix}f_1&f_2\end{bmatrix},\\[0.6em]
J
&=\begin{bmatrix}
\dfrac{1}{S^2}\displaystyle\sum_{\ell=1}^{K}
\dfrac{r_\ell\pi_\ell}{\theta_\ell(1-\theta_\ell)}
&-\dfrac{r_0\pi_0}{S}\\[0.8em]
-\dfrac{r_0\pi_0}{S}
&-r_0\pi_0\theta_0(1-\theta_0)
\end{bmatrix},\quad \text{and}\quad 
\mathbf{h}^{T}D_1^{-1}\mathbf{h}
=\sum_{\ell=1}^{K}
\frac{r_\ell\pi_\ell}{\theta_\ell(1-\theta_\ell)}(\theta_\ell-\theta^{*})^{2}.
\end{align*}
After simplification, we obtain
\begin{align*}
\phi
&=\frac{m}{\displaystyle\sum_{\ell=0}^{K}
\frac{r_\ell\pi_\ell}{\theta_\ell(1-\theta_\ell)}}
\left\{
\left(\sum_{\ell=0}^{K}
\frac{r_\ell\pi_\ell}{\theta_\ell(1-\theta_\ell)}\right)
\left(\sum_{\ell=0}^{K}
\frac{r_\ell\pi_\ell}{\theta_\ell(1-\theta_\ell)}\theta_\ell^{2}\right)-
\left[\sum_{\ell=0}^{K}
\frac{r_\ell\pi_\ell}{\theta_\ell(1-\theta_\ell)}\theta_\ell\right]^{2}
\right\}\\[0.8em]
&=m\left\{
\sum_{\ell=0}^{K}
\frac{r_\ell\pi_\ell}{\theta_\ell(1-\theta_\ell)}(\theta_\ell-\theta_K)^{2}
-\frac{
\left[\displaystyle\sum_{\ell=0}^{K}
\frac{r_\ell\pi_\ell}{\theta_\ell(1-\theta_\ell)}(\theta_\ell-\theta_K)\right]^{2}
}{\displaystyle\sum_{\ell=0}^{K}
\frac{r_\ell\pi_\ell}{\theta_\ell(1-\theta_\ell)}}
\right\}.
\end{align*}
This completes the proof.
\end{proof}
\vspace{0.6cm}
\section*{Proof of the Theorem \ref{binary.mm.theorem.1.1}}
The following Lemma is required to prove the Theorem~\ref{binary.mm.theorem.1.1}.
\begin{lemma}
\label{binary.lemma.lfc}
Fix distinct $s,t\in\mathcal S$, and define $\mathcal I_{st}(\delta)
=\{\boldsymbol\theta\in(0,1)^{K+1}:
\theta_s-\theta_t=\delta\}.$
Let $w_{\ell}=r_{\ell}\pi_{\ell},\;\ell \in \mathcal{S}$ , $G=w_s+w_t$, and $\delta_{st}^{\mathrm{crit}}=\frac{G}{2\max(w_s,w_t)}$. Then the following statements hold.

\textbf{Case I:} If $0<\delta<\delta_{st}^{\mathrm{crit}}$, the NCP (as a function of $\boldsymbol{\theta}$) given in equation ~\eqref{binary.ncp} attains a unique global
minimizer over $\mathcal I_{st}(\delta)$, given by
\begin{equation}
\label{eq:binary_lfc_interior}
\begin{aligned}
\theta_s^*=\frac12+\frac{w_s}{G}\delta,\quad\theta_t^*=\frac12-\frac{w_t}{G}\delta,
\quad\text{and}\quad
\theta_\ell^*=\frac12+
\frac{\delta G(w_s-w_t)}{G^2-4\delta^2w_sw_t},
\quad\ell \in\mathcal{S}\setminus\{s,t\}.
\end{aligned}
\end{equation}
The minimum NCP equals
\begin{equation}
\label{eq:binary_lfc_min_interior}
\min_{\boldsymbol\theta\in\mathcal I_{st}(\delta)}
\phi(\boldsymbol{\pi},\boldsymbol{\theta},m,\boldsymbol{n},\boldsymbol{\rho})
=\frac{4m\delta^2w_sw_t G}{G^2-4\delta^2w_sw_t}.
\end{equation}

\textbf{Case II:} If $\delta_{st}^{\mathrm{crit}}\leq\delta<1$, the infimum is not
attained in $(0,1)^{K+1}$, and
\begin{equation}
\label{eq:binary_lfc_min_boundary}
\inf_{\boldsymbol\theta\in\mathcal I_{st}(\delta)}
\phi(\boldsymbol{\pi},\boldsymbol{\theta},m,\boldsymbol{n},\boldsymbol{\rho})
=\frac{m\delta\min(w_s,w_t)}{1-\delta}.
\end{equation}
In particular, if $w_t>w_s$, and this infimum is approached by
\[
\theta_t=a,\;\theta_s=a+\delta,\; \theta_\ell=a+\frac{w_sa(1-a)}{w_sa(1-a)+w_t(a+\delta)(1-a-\delta)}\delta, \quad\ell\notin\{s,t\},
\]
as $a\downarrow0$. If $w_s>w_t$, it is approached by the same construction as
$a\uparrow1-\delta$. 

\end{lemma}

\begin{proof}
For each fixed $\boldsymbol\theta$, we have 
\begin{equation}
\label{binary.ncp.weighted.eqn}
\frac{\phi(\boldsymbol{\pi},\boldsymbol{\theta},m,\boldsymbol{n},\boldsymbol{\rho})}{m}
=\sum_{\ell=0}^{K}\frac{w_\ell(\theta_\ell-\bar\theta)^2}{\gamma_{\ell}}= \min\limits_{c}\sum_{\ell=0}^{K}\frac{w_\ell(\theta_\ell-c)^2}{\gamma_{\ell}},
\end{equation}
where $\bar{\theta}=\frac{\sum_{\ell=0}^{K}w_{\ell}\theta_{\ell}/\gamma_{\ell}}{\sum_{\ell=0}^{K}w_{\ell}/\gamma_{\ell}}$. Note that,
\begin{eqnarray}
\begin{aligned}
 \frac{\phi(\boldsymbol{\pi},\boldsymbol{\theta},m,\boldsymbol{n},\boldsymbol{\rho})}{m}
=&\sum_{\ell=0}^{K}\frac{w_\ell(\theta_\ell-\bar\theta)^2}{\gamma_{\ell}}\\
=&\frac{w_0(\theta_0-\bar\theta)^2}{\gamma_{0}}+\frac{w_1(\theta_1-\bar\theta)^2}{\gamma_{1}}+\ldots+\frac{w_K(\theta_K-\bar\theta)^2}{\gamma_{K}}\\
\ge& \frac{w_s(\theta_s-\bar\theta)^2}{\gamma_{s}}+\frac{w_t(\theta_t-\bar\theta)^2}{\gamma_{t}}\\
\ge&\min\limits_{c}\{\frac{w_s(\theta_s-c)^2}{\gamma_{s}}+\frac{w_t(\theta_t-c)^2}{\gamma_{t}}\}
\end{aligned}
\end{eqnarray}
The minimum of $\frac{w_s(\theta_s-c)^2}{\gamma_{s}}+\frac{w_\ell(\theta_t-c)^2}{\gamma_{t}}$ is attained at $c=\frac{w_s\theta_s\gamma_t+w_t \theta_t \gamma_s}{w_s\gamma_t+w_t\gamma_s}$. Under $\theta_s-\theta_t=\delta$, write $\theta_t=a$ and $\theta_s=a+\delta$, with $0<a<1-\delta$. Therefore $c=\frac{w_s\theta_s\gamma_t+w_t \theta_t \gamma_s}{w_s\gamma_t+w_t\gamma_s}=a+\frac{w_sa(1-a)}{w_sa(1-a)+w_t(a+\delta)(1-a-\delta)}\delta
=:c(a)$ and \begin{eqnarray*}
 \min\limits_{c \in \mathbb{R}}\{\frac{w_s(\theta_s-c)^2}{\gamma_{s}}+\frac{w_\ell(\theta_t-c)^2}{\gamma_{t}}\}=\frac{\delta^2}{a(1-a)/w_t+(a+\delta)(1-a-\delta)/w_s}=:\frac{\delta^2}{D(a)}
\end{eqnarray*}
Since $a<c(a)<a+\delta$, set $\theta_\ell=c(a)$ for all $\ell\notin\{s,t\}$. These probabilities are feasible, and $\bar\theta=c(a).$
Therefore, from 
\begin{equation}
\inf_{\boldsymbol\theta\in\mathcal I_{st}(\delta)}\phi
=\frac{m\delta^2}{\displaystyle\sup_{0<a<1-\delta}D(a)}\label{global condition}.
\end{equation}
Differentiation gives
\[
D'(a)=\frac{1-2a-2\delta}{w_s}+\frac{1-2a}{w_t},
\qquad
D''(a)=-2\left(\frac1{w_s}+\frac1{w_t}\right)<0.
\]
Solving $D'(a)=0$, gives $ a*=\frac12-\frac{w_t}{G}\delta$. The strict concavity shows that $a^*$ is the unique unconstrained
maximizer. It lies in the required open interval precisely when
\[
a^*>0\ \Longleftrightarrow\ \delta<\frac{G}{2w_t},
\qquad
a^*<1-\delta\ \Longleftrightarrow\ \delta<\frac{G}{2w_s}.
\]
and both these inequalities hold exactly when
$\delta<\delta_{st}^{\mathrm{crit}}$.
Equivalently, the maximizer over the closed interval is $\widetilde a=\min\{1-\delta,\max(0,a^*)\}$.
 Thus, when $a*$ is interior,
\[
\theta_s^*=\frac12+\frac{w_s\delta}{G},
\quad \mbox{and }\quad \theta_t^*=\frac12-\frac{w_t\delta}{G},\quad
\implies \gamma_s^*=\frac14-\frac{w_s^2\delta^2}{G^2},
\quad \mbox{and}\quad
\gamma_t^*=\frac14-\frac{w_t^2\delta^2}{G^2}.
\]
Consequently,
\begin{align*}
D(a^*)=\frac{\gamma_s^*}{w_s}+\frac{\gamma_t^*}{w_t}
=\frac{G}{4w_sw_t}-\frac{\delta^2}{G}
=\frac{G^2-4\delta^2w_sw_t}{4w_sw_t G},\quad \mbox{and}\quad
c(a^*)
=\frac12+\frac{\delta G(w_s-w_t)}{G^2-4\delta^2w_sw_t},
\end{align*}
which proves \eqref{eq:binary_lfc_interior}. Strict concavity of $D$
and the equality condition~\eqref{global condition} establish uniqueness and global
optimality.

\noindent\textbf{Case 2.}
Suppose first that 
$\delta\geq G/(2w_t)$ and $w_t>w_s$. Then $a^*\leq0$, and strict concavity implies
$D(a)<D(0)$ for every $a\in(0,1-\delta)$. Since $
D(0)=\frac{\delta(1-\delta)}{w_s},$
\eqref{global condition} gives the infimum
$m w_s\delta/(1-\delta)$. The infimum is not attained because $D(a)<D(0)$ for every interior $a$.
 For every positive $a$ in the interval,
$\theta_s=a+\delta, \theta_t=a, \theta_\ell=c(a),\ell\in \mathcal{S}\setminus\{s,t\}$ gives a feasible vector attaining $m\delta^2/D(a)$.
Letting $a\downarrow0$ proves attainability as a limit. Also, if instead $w_s>w_t$ and $\delta\geq G/(2w_s)$, then
$a^{*}\geq1-\delta$. In this case
\[
D(a)<D(1-\delta)=\frac{\delta(1-\delta)}{w_t}
\quad\text{for }0<a<1-\delta.
\]
The infimum is $m w_t\delta/(1-\delta)$, approached by $\theta_t=a, \theta_s=a+\delta, \theta_\ell=c(a),\ell\in \mathcal{S}\setminus\{s,t\}$ as 
$a\uparrow1-\delta$.
This also is not an attained minimum. This completes the proof.
\end{proof}

Now we will proceed with the proof of Theorem~\ref{binary.mm.theorem.1.1}.
\begin{proof}
As in the case of normal responses, we will prove the theorem in two parts. First, taking the proposed LFCs set $\Theta_{\rm LFC}$ and its associated LFDs set $\mathcal Q_{\rm LFD}$ as candidates, we show that the average
NCP is maximized at the stated design $\boldsymbol \pi_{\rm MM}$. Second, fixing design to be $\boldsymbol{\pi}_{\rm MM}$, we show that the $\boldsymbol{\Theta}_{\rm LFC}$ is the set of LFC and $\mathcal Q$ is the corresponding set of LFDs. The maximin property is then established using the fact that the  Nash equilibrium is the solution of this mixed game. \\
\noindent\textbf{Part I:} The NCP in~\eqref{binary.ncp} can be written as
\begin{equation*}
\phi(\boldsymbol{\pi},\boldsymbol{\theta},m,\boldsymbol{n},\boldsymbol{\rho})
=m\sum_{\ell=0}^{K}\frac{w_\ell(\theta_\ell-\bar\theta)^2}{\gamma_{\ell}}= \min\limits_{c}m\sum_{\ell=0}^{K}\frac{w_\ell(\theta_\ell-c)^2}{\gamma_{\ell}},
\end{equation*}
Therefore for any fixed c, we have
\begin{equation}
\phi(\boldsymbol{\pi},\boldsymbol\theta)
\leq m\sum_{\ell=0}^{K}\frac{w_\ell(\theta_\ell-c)^2}{\gamma_{\ell}},\label{binary.ncp.upperbound}
\end{equation}
\textbf{Case 1.} The average NCP function with respect to $\boldsymbol{\Theta}_{\rm LFC}$ and any $\boldsymbol{q}^{*} \in \mathcal Q_{\rm LFD}$ is defined as 
\begin{eqnarray*}
 \Psi(\boldsymbol{\pi}, \boldsymbol{q^*})= \sum\limits_{0\leq i<j \leq K} q_{ij}^{*}\phi(\boldsymbol{\pi},\boldsymbol{\theta}^{(ij)})
\end{eqnarray*}
Since $\ell$th component of $\boldsymbol\theta^{(ij)}\in \Theta_{\rm LFC}$ is
\begin{equation}
\label{eq:binary_mm_lfc_components_A}
\theta_\ell^{(ij)}
=
\begin{cases}
(1+\delta)/2, & \ell=i,\\
(1-\delta)/2, & \ell=j,\\
1/2, & \ell\notin\{i,j\}.
\end{cases}
\end{equation}
Taking $c=1/2$ in \eqref{binary.ncp.upperbound} therefore gives
\[
\phi(\boldsymbol\pi,\boldsymbol\theta^{(ij)})
\leq
\frac{m\delta^2}{1-\delta^2}
(\pi_i r_i+\pi_j r_j).
\]
and consequently,
\begin{align}
\Psi(\boldsymbol{\pi}, \boldsymbol{q}^{*})
&\leq\frac{m\delta^2}{1-\delta^2}\sum\limits_{0\leq i<j \leq K} q_{ij}^{*}(\pi_i r_i+\pi_j r_j)=
\frac{m\delta^2}{1-\delta^2}
\sum_{\ell=0}^{K}
\pi_\ell r_\ell
\left(
\sum_{i<\ell}q_{i\ell}^{*}
+
\sum_{j>\ell}q_{\ell j}^{*}
\right)\notag\\
&=
\frac{m\delta^2}{1-\delta^2}
\sum_{\ell=0}^{K}
\pi_\ell r_\ell
\frac{2/r_\ell}{\sum_{h=0}^{K}1/r_h}=
\frac{2m\delta^2}
{(1-\delta^2)\sum_{h=0}^{K}1/r_h}.
\label{binary.average.ncp.upperbound.eqn}
\end{align}
It can be easily verified that the maximin design $\boldsymbol{\pi}_{\rm MM}$ given in~\eqref{binary.mm.eqn.1.case1} attains this upper bound. Thus $\boldsymbol{\pi}_{\rm MM}$ is the global maximizer of the average NCP. \\

\noindent\textbf{Case 2.} Now suppose that for treatment arm $ s\in \mathcal{S}$, we have $r_{s}^{-1}>A_s$. This $s$ is unique. The set of LFCs in this case is given by $\boldsymbol{\Theta}_{\rm LFC}
=
\left\{
\boldsymbol\theta^{(sj)},\boldsymbol{1}-\boldsymbol\theta^{(sj)},\;
j\in \mathcal{S}\setminus\{s\}
\right\}$, where the $\ell$th component of $\boldsymbol\theta^{(sj)}$is
\begin{equation*}
\theta_\ell^{(sj)}
=
\begin{cases}
\frac12+\frac {\delta}{1+z},
& \ell=s,\\
\frac12-\frac {\delta\,z}{1+z},
& \ell=j,\\
\frac12-\frac{\delta(z^2-1)}{(1+z)^2-4\delta^2\,z},
& \ell\notin\{s,j\},
\end{cases}
\end{equation*}
Let $\gamma_{s}^{(sj)}=\theta_{s
}^{(sj)}(1-\theta_{s}^{(sj)})=\frac14-\frac{\delta^2}{(1+z)^2}$, $\gamma_{j}^{(sj)}=\theta_{j}^{(sj)}(1-\theta_{j}^{(sj)})=\frac14-\frac{\delta^2z^2}{(1+z)^2}$ and $c
=
\frac{
\theta_s^{(sj)}/\gamma_{s}^{(sj)}
+
z\theta_j^{(sj)}/\gamma_{j}^{(sj)}
}{
1/\gamma_{s}^{(sj)}+z/\gamma_{j}^{(sj)}
}$. 
Multiplying \eqref{binary.mm.root.eqn.1.case2} by $z^2$ and collecting
terms gives
\[
A_{s}z^2\{(1+z)^2-4\delta^2\}
=r_s^{-1}\{(1+z)^2-4\delta^2z^2\}.
\]
Equivalently, $ A_{s}r_sz^2\gamma_{s}^{(sj)}=\gamma_{j}^{(sj)}$. Since $\theta_s^{(sj)}-\theta_j^{(sj)}=\delta$, direct subtraction gives
\[
\theta_s^{(sj)}-c
=
\frac{z\gamma_{s}^{(sj)}\delta}{\gamma_{j}^{(sj)}+z\gamma_{s}^{(sj)}},
\qquad
\theta_j^{(sj)}-c
=
-\frac{\gamma_{j}^{(sj)}\delta}{\gamma_{j}^{(sj)}+z\gamma_{s}^{(sj)}}.
\]
Also, $\theta_{\ell}=\displaystyle
\frac12-
\frac{\delta(z^2-1)}
{(1+z)^2-4\delta^2z},
=c, \ell \in \mathcal{S}\setminus\{s,j\},$ thus from~\eqref{binary.ncp.upperbound}, we obtain
\begin{eqnarray*}
\begin{aligned}
\phi(\boldsymbol\pi,\boldsymbol\theta^{(sj)})&\leq
\frac{m\delta^2}{(\gamma_{j}^{(sj)}+z\gamma_{s}^{(sj)})^2}
\left(
r_s\pi_s z^2\gamma_{s}^{(sj)}
+
r_j\pi_j\gamma_{j}^{(sj)}
\right)\\
&=\frac{m\delta^2}{(A_{s}r_sz^2\gamma_{s}^{(sj)}+z\gamma_{s}^{(sj)})^2}
\left(
r_s\pi_s z^2\gamma_{s}^{(sj)}
+
r_j\pi_jA_{s}r_sz^2\gamma_{s}^{(sj)}
\right)=\frac{m\delta^2r_{s}^{-1}}{\gamma_{s}^{(sj)}(A_{s}z+r_s^{-1})^2}
\left(\pi_s+ r_j\pi_jA_{s}
\right).
\end{aligned}
\end{eqnarray*}
For any $\boldsymbol q^{*}\in \mathcal Q_{\rm LFD}$ , the average NCP is given by 
\begin{eqnarray}
\begin{aligned}
\Psi(\boldsymbol\pi,\boldsymbol q^{*})=
\sum_{j\neq s}^{K}
q_{sj}^*\phi(\boldsymbol\pi,\boldsymbol\theta^{(sj)})
\notag\leq\frac{m\delta^2r_{s}^{-1}}{\gamma_{s}^{(sj)}(A_{s}z+r_s^{-1})^2}
\left(\pi_s+\sum_{j\neq s}\pi_{j})
\right)
\notag&=\frac{m\delta^2r_{s}^{-1}}{\gamma_{s}^{(sj)}(A_{s}z+r_s^{-1})^2}\notag\\
&=\frac{4m\delta^2r_{s}^{-1}(1+z)^{2}}{(r_s^{-1}+zA_{s})^{2}((1+z)^{2}-4\delta^2)}.
\label{binary.average.ncp.upperbound.2}
\end{aligned}
\end{eqnarray}
It can verified that the design $\boldsymbol{\pi}_{\rm MM}$ given in \eqref{binary.mm.eqn.1.case2} attains this upper bound and is therefore the global maximizer of the average NCP. 

\noindent\textbf{Part II:} In this part we show that at fixed design $\boldsymbol{\pi}_{\rm MM}$, the set of LFC is $\Theta_{\rm LFC}$ with the associated set of LFDs $\mathcal Q_{\rm LFD}$. 
First, suppose that $\boldsymbol{\theta}$ is such that its range $d$ exceeds $\delta$, that is $d>\delta$, then it can not be the minimizer of the NCP. To realise it, let $t=\delta/d\in(0,1)$ and define $\boldsymbol \theta^{\prime}$ with components $
\theta_\ell'=\frac12+t(\theta_\ell-1/2),\quad \ell \in \mathcal S $.
The new vector has range $\delta$, and $
\theta_\ell'(1-\theta_\ell')
=\frac14-t^2(\theta_\ell-1/2)^2
\geq\theta_\ell(1-\theta_\ell)$.
Taking $c'=1/2+t(\bar\theta-1/2)$ in \eqref{binary.ncp.weighted.eqn} gives $
\phi(\boldsymbol\pi,\boldsymbol\theta')
\leq t^2\phi(\boldsymbol\pi,\boldsymbol\theta)
<\phi(\boldsymbol\pi,\boldsymbol\theta)$. Thus, for minimization with respect to the $\boldsymbol{\theta}$, it suffices to only consider vectors with range exactly $\delta$, and Lemma~\ref{binary.lemma.lfc} applies.
Let $\boldsymbol{\theta}^{(ij)}$ be the minimizer of NCP given by Lemma~\ref{binary.lemma.lfc} when the maximum separation of $\delta$ is in pair \{i,j\} with i<j . Furthermore, if $\delta<\delta^{crit}$ defined in the Lemma \ref{binary.lemma.lfc} , then the $\ell$--th component of $\boldsymbol \theta^{(ij)}$ is given by 
\begin{eqnarray}\label{binary.lfc.eqn.from.lemma}
 \theta_{\ell}^{(ij)}=\begin{cases}
 &\frac12+ \frac {\delta r_{i}\pi_{i}}{G^{
 (ij)}}, \qquad \ell=i\\
 & \frac 12-\frac {\delta r_{j}\pi_{j}}{G^{
 (ij)}}, \qquad \ell =j,\\
 & \frac12+\frac{\delta G^{
 (ij)}(r_{i}\pi_{i}-r_{j}\pi_{j})}{(G^{(ij)})^{2}-4\delta^2r_{i}\pi_{i}r_{j}\pi_{j}}, \qquad \ell\in \mathcal{S}\setminus\{i,j\}.
 \end{cases}
\end{eqnarray}
Here, $G^{(ij)}=r_{i}\pi_{i}+r_{j}\pi_{j}$.\\
\noindent\textbf{Case 1.} The $\boldsymbol{\pi}_{\rm MM}$ in this case is such that all $\pi_\ell^*r_\ell$ are equal and therefore, $\delta<\delta^{crit}=1$. Thus, using ~\eqref{binary.lfc.eqn.from.lemma}, the set of minimizers of NCP is given by $\Theta=\{\boldsymbol{\theta}^{(ij)},\boldsymbol 1-\boldsymbol{\theta}^{(ij)}:i<j, i,j \in \mathcal S \}$, where the $\ell$--th component of $\boldsymbol \theta^{(ij)}$ is given by 
\begin{eqnarray*}
 \theta_{\ell}^{(ij)}=\begin{cases}
 & \frac {(1+\delta)}{2}, \qquad \ell=i\\
 & \frac {(1-\delta)}{2}, \qquad \ell =j,\\
 & \frac12, \qquad \ell\in \mathcal{S}\setminus\{i,j\}.
 \end{cases}
\end{eqnarray*}
which is same as that of the set given in~\eqref{binary.lfc.eqn.1.1}, thus the set of LFC is $\boldsymbol{\Theta}_{\rm LFC}$. Also, for any $\boldsymbol{\theta}^{(ij)}\in \Theta_{\rm LFC}$ , we have 
\[
\phi(\boldsymbol\pi_{\rm MM},\theta^{(ij)})
=
\frac{2m\delta^2}
{(1-\delta^2)A}.
\]
The average NCP for any distribution $\boldsymbol{q}$ on the set of LFCs is 
\begin{eqnarray*}
 \Psi(\boldsymbol{\pi}_{\rm MM}, \boldsymbol{q})= \sum\limits_{0\leq i<j \leq K} q_{ij}\phi(\boldsymbol{\pi}_{\rm MM},\boldsymbol{\theta}^{(ij)})=
\frac{2m\delta^2}
{(1-\delta^2)A}.
\end{eqnarray*}
Since this is a constant, therefore, every proposed
$\boldsymbol q\in\mathcal Q_{\rm LFD}$ is a minimizing response.\\\medskip
\noindent
\textbf{Case 2.}
For each pair $(s,j)$, the relation
$\pi_j^*r_j=z\pi_s^*r_s$ and again it can be verified that $\delta<\delta^{crit}$. Therefore, using ~\eqref{binary.lfc.eqn.from.lemma}, the set of candidate minimizers of NCP is given by $\Theta=\{\boldsymbol{\theta}^{(ij)},\boldsymbol 1-\boldsymbol{\theta}^{(ij)}:i<j, i,j \in \mathcal S \}$, where the $\ell$--th component of $\boldsymbol \theta^{(ij)}$ is given by 

\begin{equation*}
\theta_\ell^{(ij)}
=
\begin{cases}
\displaystyle
\frac{1+\delta}{2},
& \ell=i,\quad s\notin\{i,j\},\\[0.6em]
\displaystyle
\frac{1-\delta}{2},
& \ell=j,\quad s\notin\{i,j\},\\[0.6em]
\displaystyle
\frac12,
& \ell\notin\{i,j\},\quad s\notin\{i,j\},\\[1em]
\displaystyle
\frac12+\frac{\delta}{1+z},
& \ell=i,\quad i=s,\\[0.6em]
\displaystyle
\frac12-\frac{\delta z}{1+z},
& \ell=j,\quad i=s,\\[0.6em]
\displaystyle
\frac12-
\frac{\delta(z^2-1)}{(1+z)^2-4\delta^2z},
& \ell\notin\{i,j\},\quad i=s,\\[1em]
\displaystyle
\frac12+\frac{\delta z}{1+z},
& \ell=i,\quad j=s,\\[0.6em]
\displaystyle
\frac12-\frac{\delta}{1+z},
& \ell=j,\quad j=s,\\[0.6em]
\displaystyle
\frac12+
\frac{\delta(z^2-1)}{(1+z)^2-4\delta^2z},
& \ell\notin\{i,j\},\quad j=s,
\end{cases}
\qquad i<j,\quad i,j\in\mathcal S.
\end{equation*}
Also, $\phi(\boldsymbol\pi_{\rm MM},\boldsymbol\theta^{(sj)}) =\frac{m\delta^2}{(r_s^{-1}+zA_{s})(\gamma_{s}^{(sj)}+\gamma_{j}^{(sj)}/z)}=\frac{m\delta^2r_{s}^{-1}}{(r_s^{-1}+zA_{s})^{2}\gamma_{s}^{(sj)}}=\frac{4m\delta^2r_{s}^{-1}(1+z)^{2}}{(r_s^{-1}+zA_{s})^{2}((1+z)^{2}-4\delta^2)}$. For a pair $i,j\in \mathcal{S}\setminus\{s\}, i<j$,  
\[
\phi(\boldsymbol\pi_{\rm MM},\boldsymbol{\theta}^{(ij)})
=\frac{2m\delta^2z}{(r_s^{-1}+zA_s)(1-\delta^2)}.
\]
Taking the ratio, we obtain
\[
\frac{\phi(\boldsymbol\pi_{\rm MM},\boldsymbol{\theta}^{(sj)})}
{\phi(\boldsymbol\pi_{\rm MM},\boldsymbol{\theta}^{(ij)})}
=
\frac{2(1+z)(1-\delta^2)}
{(1+z)^2-4\delta^2z}
<1.
\]
Thus, only pairs involving arm $s$ attain the global
minimum. Thus the set of LFCs is $\boldsymbol{\Theta}_{\rm LFC}=\left\{
\boldsymbol\theta^{(sj)},\boldsymbol{1}-\boldsymbol\theta^{(sj)},
j\in \mathcal{S}\setminus\{s\}
\right\}$. The average NCP for any distribution $\boldsymbol{q}$ on this set of LFCs is 
\begin{eqnarray*}
 \Psi(\boldsymbol{\pi}_{\rm MM}, \boldsymbol{q})= \sum\limits_{j\neq s} q_{sj}\phi(\boldsymbol{\pi}_{\rm MM},\boldsymbol{\theta}^{(sj)})=\frac{4m\delta^2r_{s}^{-1}(1+z)^{2}}{(r_s^{-1}+zA_{s})^{2}((1+z)^{2}-4\delta^2)}.
\end{eqnarray*}
Since this is a constant, therefore, every proposed
$\boldsymbol q\in\mathcal Q_{\rm LFD}$ is a minimizing response. In either case, we established that, 
\[
\max_{\boldsymbol\pi\in\Pi}
\min_{\boldsymbol q \in Q}
\Psi(\boldsymbol\pi,\boldsymbol q)
=
\min_{\boldsymbol q \in Q}
\max_{\boldsymbol\pi\in\Pi}
\Psi(\boldsymbol\pi,\boldsymbol q)
=\Psi(\boldsymbol\pi_{\rm MM},\boldsymbol q^{*})
\]
where $\boldsymbol q^{*}\in\mathcal Q_{\rm LFD}$. Thus $(\boldsymbol\pi_{\rm MM},\boldsymbol q^{*})$ is the saddle point and therefore, $\boldsymbol\pi_{\rm MM}$ is the required maximin design. This completes the proof.
\end{proof}

\section*{Proof of Theorem~\ref{binary.mm.theorem.joint}}
The proof of Theorem~\ref{binary.mm.theorem.joint} requires following Lemma.
\begin{lemma}\label{binary.mm.root.lemma}
Under the notations established in the statement of theorem~\ref{binary.mm.theorem.joint}, suppose that $e_s^2>1/2$ for some $s\in\mathcal S$. For $x>\Lambda_{s}$, define $\boldsymbol t(x)=(x\boldsymbol I_K-\mathbf H_{ss})^{-1}\boldsymbol h_s,
\quad
\tau_s(x)=h_{ss}+\boldsymbol h_s^{\rm T}\boldsymbol t(x)$, then the equation
\begin{equation*}
\|\boldsymbol t(x)\|_2^2
=
\frac{
\{\tau_s(x)+x\}^2-4\delta^2\tau_s(x)^2
}{
\{\tau_s(x)+x\}^2-4\delta^2x^2
},
\end{equation*}
has exactly one solution $z$ in
$(\Lambda_s,\Lambda)$.
It has no additional solution for $x\geq\Lambda$ at
which its right-hand side is defined.
Moreover,
\[
\tau_s(z)>z>0,
\qquad
0<\|\boldsymbol t(z)\|_2^2<1.
\]
\end{lemma}
\begin{proof}Note that, the matrix $\mathbf H$ is symmetric with strictly positive entries, thus we have $0<\Lambda_s<\Lambda$. Also, $
(\Lambda\boldsymbol I_{K}-\mathbf H_{ss})
\boldsymbol e_{-s}=e_s\boldsymbol h_s,
\quad\mbox{and}\quad
h_{ss}e_s+\boldsymbol h_s^{\rm T}\boldsymbol e_{-s}
=\Lambda e_s$. Consequently,$\boldsymbol t(\Lambda)=\frac{\boldsymbol e_{-s}}{e_s},
\quad
\tau_s(\Lambda)=\Lambda,
\quad\|\boldsymbol t(\Lambda)\|_2^2=\frac{1-e_s^2}{e_s^2}<1.$ For $x>\Lambda_s$, the matrix
$x\boldsymbol I_K-\mathbf H_{ss}$ is positive definite. Define $f(x)=\|\boldsymbol t(x)\|_2^2>0$. Differentiating $(x\boldsymbol I_K-\mathbf H_{ss})\boldsymbol t(x)=\boldsymbol h_s$ with respect x, we get $\boldsymbol t'(x)=
-(x\boldsymbol I_K-\mathbf H_{ss})^{-1}
\boldsymbol t(x)$. Therefore,
\[
\begin{aligned}
f'(x)
&=
-2\boldsymbol t(x)^{\rm T}
(x\boldsymbol I_K-\mathbf H_{ss})^{-1}
\boldsymbol t(x)<0,\\
\tau_s'(x)
&=
-\boldsymbol h_s^{\rm T}
(x\boldsymbol I_K-\mathbf H_{ss})^{-2}
\boldsymbol h_s
=-f(x).
\end{aligned}
\]
Thus, $f$ is strictly decreasing, and
$(\tau_s(x)-x)'=-f(x)-1<0$.
Since $\tau_s(\Lambda)=\Lambda$, it follows that
$\tau_s(x)>x$ for $\Lambda_s<x<\Lambda$
and $\tau_s(x)<x$ for $x>\Lambda$. For $\Lambda_s<x<\Lambda$, let
\[
R(x)=\frac{\tau_s(x)}x>1,
\qquad
g(R)=
\frac{(1+R)^2-4\delta^2R^2}
     {(1+R)^2-4\delta^2}.
\]
The function $R$ is strictly decreasing and $g(R(x))$ is strictly increasing. Let $S(x)=f(x)-g(R(x))$. $S(x)$ is continuous and strictly decreasing over $(\Lambda_s,\Lambda)$ 
with $\lim_{x\downarrow\Lambda_s}S(x)=+\infty,
\; \mbox{and}\;\\
S(\Lambda)=e_s^{-2}(1-2e_s^2)<0.$ The continuity and strict monotonicity therefore imply
that there is exactly one $z\in(\Lambda_s,\Lambda)$
satisfying
\begin{equation}
\label{binary.joint.proof.root}
\|\boldsymbol t(z)\|_2^2
=
\frac{
\{\tau_s(z)+z\}^2-4\delta^2\tau_s(z)^2
}{
\{\tau_s(z)+z\}^2-4\delta^2z^2
}.
\end{equation}
Finally, $z<\Lambda$ implies $\tau_s(z)>z>0$.
Moreover, $R(z)>1$ and the strict decrease of $g$
give
\[
0<\|\boldsymbol t(z)\|_2^2
=g(R(z))<g(1)=1.
\]
This completes the proof.
\end{proof}

We will now proceed with the proof of Theorem~\ref{binary.mm.theorem.joint}.
\begin{proof}
As in the preceding maximin results, the proof proceeds
in two parts: \textbf{Part I} and \textbf{Part II}. In Part I, we hold the proposed LFC set and LFDs set fixed and show that the
average NCP is maximized at the proposed joint design $(\boldsymbol{\pi}_{\rm MM},\boldsymbol{n}_{\rm MM})$.
In Part II, we fix the joint design at $(\boldsymbol{\pi}_{\rm MM},\boldsymbol{n}_{\rm MM})$, and show that the LFC
set is $\Theta_{\rm LFC}$ with associated set of LFDs $\mathcal Q_{\rm LFD}$. In both parts, Case 1 follows from the same way as in Case 1 of the Theorem~\ref{cont.mm.theorem.joint},
with adjustments for the binary response. Therefore, we omit Case 1 calculations here and give the calculations for Case 2 only.

\medskip
\noindent\textbf{Part I:}

\medskip
\noindent\textbf{Case 2.} Let $\boldsymbol{\theta}^{(sj)}\in \Theta_{\rm LFC}$, and its $\ell$--th component is given by 
\[
\theta_\ell^{(sj)}
=
\begin{cases}
\displaystyle
\frac12+\frac{\delta z}{z+\tau_s},
& \ell=s,\\[0.8em]
\displaystyle
\frac12-\frac{\delta\tau_s}{z+\tau_s},
& \ell=j,\\[0.8em]
\displaystyle
\frac12-
\frac{\delta(\tau_s^2-z^2)}
{(z+\tau_s)^2-4\delta^2z\tau_s},
& \ell\notin\{s,j\}.
\end{cases}
\]
Let $\gamma_s^{(sj)}=
\theta_s^{(sj)}(1-\theta_s^{(sj)})
=
\frac14-\frac{\delta^2z^2}{(\tau_s+z)^2},\quad
\gamma_j^{(sj)}=
\theta_j^{(sj)}(1-\theta_j^{(sj)})
 =\frac14-\frac{\delta^2\tau_s^2}{(\tau_s+z)^2}$. 
Although indexed by $j$, these two variance values do not depend on which arm $j\neq s$ is selected. Dividing the two expressions and using
\eqref{binary.joint.proof.root} gives
\begin{equation}
\label{binary.joint.proof.variance.balance}
\sum_{\ell\neq s}v_\ell^2
=
\frac{\gamma_j^{(sj)}}{\gamma_s^{(sj)}}.
\end{equation}
Let $
T=\tau_s\gamma_s^{(sj)}
+
z\gamma_j^{(sj)}=
\frac{(\tau_s+z)^2-4\delta^2\tau_s z}
 {4(\tau_s+z)}
>0$ and $c=
\frac{
\theta_s^{(sj)}/(\tau_s\gamma_s^{(sj)})
+
\theta_j^{(sj)}/(z\gamma_j^{(sj)})
}{
1/(\tau_s\gamma_s^{(sj)})
+
1/(z\gamma_j^{(sj)})
}=
\frac12-
\frac{\delta(\tau_s^2-z^2)}
 {(\tau_s+z)^2-4\delta^2\tau_s z}$. Thus we have, $
\theta_s^{(sj)}-c=
\frac{\delta\tau_s\gamma_s^{(sj)}}{T},\quad\mbox{and}\quad
\theta_j^{(sj)}-c=
-\frac{\delta z\gamma_j^{(sj)}}{T}$.
 Also, $\theta_\ell^{(sj)}-c=0,\,\ell \notin\{s,j\}$. Therefore,~\eqref{binary.ncp.upperbound} gives
\begin{align*}
\phi(\boldsymbol\pi,\boldsymbol\theta^{(sj)},
 m,\boldsymbol n,\boldsymbol\rho)&\leq
m\left\{
\frac{\pi_s r_s}{\gamma_s^{(sj)}}
\left(
\frac{\delta\tau_s\gamma_s^{(sj)}}{T}
\right)^2
+
\frac{\pi_j r_j}{\gamma_j^{(sj)}}
\left(
-\frac{\delta z\gamma_j^{(sj)}}{T}
\right)^2
\right\}\\&=
\frac{m\delta^2}{T^2}
\left\{
\tau_s^2\gamma_s^{(sj)}\pi_s r_s
+
z^2\gamma_j^{(sj)}\pi_j r_j
\right\}.
\end{align*}
For any $\boldsymbol q \in \mathcal Q_{\rm LFD}$, we have $q_{sj}^*
=
\frac{v_j^2}{\sum_{h\neq s}v_h^2}=v_j^2
\frac{\gamma_s^{(sj)}}{\gamma_j^{(sj)}}$.
 Therefore, the average NCP function is given by 
\begin{eqnarray}
\begin{aligned}
\Psi(\boldsymbol\pi,\boldsymbol n,\boldsymbol q^*)=
\sum_{j\neq s}q_{sj}^*
\phi(\boldsymbol\pi,\boldsymbol\theta^{(sj)},
 m,\boldsymbol n,\boldsymbol\rho)&\leq
\frac{m\delta^2}{T^2}
\left\{
\tau_s^2\gamma_s^{(sj)}\pi_s r_s
\sum_{j\neq s}q_{sj}^*
+
z^2\gamma_j^{(sj)}
\sum_{j\neq s}q_{sj}^*\pi_j r_j
\right\}\\&=
\frac{m\delta^2\gamma_s^{(sj)}}{T^2}
\left\{
\tau_s^2\pi_s r_s
+
z^2\sum_{j\neq s}v_j^2\pi_j r_j
\right\}.\label{binary.joint.proof.average.firstbound}
\end{aligned}
\end{eqnarray}
The defining equation for $\boldsymbol t$ gives $\mathbf H_{ss}\boldsymbol t+\boldsymbol h_s
=z\boldsymbol t.$ Consequently,
\begin{equation}
\label{binary.joint.proof.Hv}
(\mathbf H\boldsymbol v)_\ell
=
\sqrt{\rho_\ell}\,U_3
+
\frac{\sqrt{1-\rho_\ell}}{\eta}\,U_4
=
\begin{cases}
\tau_s, & \ell=s,\\
zv_\ell, & \ell\neq s.
\end{cases}
\end{equation}
For every $n_\ell>0$, Cauchy--Schwarz gives
\begin{align*}
\{(\mathbf H\boldsymbol v)_\ell\}^2=
\left(
\sqrt{\rho_\ell}\,U_3
+
\sqrt{\frac{1-\rho_\ell}{n_\ell}}\,
\frac{\sqrt{n_\ell}\,U_4}{\eta}
\right)^2\leq
\left(
\rho_\ell+\frac{1-\rho_\ell}{n_\ell}
\right)
\left(
U_3^2+\frac{n_\ell U_4^2}{\eta^2}
\right)=
\frac1{r_\ell}
\left(
U_3^2+\frac{n_\ell U_4^2}{\eta^2}
\right).
\end{align*}
Multiplying by $\pi_\ell r_\ell$ and summing gives
\begin{align*}
\tau_s^2\pi_s r_s
+
z^2\sum_{j\neq s}v_j^2\pi_j r_j=
\sum_{\ell=0}^{K}
\pi_\ell r_\ell\{(\mathbf H\boldsymbol v)_\ell\}^2 & \leq
U_3^2\sum_{\ell=0}^{K}\pi_\ell
+
\frac{U_4^2}{\eta^2}
\sum_{\ell=0}^{K}\pi_\ell n_\ell\\&=
U_3^2+\frac{U_4^2}{\eta}
=
\boldsymbol v^{\rm T}\mathbf H\boldsymbol v=
\tau_s+z\sum_{j\neq s}v_j^2=
\tau_s+
z\frac{\gamma_j^{(sj)}}{\gamma_s^{(sj)}}\\
 &=
\frac{\tau_s\gamma_s^{(sj)}
 +z\gamma_j^{(sj)}}{\gamma_s^{(sj)}}=
\frac{T}{\gamma_s^{(sj)}}.
\end{align*}
Using this in 
\eqref{binary.joint.proof.average.firstbound}, we obtain
\begin{equation}
\label{eq:average_boundB}
\Psi(\boldsymbol\pi,\boldsymbol n,\boldsymbol q^*)
\leq
\frac{m\delta^2\gamma_s^{(sj)}}{T^2}
\frac{T}{\gamma_s^{(sj)}}
=
\frac{m\delta^2}{T}.
\end{equation}
Now, for the proposed joint design $(\boldsymbol{\pi}_{\rm MM},\boldsymbol{n}_{\rm MM})$, we have
\[
\pi_\ell
=
\frac{\sqrt{\rho_\ell}\,v_\ell}{U_3},
\qquad
n_\ell
=
\frac{\eta U_3}{U_4}
\sqrt{\frac{1-\rho_\ell}{\rho_\ell}}.
\]
Therefore,
\begin{equation}
\label{binary.joint.proof.information}
\pi_sr_s=\frac1{\tau_s},
\qquad
\pi_jr_j=\frac1z,
\qquad j\neq s.
\end{equation}
Substitution into the original NCP gives
\begin{align*}
\phi(\boldsymbol\pi_{\rm MM},\boldsymbol\theta^{(sj)},
 m,\boldsymbol n_{\rm MM},\boldsymbol\rho)&=
m\left\{
\frac1{\tau_s\gamma_s^{(sj)}}
\left(
\frac{\delta\tau_s\gamma_s^{(sj)}}{T}
\right)^2
+
\frac1{z\gamma_j^{(sj)}}
\left(
-\frac{\delta z\gamma_j^{(sj)}}{T}
\right)^2
\right\}\\&=
\frac{m\delta^2}{T^2}
\left(
\tau_s\gamma_s^{(sj)}
+
z\gamma_j^{(sj)}
\right)=
\frac{m\delta^2}{T}.
\end{align*}
Also $\Psi(\boldsymbol\pi_{\rm MM},\boldsymbol n_{\rm MM},\boldsymbol q^{*})=
\frac{m\delta^2}{T}$. Thus the proposed design attains the upper bound and is therefore, the global maximizer of the average NCP. 

\medskip
\noindent\textbf{Part II:} In this part, we fix the joint design at $
(\boldsymbol\pi_{\rm MM},\boldsymbol n_{\rm MM})$ and show that the set of LFCs is $\Theta_{\rm LFC}$ with associated set of LFDs as $\mathcal Q_{\rm LFD}$. Let $\boldsymbol{\theta}^{(ij)}$ be the minimizer of NCP given by Lemma~\ref{binary.lemma.lfc} when the maximum separation of $\delta$ is in pair \{i,j\} with i<j . Furthermore, if $\delta<\delta^{crit}$ defined in the Lemma \ref{binary.lemma.lfc} , then the $\ell$--th component of $\boldsymbol \theta^{(ij)}$ is given by 
\begin{equation}
\label{binary.lfc.eqn.from.lemma.joint}
\theta_\ell^{(ij)}
=
\begin{cases}
\displaystyle
\frac12+
\frac{\delta\pi_i^*r_i^*}{G^{(ij)}},
& \ell=i,\\[0.9em]
\displaystyle
\frac12-
\frac{\delta\pi_j^*r_j^*}{G^{(ij)}},
& \ell=j,\\[0.9em]
\displaystyle
\frac12+
\frac{
\delta G^{(ij)}
(\pi_i^*r_i^*-\pi_j^*r_j^*)
}{
(G^{(ij)})^2
-
4\delta^2\pi_i^*r_i^*\pi_j^*r_j^*
},
& \ell\notin\{i,j\}.
\end{cases}
\end{equation}
Here, $ G^{(ij)}=\pi_ir_i+\pi_jr_j$.

\medskip
\noindent\textbf{Case 2.} For the joint design $
(\boldsymbol\pi_{\rm MM},\boldsymbol n_{\rm MM})$ given in~\eqref{binary.mm.eqn.joint.case2}, we have $\pi_s^*r_s^*=\frac1{\tau_s},
\qquad
\pi_j^*r_j^*=\frac1z,
\quad j\neq s$,with $\tau_s>z$. Therefore, it can be easily verifed that $\delta_{sj}^{\rm crit}<1 \quad \mbox{and}\quad \delta
<\delta_{sj}^{\rm crit}$. Therefore, using ~\eqref{binary.lfc.eqn.from.lemma}, the set of minimizers of NCP is given by $\Theta=\{\boldsymbol{\theta}^{(ij)},\boldsymbol 1-\boldsymbol{\theta}^{(ij)}:i<j, i,j \in \mathcal S \}$, where the $\ell$--th component of $\boldsymbol \theta^{(ij)}$ is given by 

\begin{equation*}
\theta_\ell^{(ij)}
=
\begin{cases}
\displaystyle
\frac{1+\delta}{2},
& \ell=i,\quad s\notin\{i,j\},\\[0.6em]
\displaystyle
\frac{1-\delta}{2},
& \ell=j,\quad s\notin\{i,j\},\\[0.6em]
\displaystyle
\frac12,
& \ell\notin\{i,j\},\quad s\notin\{i,j\},\\[1em]
\displaystyle
\frac12+\frac{\delta z}{\tau_{s}+z},
& \ell=i,\quad i=s,\\[0.6em]
\displaystyle
\frac12-\frac{\delta \tau_{s}}{\tau_{s}+z},
& \ell=j,\quad i=s,\\[0.6em]
\displaystyle
\frac12-
\frac{\delta(\tau_{s}^{2}-z^2)}{(z+\tau_{s})^2-4\delta^2\tau_{s}z},
& \ell\notin\{i,j\},\quad i=s,\\[1em]
\displaystyle
\frac12+\frac{\delta \tau_{s}}{\tau_{s}+z},
& \ell=i,\quad j=s,\\[0.6em]
\displaystyle
\frac12-\frac{\delta z}{\tau_{s}+z},
& \ell=j,\quad j=s,\\[0.6em]
\displaystyle
\frac12+
\frac{\delta(\tau_{s}^{2}-z^2)}{(z+\tau_{s})^2-4\delta^2\tau_{s}z},
& \ell\notin\{i,j\},\quad j=s,
\end{cases}
\qquad i<j,\quad i,j\in\mathcal S.
\end{equation*}
For any $\boldsymbol \theta^{(sj)}$, we have 
\begin{align*}
\phi(\boldsymbol\pi_{\rm MM},\boldsymbol\theta^{(sj)},
 m,\boldsymbol n_{\rm MM},\boldsymbol\rho)=
\frac{m\delta^2}{
\gamma_s^{(sj)}/(1/\tau_s)
+
\gamma_j^{(sj)}/(1/z)
}=
\frac{m\delta^2}{
\tau_s\gamma_s^{(sj)}+z\gamma_j^{(sj)}
}=
\frac{m\delta^2}{T}.
\end{align*}
Also, for any $\boldsymbol \theta^{(ij)},s\notin\{i,j\}$, we have 
\begin{align*}
\phi(\boldsymbol\pi_{\rm MM},\boldsymbol\theta^{(ij)},
 m,\boldsymbol n_{\rm MM},\boldsymbol\rho)=
\frac{m\delta^2}{
z(1-\delta^2)/4+z(1-\delta^2)/4
}=
\frac{m\delta^2}{z(1-\delta^2)/2}.
\end{align*}
Comparing this denominator with T, we have 
\begin{align*}
T-\frac{z(1-\delta^2)}2=
\frac{\tau_s+z}{4}
-
\frac{\delta^2\tau_s z}{\tau_s+z}
-
\frac z2+\frac{\delta^2z}{2}=
\frac{\tau_s-z}{4}
-
\frac{\delta^2z(\tau_s-z)}{2(\tau_s+z)}=
\frac{
(\tau_s-z)(\tau_s+z-2\delta^2z)
}{
4(\tau_s+z)
}
>0.
\end{align*}
The strict inequality follows from $\tau_s>z$ and $\tau_s+z-2\delta^2z
=
(\tau_s-z)+2z(1-\delta^2)>0$. Therefore, $\phi(\boldsymbol\pi_{\rm MM},\boldsymbol\theta^{(sj)},
 m,\boldsymbol n_{\rm MM},\boldsymbol\rho)<\phi(\boldsymbol\pi_{\rm MM},\boldsymbol\theta^{(ij)},
 m,\boldsymbol n_{\rm MM},\boldsymbol\rho)$. Thus, only pairs involving arm $s$ attain the global minimum and the set of LFCs is $\boldsymbol{\Theta}_{\rm LFC}=\left\{
\boldsymbol\theta^{(sj)},\boldsymbol{1}-\boldsymbol\theta^{(sj)},
j\in \mathcal{S}\setminus\{s\}
\right\}$. The average NCP for any distribution $\boldsymbol{q}$ on this set of LFCs is 
\begin{eqnarray*}
 \Psi(\boldsymbol{\pi}_{\rm MM},\boldsymbol n_{\rm MM}, \boldsymbol{q})= \sum\limits_{j\neq s} q_{sj}\phi(\boldsymbol{\pi}_{\rm MM},\boldsymbol{\theta}^{(sj)},m,\boldsymbol n_{\rm MM},\boldsymbol \rho)=\frac{m\delta^2}{T}.
\end{eqnarray*}
Since this is a constant, therefore, every proposed
$\boldsymbol q\in\mathcal Q_{\rm LFD}$ is a minimizing response. In either case, we established that, 
\[
\max_{(\boldsymbol\pi,\boldsymbol n)\in\Xi}
\min_{\boldsymbol q \in Q}
\Psi(\boldsymbol\pi,\boldsymbol n,\boldsymbol q)
=
\min_{\boldsymbol q \in Q}
\max_{(\boldsymbol\pi,\boldsymbol n)\in\Xi}
\Psi(\boldsymbol\pi,\boldsymbol n,\boldsymbol q)
=\Psi(\boldsymbol\pi_{\rm MM},\boldsymbol n_{\rm MM},\boldsymbol q^{*})
\]
where $\boldsymbol q^{*}\in\mathcal Q_{\rm LFD}$ and $\mathcal Q $ is any probability distribution on $ \Theta_{\rm LFC}$. Thus $(\boldsymbol\pi_{\rm MM},\boldsymbol n_{\rm MM},\boldsymbol q^{*})$ is the saddle point and therefore, $(\boldsymbol\pi_{\rm MM},\boldsymbol n_{\rm MM})$ is the required maximin design. This completes the proof.
\end{proof}

\section*{Algorithms and Examples}
\refstepcounter{algorithm}
\noindent\textbf{Algorithm \thealgorithm :}\label{cont.lfd.construction}
The LFDs in \eqref{cont.lfd.eqn.1.case1} need not be unique for $K\ge 3$. We outline an approach to construct one such LFD satisfying the conditions of \eqref{cont.lfd.eqn.1.case1}. Let $\boldsymbol\pi_{\rm MM}$ be the maximin allocation given by~\eqref{cont.mm.eqn.1.case1}. Since, every $\pi_\ell^*\leq1/2$. 
We define the cumulative sums 
\[
B_{-1}=0,
\qquad
B_\ell=\sum_{h=0}^{\ell}\pi_h^*,
\qquad \ell=0,\ldots,K.
\]
In particular, $B_K=1$.
One valid LFD satisfying the conditions of~\eqref{cont.lfd.eqn.1.case1} is given by
\[
q_{ij}
=
2\max\{
\min\{B_i,B_j-1/2\}
-
\max\{B_{i-1},B_{j-1}-1/2\},0\},
\qquad 0\leq i<j\leq K.
\]

This construction supplies one valid
distribution directly, without numerical optimization.

Next, we provide Algorithm \ref{algorithm.1} and Algorithm \ref{algorithm.2} to determine the required number of clusters and the corresponding total number of observations to achieve a prespecified power.
\vspace{0.5cm}

\refstepcounter{algorithm}
\noindent\textbf{Algorithm \thealgorithm:} 
\label{algorithm.1}

\begin{algorithmic}[1]
 \State Fix the target power $1-\beta$.
 
 \State Obtain the optimal-design $\boldsymbol{\pi}^{*}$ and compute
 $\phi_{1}(\boldsymbol{\pi}^{*},\boldsymbol{\theta},
 \boldsymbol{n},\boldsymbol{\rho})$.
 
 \State Set $\operatorname{Power}=1-\beta$ in the left-hand side of
 \eqref{power.eqn} and solve for $\phi$, denoted by $\phi^{*}$,
 such that
 \[
 1-F_{\chi^2_K(\phi^{*})}
 \left(\chi^2_{K,1-\alpha}\right)=1-\beta.
 \]
 
 \State Since $\phi=m\phi_{1}$, the required number of clusters is
 \[
 m^{*}
 =
 \left\lceil
 \frac{\phi^{*}}
 {\phi_{1}(\boldsymbol{\pi}^{*},\boldsymbol{\theta},
 \boldsymbol{n},\boldsymbol{\rho})}
 \right\rceil.
 \]
 where $\lceil\cdot\rceil$ denotes the ceiling function.
\end{algorithmic}

\vspace{0.5cm}

\refstepcounter{algorithm}
\noindent\textbf{Algorithm \thealgorithm:}
\label{algorithm.2}

\begin{algorithmic}[1]
 \State Fix the target power $1-\beta$ and a feasible value of $\eta$.
 
 \State Obtain the optimal-design $\boldsymbol{\pi}^{*}$ and
 $\boldsymbol{n}^{*}$ in the restricted space $\Xi$ and compute
 \[
 \phi_{1}^{*}(\eta)
 =
 \phi_{1}(\boldsymbol{\pi}^{*},\boldsymbol{\theta},
 \boldsymbol{n}^{*},\boldsymbol{\rho}).
 \]
 
 \State Set $\operatorname{Power}=1-\beta$ in the left-hand side of
 \eqref{power.eqn} and solve for $\phi$, denoted by $\phi^{*}$,
 such that
 \[
 1-F_{\chi^2_K(\phi^{*})}
 \left(\chi^2_{K,1-\alpha}\right)=1-\beta.
 \]
 
 \State Since $\phi=m\phi_{1}^{*}(\eta)$, the required number of
 clusters for the specified value of $\eta$ is
 \[
 m^{*}(\eta)
 =
 \left\lceil
 \frac{\phi^{*}}{\phi_{1}^{*}(\eta)}
 \right\rceil.
 \]
 
 \State The corresponding total number of observations is
 \[
 N^{*}(\eta)=m^{*}(\eta)\eta.
 \]
 
 \State If the objective is to minimize the total number of
 observations, repeat Steps 1--5 over the feasible values of
 $\eta$ and select the value of $\eta$ that minimizes
 $N^{*}(\eta)$.
\end{algorithmic}
\begin{example}\label{algorithm.cont.sample.size.example} 
In this example, we demonstrate the use of Algorithm ~\ref{algorithm.1} to calculate the number of clusters as well as the optimal maximin allocation scheme to achieve a target power of 0.8 for testing the hypothesis \eqref{hyp.1} in case of a normal response. For implementing Algorithm ~\ref{algorithm.1}, both $\boldsymbol{n}$ and $\boldsymbol{\rho}$ are fixed in advance. Suppose that $K=2,$ $\sigma^2=1,$ $\boldsymbol{\rho}=(0.07,0.004,0.1)$ and $\boldsymbol{n}=(50,50,50)$. Given these values, and taking $\delta=0.1$, we obtained the MM--design for the optimal cluster proportions as $\boldsymbol{\pi}_{\rm MM}=(0.4,0.1,0.5)$.
Now using $\operatorname{Power}=0.8$ on the left-hand side of \eqref{power.eqn}, and solving for $\phi$ gives the optimal value as $\phi^{*}=9.634689$. Let's suppose $\boldsymbol{\theta}=(0.1,0.15,0.2)$, then $\phi_{1}(\boldsymbol{\pi}_{\rm MM},\boldsymbol{\theta},\boldsymbol{n},\boldsymbol{\rho})= 0.02169419$. Therefore, the number of clusters $'m'$ needed to achieve a power of 0.8 through the maximin allocation is $m=\left\lceil9.634689/0.02169419\right\rceil=445$ 
\end{example}

\begin{example}\label{cont.mm.example.joint}
To illustrate Theorem~\ref{cont.mm.theorem.joint}, consider
a hypothetical CRT with four treatment arms ($K=3$) and an
exchangeable working correlation structure. Both the
cluster allocation proportions and cluster sizes are
optimized, with $\eta=25$ and $\delta=0.1$.
All calculations use unrounded quantities; only the
displayed results are rounded.

\medskip
\noindent\textbf{Case 1 illustration.}
Let $\boldsymbol\rho=(0.04,0.05,0.03,0.06)^{\rm T}$. The largest eigenvalue $\Lambda$ of the matrix $\mathbf{H}$ is approximately
$ 0.33$, and with its positive unit
eigenvector as $\boldsymbol e=
(0.486,0.515,0.453,0.541)^{\rm T}$ and $(e_0^2,e_1^2,e_2^2,e_3^2)
=
(0.236,0.265,0.205,0.293)$.
Since every $e_\ell^2<1/2$, Case 1 applies.
The sums in the theorem are
$U_1=0.423$ and $U_2=1.949$.
Using \eqref{cont.mm.eqn.joint.case1} gives
\[
\begin{aligned}
\boldsymbol\pi_{\rm MM}
&=
(0.23,0.272,0.185,0.313)^{\rm T},\\
\boldsymbol n_{\rm MM}
&=
(26.6,23.7,30.9,21.5)^{\rm T}.
\end{aligned}
\]
The set of LFCs is 
\[\Theta_{\rm LFC}
=
\left\{\boldsymbol\theta^{(ij)},\ -\boldsymbol\theta^{(ij)}:
,\quad 0\leq i<j\leq3
\right\}.\] This set consists of all permutation of the vector $(0.05,-0.05,0,0)^{\rm T}$. 
To obtain an LFD, the pair probabilities must satisfy
\[
\begin{aligned}
q_{01}+q_{02}+q_{03}&=2e_0^2,\;
q_{01}+q_{12}+q_{13}&=2e_1^2,\\
q_{02}+q_{12}+q_{23}&=2e_2^2,\;
q_{03}+q_{13}+q_{23}&=2e_3^2.
\end{aligned}
\]
There are six pair probabilities and four independent
incidence equations, so the solution need not be unique.
For these inputs, taking $q_{12}=q_{23}=0$ gives the
nonnegative solution
\[
\begin{aligned}
q_{01}&=2(e_0^2+e_1^2)-1,
&
q_{02}&=2e_2^2,\\
q_{03}&=1-2(e_1^2+e_2^2),
&
q_{13}&=1-2e_0^2.
\end{aligned}
\]
Thus, one least favorable pair distribution is
\[
\boldsymbol q
=
(0.0035,\,
 0.4107,\,
 0.0586,\,
 0,\,
 0.5273,\,
 0)^{\rm T}.
\]

\medskip
\noindent\textbf{Case 2 illustration.}
Now let $\boldsymbol\rho=(0.60,0.02,0.03,0.04)^{\rm T}$,
keeping $\eta=25$ and $\delta=0.1$ unchanged. Then the largest eigenvalue $\Lambda$ of matrix $\mathbf H$ is
$0.754$, and $\boldsymbol e=
(0.894,0.225,0.259,0.288)^{\rm T},$ with $(e_0^2,e_1^2,e_2^2,e_3^2)
=
(0.799,0.05,0.067,0.083)$.
Since $e_0^2>1/2$, Case 2 applies with $s=0$. The largest eigenvalue of $\mathbf H_{00}$ is
$\Lambda_{0}=0.205$.
Solving $\left\|
(z\boldsymbol I_3-\mathbf H_{00})^{-1}
\boldsymbol h_{0}
\right\|_2=1 $ gives $z=0.480,
\qquad
\boldsymbol t=
(0.508,0.579,0.638)^{\rm T}$. Hence $\boldsymbol v=(1,t_1,t_2,t_3)^{\rm T},
\qquad
\tau_0
=
h_{00}+\boldsymbol h_{0}^{\rm T}\boldsymbol t
=0.891>z.$
The sums are $U_3=1.074$ and
$U_4=2.33$.
Using \eqref{cont.mm.eqn.joint.case2} gives
\[
\begin{aligned}
\boldsymbol\pi_{\rm MM}
&=
(0.721,0.067,0.093,0.119)^{\rm T},\;
\boldsymbol n_{\rm MM}
=
(9.4,80.7,65.5,56.5)^{\rm T}.
\end{aligned}
\]
The set is of LFC is 
\[
\Theta_{\rm LFC}
=
\left\{\boldsymbol\theta^{(0j)},-\boldsymbol\theta^{(0j)}: \quad j=1,2,3
\right\}, 
\]
where 
\[
\begin{aligned}
\boldsymbol\theta^{(01)}
&=(0.05,-0.05,c,c)^{\rm T},\;
\boldsymbol\theta^{(02)}
&=(0.05,c,-0.05,c)^{\rm T},\;
\boldsymbol\theta^{(03)}
&=(0.05,c,c,-0.05)^{\rm T}.
\end{aligned}
\]
Here, \[
c=
\frac{\delta}{2}\frac{z-\tau_0}{z+\tau_0}
=-0.015.
\]
The least favorable pair probabilities are uniquely given by
\[
(q_{01},q_{02},q_{03})
=
(t_1^2,t_2^2,t_3^2)
=
(0.2586,0.3348,0.4065).
\]
All other pair probabilities are zero.
For the LFC matrix
\[
\left[
\boldsymbol\theta^{(01)},
\boldsymbol\theta^{(02)},
\boldsymbol\theta^{(03)},
-\boldsymbol\theta^{(01)},
-\boldsymbol\theta^{(02)},
-\boldsymbol\theta^{(03)}
\right],
\]
one symmetric LFD is approximately
\[
(0.1293,\,
 0.1674,\,
 0.2033,\,
 0.1293,\,
 0.1674,\,
 0.2033)^{\rm T}.
\]
\end{example}

\end{document}